%%%%%%%%%%%%%%%%%%     PLAIN TEX FILE
%%%%%%%%%%%%%%%%%%
 %%%%%%%%%%%%%%%%%%  %%%%%%%%%%%%%%%%%%  %%%%%%%%%%%%%%%%%%  %%%%%%%%%%%%%%%%%%
 %%%%%%%%%%%%%%%%%%  %%%%%%%%%%%%%%%%%%  %%%%%%%%%%%%%%%%%%  %%%%%%%%%%%%%%%%%%
 %%%%%%%%%%%%%%%%%%  %%%%%%%%%%%%%%%%%%  %%%%%%%%%%%%%%%%%%  %%%%%%%%%%%%%%%%%%

 %%%%%%%%%%%%%%%%%%  hex macros for preprints, cm version %%%%%%%%%%%%%%
%                     (P. Ginsparg, last updated 9/91)
%                if confused, type `b' in response to query
%
%---------------------------------------------------------------------%
%% site dependent options:
%% \unredoffs and \redoffs define horizontal and vertical offsets
%% respectively for unreduced and reduced modes. \speclscape defines
%% the \special{} call that sets printer to landscape (sideways) mode.
%% from standard set below, leave uncommented as appropriate or redefine
%
%%% next 400dpi
%\def\unredoffs{} \def\redoffs{\voffset=-.31truein\hoffset=-.48truein}
%\def\speclscape{\special{landscape}}
%
%%% apple lw
\def\unredoffs{} \def\redoffs{\voffset=-.31truein\hoffset=-.59truein}
\def\speclscape{\special{ps: landscape}}
%
%%% qms lasergrafix:
%\def\unredoffs{} \def\redoffs{\voffset=-.4truein\hoffset=.125truein}
%\def\speclscape{\special{qms: landscape}}
%
%%% saclay A4 paper:
%\def\unredoffs{\hoffset-.14truein\voffset-.2truein}
%\def\redoffs{\voffset=-.55truein\hoffset=-.1truein} \def\speclscape{}
%
%---------------------------------------------------------------------%
%
\newbox\leftpage \newdimen\fullhsize \newdimen\hstitle \newdimen\hsbody
\tolerance=1000\hfuzz=2pt
\catcode`\@=11 % This allows us to modify PLAIN macros.
\def\bigans{b }
%\message{ big or little (b/l)? }\read-1 to\answ
\def\answ{b }
\ifx\answ\bigans\message{(This will come out unreduced.}
\magnification=1200\unredoffs\baselineskip=16pt plus 2pt minus 1pt
\hsbody=\hsize \hstitle=\hsize %take default values for unreduced format
\else\message{(This will be reduced.} \let\l@r=L
\magnification=1000\baselineskip=16pt plus 2pt minus 1pt \vsize=7truein
\redoffs \hstitle=8truein\hsbody=4.75truein\fullhsize=10truein\hsize=\hsbody
\output={\ifnum\pageno=0 %%% This is the HUTP version
  \shipout\vbox{\speclscape{\hsize\fullhsize\makeheadline}
    \hbox to \fullhsize{\hfill\pagebody\hfill}}\advancepageno
  \else
  \almostshipout{\leftline{\vbox{\pagebody\makefootline}}}\advancepageno
  \fi}
\def\almostshipout#1{\if L\l@r \count1=1 \message{[\the\count0.\the\count1]}
      \global\setbox\leftpage=#1 \global\let\l@r=R
 \else \count1=2
  \shipout\vbox{\speclscape{\hsize\fullhsize\makeheadline}
      \hbox to\fullhsize{\box\leftpage\hfil#1}}  \global\let\l@r=L\fi}
\fi
%---------------------------------------------------------------------
%
\newcount\yearltd\yearltd=\year\advance\yearltd by -1900

\def\Title#1#2{\nopagenumbers\abstractfont\hsize=\hstitle\rightline{#1}%
\vskip 1in\centerline{\titlefont #2}\abstractfont\vskip .5in\pageno=0}
\def\Date#1{\vfill\leftline{#1}\tenpoint\supereject\global\hsize=\hsbody%
\footline={\hss\tenrm\folio\hss}}%      restores pagenumbers
%
%       use following instead of \Date on the preliminary draft,
%       puts date/time on each page in big mode, writes labels in margins

\def\draftmode{\message{ DRAFTMODE }\def\draftdate{{\rm preliminary draft:
\number\month/\number\day/\number\yearltd\ \ \hourmin}}%
\headline={\hfil\draftdate}\writelabels\baselineskip=20pt plus 2pt minus 2pt
 {\count255=\time\divide\count255 by 60 \xdef\hourmin{\number\count255}
  \multiply\count255 by-60\advance\count255 by\time
  \xdef\hourmin{\hourmin:\ifnum\count255<10 0\fi\the\count255}}}
%       use \nolabels to get rid of eqn, ref, and fig labels in draft mode
\def\nolabels{\def\wrlabeL##1{}\def\eqlabeL##1{}\def\reflabeL##1{}}
\def\writelabels{\def\wrlabeL##1{\leavevmode\vadjust{\rlap{\smash%
{\line{{\escapechar=` \hfill\rlap{\sevenrm\hskip.03in\string##1}}}}}}}%
\def\eqlabeL##1{{\escapechar-1\rlap{\sevenrm\hskip.05in\string##1}}}%
\def\reflabeL##1{\noexpand\llap{\noexpand\sevenrm\string\string\string##1}}}
\nolabels
%
% tagged sec numbers
\global\newcount\secno \global\secno=0
\global\newcount\meqno \global\meqno=1
\def\newsec#1{\global\advance\secno by1\message{(\the\secno. #1)}
%\ifx\answ\bigans \vfill\eject \else \bigbreak\bigskip \fi  %if desired
\global\subsecno=0\eqnres@t\noindent{\bf\the\secno. #1}
\writetoca{{\secsym} {#1}}\par\nobreak\medskip\nobreak}
\def\eqnres@t{\xdef\secsym{\the\secno.}\global\meqno=1\bigbreak\bigskip}
\def\sequentialequations{\def\eqnres@t{\bigbreak}}\xdef\secsym{}
\global\newcount\subsecno \global\subsecno=0
\def\subsec#1{\global\advance\subsecno by1\message{(\secsym\the\subsecno.
#1)}
\ifnum\lastpenalty>9000\else\bigbreak\fi
\noindent{\it\secsym\the\subsecno. #1}\writetoca{\string\quad
{\secsym\the\subsecno.} {#1}}\par\nobreak\medskip\nobreak}
\def\appendix#1#2{\global\meqno=1\global\subsecno=0\xdef\secsym{\hbox{#1.}}
\bigbreak\bigskip\noindent{\bf Appendix #1. #2}\message{(#1. #2)}
\writetoca{Appendix {#1.} {#2}}\par\nobreak\medskip\nobreak}
%
%       \eqn\label{a+b=c}       gives displayed equation, numbered
%                               consecutively within sections.
%     \eqnn and \eqna define labels in advance (of eqalign?)
%
\def\eqnn#1{\xdef #1{(\secsym\the\meqno)}\writedef{#1\leftbracket#1}%
\global\advance\meqno by1\wrlabeL#1}
\def\eqna#1{\xdef #1##1{\hbox{$(\secsym\the\meqno##1)$}}
\writedef{#1\numbersign1\leftbracket#1{\numbersign1}}%
\global\advance\meqno by1\wrlabeL{#1$\{\}$}}
\def\eqn#1#2{\xdef #1{(\secsym\the\meqno)}\writedef{#1\leftbracket#1}%
\global\advance\meqno by1$$#2\eqno#1\eqlabeL#1$$}
%
%                            footnotes
\newskip\footskip\footskip14pt plus 1pt minus 1pt %sets footnote baselineskip
\def\footnotefont{\ninepoint}\def\f@t#1{\footnotefont #1\@foot}
\def\f@@t{\baselineskip\footskip\bgroup\footnotefont\aftergroup\@foot\let\next}
\setbox\strutbox=\hbox{\vrule height9.5pt depth4.5pt width0pt}
\global\newcount\ftno \global\ftno=0
\def\foot{\global\advance\ftno by1\footnote{$^{\the\ftno}$}}
%
%say \footend to put footnotes at end
%will cause problems if \ref used inside \foot, instead use \nref before
\newwrite\ftfile
\def\footend{\def\foot{\global\advance\ftno by1\chardef\wfile=\ftfile
$^{\the\ftno}$\ifnum\ftno=1\immediate\openout\ftfile=foots.tmp\fi%
\immediate\write\ftfile{\noexpand\smallskip%
\noexpand\item{f\the\ftno:\ }\pctsign}\findarg}%
\def\footatend{\vfill\eject\immediate\closeout\ftfile{\parindent=20pt
\centerline{\bf Footnotes}\nobreak\bigskip\input foots.tmp }}}
\def\footatend{}
%
%     \ref\label{text}
% generates a number, assigns it to \label, generates an entry.
% To list the refs on a separate page,  \listrefs
%
\global\newcount\refno \global\refno=1
\newwrite\rfile
\def\ref{[\the\refno]\nref}
\def\nref#1{\xdef#1{[\the\refno]}\writedef{#1\leftbracket#1}%
\ifnum\refno=1\immediate\openout\rfile=refs.tmp\fi
\global\advance\refno by1\chardef\wfile=\rfile\immediate
\write\rfile{\noexpand\item{#1\ }\reflabeL{#1\hskip.31in}\pctsign}\findarg}
%        horrible hack to sidestep tex \write limitation
\def\findarg#1#{\begingroup\obeylines\newlinechar=`\^^M\pass@rg}
{\obeylines\gdef\pass@rg#1{\writ@line\relax #1^^M\hbox{}^^M}%
\gdef\writ@line#1^^M{\expandafter\toks0\expandafter{\striprel@x #1}%
\edef\next{\the\toks0}\ifx\next\em@rk\let\next=\endgroup\else\ifx\next\empty%
\else\immediate\write\wfile{\the\toks0}\fi\let\next=\writ@line\fi\next\relax}}
\def\striprel@x#1{} \def\em@rk{\hbox{}}
\def\lref{\begingroup\obeylines\lr@f}
\def\lr@f#1#2{\gdef#1{\ref#1{#2}}\endgroup\unskip}
\def\semi{;\hfil\break}
\def\addref#1{\immediate\write\rfile{\noexpand\item{}#1}} %now unnecessary
\def\startrefs#1{\immediate\openout\rfile=refs.tmp\refno=#1}
\def\xref{\expandafter\xr@f}\def\xr@f[#1]{#1}
\def\refs#1{\count255=1[\r@fs #1{\hbox{}}]}
\def\r@fs#1{\ifx\und@fined#1\message{reflabel \string#1 is undefined.}%
\nref#1{need to supply reference \string#1.}\fi%
\vphantom{\hphantom{#1}}\edef\next{#1}\ifx\next\em@rk\def\next{}%
\else\ifx\next#1\ifodd\count255\relax\xref#1\count255=0\fi%
\else#1\count255=1\fi\let\next=\r@fs\fi\next}
%

%
% this is ugly, but moore insists
\newwrite\ffile\global\newcount\figno \global\figno=1
\def\fig{fig.~\the\figno\nfig}
\def\nfig#1{\xdef#1{fig.~\the\figno}%
\writedef{#1\leftbracket fig.\noexpand~\the\figno}%
\ifnum\figno=1\immediate\openout\ffile=figs.tmp\fi\chardef\wfile=\ffile%
\immediate\write\ffile{\noexpand\medskip\noexpand\item{Fig.\ \the\figno. }
\reflabeL{#1\hskip.55in}\pctsign}\global\advance\figno by1\findarg}
\def\xfig{\expandafter\xf@g}\def\xf@g fig.\penalty\@M\ {}
\def\figs#1{figs.~\f@gs #1{\hbox{}}}
\def\f@gs#1{\edef\next{#1}\ifx\next\em@rk\def\next{}\else
\ifx\next#1\xfig #1\else#1\fi\let\next=\f@gs\fi\next}
\newwrite\lfile
{\escapechar-1\xdef\pctsign{\string\%}\xdef\leftbracket{\string\{}
\xdef\rightbracket{\string\}}\xdef\numbersign{\string\#}}

\def\writestop{\def\writestoppt{\immediate\write\lfile{\string\pageno%
\the\pageno\string\startrefs\leftbracket\the\refno\rightbracket%
\string\def\string\secsym\leftbracket\secsym\rightbracket%
\string\secno\the\secno\string\meqno\the\meqno}\immediate\closeout\lfile}}
\def\writestoppt{}\def\writedef#1{}
\def\seclab#1{\xdef #1{\the\secno}\writedef{#1\leftbracket#1}\wrlabeL{#1=#1}}
\def\subseclab#1{\xdef #1{\secsym\the\subsecno}%
\writedef{#1\leftbracket#1}\wrlabeL{#1=#1}}
\newwrite\tfile \def\writetoca#1{}
\def\leaderfill{\leaders\hbox to 1em{\hss.\hss}\hfill}
%        use this to write file with table of contents
\def\writetoc{\immediate\openout\tfile=toc.tmp
   \def\writetoca##1{{\edef\next{\write\tfile{\noindent ##1
   \string\leaderfill {\noexpand\number\pageno} \par}}\next}}}
%       and this lists table of contents on second pass

%
\catcode`\@=12 % at signs are no longer letters
%
%        Unpleasantness in calling in abstract and title fonts
\edef\tfontsize{\ifx\answ\bigans scaled\magstep3\else scaled\magstep4\fi}
\font\titlerm=cmr10 \tfontsize \font\titlerms=cmr7 \tfontsize
\font\titlermss=cmr5 \tfontsize \font\titlei=cmmi10 \tfontsize
\font\titleis=cmmi7 \tfontsize \font\titleiss=cmmi5 \tfontsize
\font\titlesy=cmsy10 \tfontsize \font\titlesys=cmsy7 \tfontsize
\font\titlesyss=cmsy5 \tfontsize \font\titleit=cmti10 \tfontsize
\skewchar\titlei='177 \skewchar\titleis='177 \skewchar\titleiss='177
\skewchar\titlesy='60 \skewchar\titlesys='60 \skewchar\titlesyss='60
\def\titlefont{\def\rm{\fam0\titlerm}% switch to title font
\textfont0=\titlerm \scriptfont0=\titlerms \scriptscriptfont0=\titlermss
\textfont1=\titlei \scriptfont1=\titleis \scriptscriptfont1=\titleiss
\textfont2=\titlesy \scriptfont2=\titlesys \scriptscriptfont2=\titlesyss
\textfont\itfam=\titleit \def\it{\fam\itfam\titleit}\rm}
 \ifx\answ\bigans\else scaled\magstep1\fi
\ifx\answ\bigans\def\abstractfont{\tenpoint}\else
\font\abssl=cmsl10 scaled \magstep1
\font\absrm=cmr10 scaled\magstep1 \font\absrms=cmr7 scaled\magstep1
\font\absrmss=cmr5 scaled\magstep1 \font\absi=cmmi10 scaled\magstep1
\font\absis=cmmi7 scaled\magstep1 \font\absiss=cmmi5 scaled\magstep1
\font\abssy=cmsy10 scaled\magstep1 \font\abssys=cmsy7 scaled\magstep1
\font\abssyss=cmsy5 scaled\magstep1 \font\absbf=cmbx10 scaled\magstep1
\skewchar\absi='177 \skewchar\absis='177 \skewchar\absiss='177
\skewchar\abssy='60 \skewchar\abssys='60 \skewchar\abssyss='60
\def\abstractfont{\def\rm{\fam0\absrm}% switch to abstract font
\textfont0=\absrm \scriptfont0=\absrms \scriptscriptfont0=\absrmss
\textfont1=\absi \scriptfont1=\absis \scriptscriptfont1=\absiss
\textfont2=\abssy \scriptfont2=\abssys \scriptscriptfont2=\abssyss
\textfont\itfam=\bigit \def\it{\fam\itfam\bigit}\def\footnotefont{\tenpoint}%
\textfont\slfam=\abssl \def\sl{\fam\slfam\abssl}%
\textfont\bffam=\absbf \def\bf{\fam\bffam\absbf}\rm}\fi
\def\tenpoint{\def\rm{\fam0\tenrm}% switch back to 10-point type
\textfont0=\tenrm \scriptfont0=\sevenrm \scriptscriptfont0=\fiverm
\textfont1=\teni  \scriptfont1=\seveni  \scriptscriptfont1=\fivei
\textfont2=\tensy \scriptfont2=\sevensy \scriptscriptfont2=\fivesy
\textfont\itfam=\tenit
\def\it{\fam\itfam\tenit}\def\footnotefont{\ninepoint}%
\textfont\bffam=\tenbf \def\bf{\fam\bffam\tenbf}\def\sl{\fam\slfam\tensl}\rm}
\font\ninerm=cmr9 \font\sixrm=cmr6 \font\ninei=cmmi9 \font\sixi=cmmi6
\font\ninesy=cmsy9 \font\sixsy=cmsy6 \font\ninebf=cmbx9
\font\nineit=cmti9 \font\ninesl=cmsl9 \skewchar\ninei='177
\skewchar\sixi='177 \skewchar\ninesy='60 \skewchar\sixsy='60
\def\ninepoint{\def\rm{\fam0\ninerm}% switch to footnote font
\textfont0=\ninerm \scriptfont0=\sixrm \scriptscriptfont0=\fiverm
\textfont1=\ninei \scriptfont1=\sixi \scriptscriptfont1=\fivei
\textfont2=\ninesy \scriptfont2=\sixsy \scriptscriptfont2=\fivesy
\textfont\itfam=\ninei \def\it{\fam\itfam\nineit}\def\sl{\fam\slfam\ninesl}%
\textfont\bffam=\ninebf \def\bf{\fam\bffam\ninebf}\rm}
%
%---------------------------------------------------------------------
%

\hyphenation{anom-aly anom-alies coun-ter-term coun-ter-terms}
\def\inv{^{\raise.15ex\hbox{${\scriptscriptstyle -}$}\kern-.05em 1}}

\def\Dsl{\,\raise.15ex\hbox{/}\mkern-13.5mu D} %this one can be subscripted
\def\dsl{\raise.15ex\hbox{/}\kern-.57em\partial}

\def\tr{{\rm tr}} 
\font\bigit=cmti10 scaled \magstep1
 %pound sterling
\def\lspace{\ifx\answ\bigans{}\else\qquad\fi}
\def\lbspace{\ifx\answ\bigans{}\else\hskip-.2in\fi} % $$\lbspace...$$
\def\boxeqn#1{\vcenter{\vbox{\hrule\hbox{\vrule\kern3pt\vbox{\kern3pt
           \hbox{${\displaystyle #1}$}\kern3pt}\kern3pt\vrule}\hrule}}}
\def\mbox#1#2{\vcenter{\hrule \hbox{\vrule height#2in
               \kern#1in \vrule} \hrule}}  %e.g. \mbox{.1}{.1}
%       matters of taste
%\def\tilde{\widetilde} \def\bar{\overline} \def\hat{\widehat}
%
% some sample definitions
  %     curly letters

\def\e#1{{\rm e}^{^{\textstyle#1}}}

\def\darr#1{\raise1.5ex\hbox{$\leftrightarrow$}\mkern-16.5mu #1}
\def\lie{\hbox{\it\$}} %pound sterling

 %puts a small half in a displayed eqn
\def\roughly#1{\raise.3ex\hbox{$#1$\kern-.75em\lower1ex\hbox{$\sim$}}}

%\input harvmac.tex

%%temporary additional macros
% \input macros.tex
% April 16 -- NN

%%%%%%%%%%%%%%%%%%%%%  Rublenye bukvy   %%%%%%%%%%%%%%%%%%%%%%%%
\def\IB{\relax\hbox{$\inbar\kern-.3em{\rm B}$}}
\def\IC{\relax\hbox{$\inbar\kern-.3em{\rm C}$}}
\def\ID{\relax\hbox{$\inbar\kern-.3em{\rm D}$}}
\def\IE{\relax\hbox{$\inbar\kern-.3em{\rm E}$}}
\def\IF{\relax\hbox{$\inbar\kern-.3em{\rm F}$}}
\def\IG{\relax\hbox{$\inbar\kern-.3em{\rm G}$}}
\def\IGa{\relax\hbox{${\rm I}\kern-.18em\Gamma$}}
\def\IH{\relax{\rm I\kern-.18em H}}
\def\IK{\relax{\rm I\kern-.18em K}}
\def\II{\relax{\rm I\kern-.18em I}}
\def\IL{\relax{\rm I\kern-.18em L}}
\def\IP{\relax{\rm I\kern-.18em P}}
\def\IR{\relax{\rm I\kern-.18em R}}
\def\IZ{\relax\ifmmode\mathchoice {\hbox{\cmss Z\kern-.4em Z}}{\hbox{\cmss
Z\kern-.4em Z}} {\lower.9pt\hbox{\cmsss Z\kern-.4em Z}}
{\lower1.2pt\hbox{\cmsss Z\kern-.4em Z}}\else{\cmss Z\kern-.4em Z}\fi}

\def\IB{\relax{\rm I\kern-.18em B}}
\def\IC{{\relax\hbox{$\inbar\kern-.3em{\rm C}$}}}
\def\ID{\relax{\rm I\kern-.18em D}}
\def\IE{\relax{\rm I\kern-.18em E}}
\def\IF{\relax{\rm I\kern-.18em F}}

%%%%%%%%%%%%%%%%%%%% Calligraphic letters  %%%%%%%%%%%%%%%%%%%%%%%

\def\CW {{\cal W}}

%%%%%%%%%%%%%%%%%%%%%%%%%% Derivatives  %%%%%%%%%%%%%%%%%%%%%%%%
\def\p{\partial}

%%Beltrami

%%%%%%%%%%%%%%%%%%%% letters with bar %%%%%%%%%%%%%%%%%%%%%%%%%%

%%%%%%%%%%%%%%%%%%%%%%%%%%% Math symbols %%%%%%%%%%%%%%%%%%%%%%%

\def\s{\lies}

%%%%%%%%%%%%%%%%%%%%% Short Cuts %%%%%%%%%%%%%%%%%%%%%%%

\def\c{\cdot}

%%%%%%%%%%%%%%%%%% Greek %%%%%%%%%%%%%%%%%%%%%%

\def\a{\alpha}
\def\b{\beta}
\def\g{\gamma}  \def\G{\Gamma}
\def\d{\delta}  
\def\m{\mu}
\def\n{\nu}
\def\r{\rho}
\def\l{\lambda} \def\L{\Lambda}
\def\k{\kappa}
\def\e{\epsilon}

%%%%%%%%%%%%%%%%%% Big ( )  %%%%%%%%%%%%%%%%%%%%%%
\def\|{\Big|}
\def\({\Big(}   \def\){\Big)}
\def\[{\Big[}   \def\]{\Big]}

%%%%%%%%%%%%%%%%%% Text %%%%%%%%%%%%%%%%%%%%%%

%%%%%%%%%%%%% References %%%%%%%%%%%%%%%%%%%%

\def\paper#1#2#3#4{#1, {\sl #2}, #3 {\tt #4}}
% refs with #1=authors, #2=title, #3=publ.ref, #4=hep no :
%\lref\NAME{\paper
%{Authors}{Title(in \it)}{\PLB{No.}{Year}{page},}
%{\hh 0303028 (in\tt)}.}

%\def\hh#1{hep-th/{\it #1}}
\def\hh{hep-th/}

% journal~{\bf no.} (year) page

\def\PLB#1#2#3{Phys. Lett.~{\bf B#1} (#2) #3}
\def\NPB#1#2#3{Nucl. Phys.~{\bf B#1} (#2) #3}
\def\PRL#1#2#3{Phys. Rev. Lett.~{\bf #1} (#2) #3}
\def\CMP#1#2#3{Comm. Math. Phys.~{\bf #1} (#2) #3}
\def\PRD#1#2#3{Phys. Rev.~{\bf D#1} (#2) #3}
\def\MPL#1#2#3{Mod. Phys. Lett.~{\bf #1} (#2) #3}
\def\IJMP#1#2#3{Int. Jour. Mod. Phys.~{\bf #1} (#2) #3}

%%%%%%%%%%%%%%%%%%% Something to deal with sub-sub-sections
%%%%%%%%%%%%%%%%%%%%%%%%%%%%%%%%%%%%%%%%%%%%%%%

\def\unlockat{\catcode`\@=11}
\def\lockat{\catcode`\@=12}

\unlockat

% Something to deal with sub-sub-sections

\def\newsec#1{\global\advance\secno by1\message{(\the\secno. #1)}
\global\subsecno=0\global\subsubsecno=0\eqnres@t\noindent {\bf\the\secno. #1}
\writetoca{{\secsym} {#1}}\par\nobreak\medskip\nobreak}
\global\newcount\subsecno \global\subsecno=0
\def\subsec#1{\global\advance\subsecno by1\message{(\secsym\the\subsecno.
#1)}
\ifnum\lastpenalty>9000\else\bigbreak\fi\global\subsubsecno=0
\noindent{\it\secsym\the\subsecno. #1}
\writetoca{\string\quad {\secsym\the\subsecno.} {#1}}
\par\nobreak\medskip\nobreak}
\global\newcount\subsubsecno \global\subsubsecno=0
\def\subsubsec#1{\global\advance\subsubsecno by1
\message{(\secsym\the\subsecno.\the\subsubsecno. #1)}
\ifnum\lastpenalty>9000\else\bigbreak\fi
\noindent\quad{\secsym\the\subsecno.\the\subsubsecno.}{#1}
\writetoca{\string\qquad{\secsym\the\subsecno.\the\subsubsecno.}{#1}}
\par\nobreak\medskip\nobreak}

\def\subsubseclab#1{\DefWarn#1\xdef #1{\noexpand\hyperref{}{subsubsection}%
{\secsym\the\subsecno.\the\subsubsecno}%
{\secsym\the\subsecno.\the\subsubsecno}}%
\writedef{#1\leftbracket#1}\wrlabeL{#1=#1}}% Macros for boxes
\lockat

%why???\font\manual=manfnt
\def\dbend{\lower3.5pt\hbox{\manual\char127}}

%%%%%%%%%%%%%%%%%%% Macros for boxes %%%%%%%%%%%%%%%%%%

\def\boxit#1{\vbox{\hrule\hbox{\vrule\kern8pt
\vbox{\hbox{\kern8pt}\hbox{\vbox{#1}}\hbox{\kern8pt}}
\kern8pt\vrule}\hrule}}

\def\mathboxit#1{\vbox{\hrule\hbox{\vrule\kern8pt\vbox{\kern8pt
\hbox{$\displaystyle #1$}\kern8pt}\kern8pt\vrule}\hrule}}

%%%%%%%%%%%%%%%%%%%% ANOTHER SET OF MACROS %%%%%%%%%%%%%%%%%%

\def\inbar{\,\vrule height1.5ex width.4pt depth0pt}

\font\cmss=cmss10 \font\cmsss=cmss10 at 7pt

%REFERENCES
%%%%%%%%%%%%%%%%%%%%%%%%%%%%%%%%%%%%%%%%%%%%%%%%%%%

\lref\simons{ J. Cheeger and J. Simons, {\it Differential Characters and
Geometric Invariants},  Stony Brook Preprint, (1973), unpublished.}

\lref\cargese{ L.~Baulieu, {\it Algebraic quantization of gauge theories},
Perspectives in fields and particles, Plenum Press, eds. Basdevant-Levy,
Cargese Lectures 1983}

\lref\antifields{ L. Baulieu, M. Bellon, S. Ouvry, C.Wallet, Phys.Letters
B252 (1990) 387; M.  Bocchichio, Phys. Lett. B187 (1987) 322;  Phys. Lett. B
192 (1987) 31; R.  Thorn    Nucl. Phys.   B257 (1987) 61. }

\lref\thompson{ George Thompson,  Annals Phys. 205 (1991) 130; J.M.F.
Labastida, M. Pernici, Phys. Lett. 212B  (1988) 56; D. Birmingham, M.Blau,
M. Rakowski and G.Thompson, Phys. Rept. 209 (1991) 129.}

\lref\tonin{ Tonin}

\lref\wittensix{ E.  Witten, {\it New  Gauge  Theories In Six Dimensions},
\hh{9710065}. }

\lref\orlando{ O. Alvarez, L. A. Ferreira and J. Sanchez Guillen, {\it  A New
Approach to Integrable Theories in any Dimension}, hep-th/9710147.}

\lref\wittentopo{ E.  Witten,  {\it  Topological Quantum Field Theory},
\hh9403195, Commun.  Math. Phys.  {117} (1988)353.  }

\lref\wittentwist{ E.  Witten, {\it Supersymmetric Yang--Mills theory on a
four-manifold}, J.  Math.  Phys.  {35} (1994) 5101.}

\lref\west{ L.~Baulieu, P.~West, {\it Six Dimensional TQFTs and  Self-dual
Two-Forms,} Phys.Lett. B {\bf 436 } (1998) 97, /hep-th/9805200}

\lref\bv{ I.A. Batalin and V.A. Vilkowisky,    Phys. Rev.   D28  (1983)
2567\semi M. Henneaux,  Phys. Rep.  126   (1985) 1\semi M. Henneaux and C.
Teitelboim, {\it Quantization of Gauge Systems}
  Princeton University Press,  Princeton (1992).}

\lref\kyoto{ L. Baulieu, E. Bergschoeff and E. Sezgin, Nucl. Phys.
B307(1988)348\semi L. Baulieu,   {\it Field Antifield Duality, p-Form Gauge
Fields
   and Topological Quantum Field Theories}, hep-th/9512026,
   Nucl. Phys. B478 (1996) 431.  }

\lref\sourlas{ G. Parisi and N. Sourlas, {\it Random Magnetic Fields,
Supersymmetry and Negative Dimensions}, Phys. Rev. Lett.  43 (1979) 744;
Nucl.  Phys.  B206 (1982) 321.  }

\lref\SalamSezgin{ A.  Salam  and  E.  Sezgin, {\it Supergravities in
diverse dimensions}, vol.  1, p. 119\semi P.  Howe, G.  Sierra and P.
Townsend, Nucl Phys B221 (1983) 331.}

\lref\nekrasov{ A. Losev, G. Moore, N. Nekrasov, S. Shatashvili, {\it
Four-Dimensional Avatars of Two-Dimensional RCFT},  hep-th/9509151, Nucl.
Phys.  Proc.  Suppl.   46 (1996) 130\semi L.  Baulieu, A.  Losev,
N.~Nekrasov  {\it Chern-Simons and Twisted Supersymmetry in Higher
Dimensions},  hep-th/9707174, to appear in Nucl.  Phys.  B.  }

\lref\WitDonagi{R.~ Donagi, E.~ Witten, ``Supersymmetric Yang--Mills Theory
and Integrable Systems'', hep-th/9510101, Nucl. Phys.{\bf B}460 (1996)
299-334}
\lref\Witfeb{E.~ Witten, ``Supersymmetric Yang--Mills Theory On A
Four-Manifold,''  hep-th/9403195; J. Math. Phys. {\bf 35} (1994) 5101.}
\lref\Witgrav{E.~ Witten, ``Topological Gravity'', Phys.Lett.206B:601, 1988}
\lref\witaffl{I. ~ Affleck, J.A.~ Harvey and E.~ Witten,
        ``Instantons and (Super)Symmetry Breaking
        in $2+1$ Dimensions'', Nucl. Phys. {\bf B}206 (1982) 413}
\lref\wittabl{E.~ Witten,  ``On $S$-Duality in Abelian Gauge Theory,''
hep-th/9505186; Selecta Mathematica {\bf 1} (1995) 383}
\lref\wittgr{E.~ Witten, ``The Verlinde Algebra And The Cohomology Of The
Grassmannian'',  hep-th/9312104}
\lref\wittenwzw{E. Witten, ``Non abelian bosonization in two dimensions,''
Commun. Math. Phys. {\bf 92} (1984)455 }
\lref\witgrsm{E. Witten, ``Quantum field theory, grassmannians and algebraic
curves,'' Commun.Math.Phys.113:529,1988}
\lref\wittjones{E. Witten, ``Quantum field theory and the Jones
polynomial,'' Commun.  Math. Phys., 121 (1989) 351. }
\lref\witttft{E.~ Witten, ``Topological Quantum Field Theory", Commun. Math.
Phys. {\bf 117} (1988) 353.}
\lref\wittmon{E.~ Witten, ``Monopoles and Four-Manifolds'', hep-th/9411102}
\lref\Witdgt{ E.~ Witten, ``On Quantum gauge theories in two dimensions,''
Commun. Math. Phys. {\bf  141}  (1991) 153}
\lref\witrevis{E.~ Witten,
 ``Two-dimensional gauge theories revisited'', hep-th/9204083; J. Geom.
Phys. 9 (1992) 303-368}
\lref\Witgenus{E.~ Witten, ``Elliptic Genera and Quantum Field Theory'',
Comm. Math. Phys. 109(1987) 525. }
\lref\OldZT{E. Witten, ``New Issues in Manifolds of SU(3) Holonomy,'' {\it
Nucl. Phys.} {\bf B268} (1986) 79 \semi J. Distler and B. Greene, ``Aspects
of (2,0) String Compactifications,'' {\it Nucl. Phys.} {\bf B304} (1988) 1
\semi B. Greene, ``Superconformal Compactifications in Weighted Projective
Space,'' {\it Comm. Math. Phys.} {\bf 130} (1990) 335.}
\lref\bagger{E.~ Witten and J. Bagger, Phys. Lett. {\bf 115B}(1982) 202}
\lref\witcurrent{E.~ Witten,``Global Aspects of Current Algebra'',
Nucl.Phys.B223 (1983) 422\semi ``Current Algebra, Baryons and Quark
Confinement'', Nucl.Phys. B223 (1993) 433}
\lref\Wittreiman{S.B. Treiman, E. Witten, R. Jackiw, B. Zumino, ``Current
Algebra and Anomalies'', Singapore, Singapore: World Scientific ( 1985) }
\lref\Witgravanom{L. Alvarez-Gaume, E.~ Witten, ``Gravitational Anomalies'',
Nucl.Phys.B234:269,1984. }

\lref\nicolai{\paper {H.~Nicolai}{New Linear Systems for 2D Poincar\'e
Supergravities}{\NPB{414}{1994}{299},}{\hh 9309052}.}

%%%%%%
%% References herein
%%%%%%%%%%%%%%%%%%%%%%%%%%%%%%%%%%%%%%%%%%%%%%%%

%%\lref\NAME{\paper
%%{Authors}{Title(in \sl)}{\PLB{No.}{Year}{page},}
%%{\hh 0303028 (in\tt)}.}

\lref\baex{\paper {L.~Baulieu, B.~Grossman}{Monopoles and Topological Field
Theory}{\PLB{214}{1988}{223}.}{}\paper {L.~Baulieu}{Chern-Simons
Three-Dimensional and
Yang--Mills-Higgs Two-Dimensional Systems as Four-Dimensional Topological
Quantum Field Theories}{\PLB{232}{1989}{473}.}{}}

\lref\bg{\paper {L.~Baulieu, B.~Grossman}{Monopoles and Topological Field
Theory}{\PLB{214}{1988}{223}.}{}}

\lref\seibergsix{\paper {N.~Seiberg}{Non-trivial Fixed Points of The
Renormalization Group in Six
 Dimensions}{\PLB{390}{1997}{169}}{\hh 9609161}\semi
\paper {O.J.~Ganor, D.R.~Morrison, N.~Seiberg}{
  Branes, Calabi-Yau Spaces, and Toroidal Compactification of the N=1
  Six-Dimensional $E_8$ Theory}{\NPB{487}{1997}{93}}{\hh 9610251}\semi
\paper {O.~Aharony, M.~Berkooz, N.~Seiberg}{Light-Cone
  Description of (2,0) Superconformal Theories in Six
  Dimensions}{Adv. Theor. Math. Phys. {\bf 2} (1998) 119}{\hh 9712117.}}

\lref\cs{\paper {L.~Baulieu}{Chern-Simons Three-Dimensional and
Yang--Mills-Higgs Two-Dimensional Systems as Four-Dimensional Topological
Quantum Field Theories}{\PLB{232}{1989}{473}.}{}}

\lref\beltrami{\paper {L.~Baulieu, M.~Bellon}{Beltrami Parametrization and
String Theory}{\PLB{196}{1987}{142}}{}\semi
\paper {L.~Baulieu, M.~Bellon, R.~Grimm}{Beltrami Parametrization For
Superstrings}{\PLB{198}{1987}{343}}{}\semi
\paper {R.~Grimm}{Left-Right Decomposition of Two-Dimensional Superspace
Geometry and Its BRS Structure}{Annals Phys. {\bf 200} (1990) 49.}{}}

\lref\bbg{\paper {L.~Baulieu, M.~Bellon, R.~Grimm}{Left-Right Asymmetric
Conformal Anomalies}{\PLB{228}{1989}{325}.}{}}

\lref\bonora{\paper {G.~Bonelli, L.~Bonora, F.~Nesti}{String Interactions
from Matrix String Theory}{\NPB{538}{1999}{100},}{\hh 9807232}\semi
\paper {G.~Bonelli, L.~Bonora, F.~Nesti, A.~Tomasiello}{Matrix String Theory
and its Moduli Space}{}{\hh 9901093.}}

\lref\corrigan{\paper {E.~Corrigan, C.~Devchand, D.B.~Fairlie,
J.~Nuyts}{First Order Equations for Gauge Fields in Spaces of Dimension
Greater Than Four}{\NPB{214}{452}{1983}.}{}}

\lref\acha{\paper {B.S.~Acharya, M.~O'Loughlin, B.~Spence}{Higher
Dimensional Analogues of Donaldson-Witten Theory}{\NPB{503}{1997}{657},}{\hh
9705138}\semi
\paper {B.S.~Acharya, J.M.~Figueroa-O'Farrill, M.~O'Loughlin,
B.~Spence}{Euclidean
  D-branes and Higher-Dimensional Gauge   Theory}{\NPB{514}{1998}{583},}{\hh
  9707118.}}

\lref\Witr{\paper{E.~Witten}{Introduction to Cohomological Field   Theories}
{Lectures at Workshop on Topological Methods in Physics (Trieste, Italy, Jun
11-25, 1990), \IJMP{A6}{1991}{2775}.}{}}

\lref\ohta{\paper {L.~Baulieu, N.~Ohta}{Worldsheets with Extended
Supersymmetry} {\PLB{391}{1997}{295},}{\hh 9609207}.}

\lref\gravity{\paper {L.~Baulieu}{Transmutation of Pure 2-D Supergravity
Into Topological 2-D Gravity and Other Conformal Theories}
{\PLB{288}{1992}{59},}{\hh 9206019.}}

\lref\wgravity{\paper {L.~Baulieu, M.~Bellon, R.~Grimm}{Some Remarks on  the
Gauging of the Virasoro and   $w_{1+\infty}$
Algebras}{\PLB{260}{1991}{63}.}{}}

\lref\fourd{\paper {E.~Witten}{Topological Quantum Field
Theory}{\CMP{117}{1988}{353}}{}\semi
\paper {L.~Baulieu, I.M.~Singer}{Topological Yang--Mills Symmetry}{Nucl.
Phys. Proc. Suppl. {\bf 15B} (1988) 12.}{}}

\lref\topo{\paper {L.~Baulieu}{On the Symmetries of Topological Quantum Field
Theories}{\IJMP{A10}{1995}{4483},}{\hh 9504015}\semi
\paper {R.~Dijkgraaf, G.~Moore}{Balanced Topological Field
Theories}{\CMP{185}{1997}{411},}{\hh 9608169.}}

\lref\wwgravity{\paper {I.~Bakas} {The Large $N$ Limit   of Extended
Conformal Symmetries}{\PLB{228}{1989}{57}.}{}}

\lref\wwwgravity{\paper {C.M.~Hull}{Lectures on $\CW$-Gravity,
$\CW$-Geometry and
$\CW$-Strings}{}{\hh 9302110}, and~references therein.}

\lref\wvgravity{\paper {A.~Bilal, V.~Fock, I.~Kogan}{On the origin of
$W$-algebras}{\NPB{359}{1991}{635}.}{}}

\lref\surprises{\paper {E.~Witten} {Surprises with Topological Field
Theories} {Lectures given at ``Strings 90'', Texas A\&M, 1990,}{Preprint
IASSNS-HEP-90/37.}}

\lref\stringsone{\paper {L.~Baulieu, M.B.~Green, E.~Rabinovici}{A Unifying
Topological Action for Heterotic and  Type II Superstring  Theories}
{\PLB{386}{1996}{91},}{\hh 9606080.}}

\lref\stringsN{\paper {L.~Baulieu, M.B.~Green, E.~Rabinovici}{Superstrings
from   Theories with $N>1$ World Sheet Supersymmetry}
{\NPB{498}{1997}{119},}{\hh 9611136.}}

\lref\bks{\paper {L.~Baulieu, H.~Kanno, I.~Singer}{Special Quantum Field
Theories in Eight and Other Dimensions}{\CMP{194}{1998}{149},}{\hh
9704167}\semi
\paper {L.~Baulieu, H.~Kanno, I.~Singer}{Cohomological Yang--Mills Theory
  in Eight Dimensions}{ Talk given at APCTP Winter School on Dualities in
String Theory (Sokcho, Korea, February 24-28, 1997),} {\hh 9705127.}}

\lref\witdyn{\paper {P.~Townsend}{The eleven dimensional supermembrane
revisited}{\PLB{350}{1995}{184},}{\hh9501068}\semi
\paper{E.~Witten}{String Theory Dynamics in Various Dimensions}
{\NPB{443}{1995}{85},}{\hh 9503124}.}

\lref\bfss{\paper {T.~Banks, W.Fischler, S.H.~Shenker,
L.~Susskind}{$M$-Theory as a Matrix Model~:
A~Conjecture}{\PRD{55}{1997}{5112},} {\hh9610043.}}

\lref\seiberg{\paper {N.~Seiberg}{Why is the Matrix Model
Correct?}{\PRL{79}{1997}{3577},} {\hh 9710009.}}

\lref\sen{\paper {A.~Sen}{$D0$ Branes on $T^n$ and Matrix Theory}{Adv.
Theor. Math. Phys.~{\bf 2} (1998) 51,} {\hh 9709220.}}

\lref\laroche{\paper {L.~Baulieu, C.~Laroche} {On Generalized Self-Duality
Equations Towards Supersymmetric   Quantum Field Theories Of
Forms}{\MPL{A13}{1998}{1115},}{\hh  9801014.}}

\lref\bsv{\paper {M.~Bershadsky, V.~Sadov, C.~Vafa} {$D$-Branes and
Topological Field Theories}{\NPB{463} {1996}{420},}{\hh9511222.}}

\lref\vafapuzz{\paper {C.~Vafa}{Puzzles at Large N}{}{\hh 9804172.}}

\lref\dvv{\paper {R.~Dijkgraaf, E.~Verlinde, H.~Verlinde} {Matrix String
Theory}{\NPB{500}{1997}{43},} {\hh9703030.}}

\lref\wynter{\paper {T.~Wynter}{Gauge Fields and Interactions in Matrix
String Theory}{\PLB{415}{1997}{349},}{\hh9709029.}}

\lref\kvh{\paper {I.~Kostov, P.~Vanhove}{Matrix String Partition
Functions}{}{\hh9809130.}}

\lref\ikkt{\paper {N.~Ishibashi, H.~Kawai, Y.~Kitazawa, A.~Tsuchiya} {A
Large $N$ Reduced Model as Superstring}{\NPB{498} {1997}{467},}{\hh
9612115.}}

\lref\ss{\paper {S.~Sethi, M.~Stern} {$D$-Brane Bound States
Redux}{\CMP{194}{1998} {675},}{\hh 9705046.}}

\lref\mns{\paper {G.~Moore, N.~Nekrasov, S.~Shatashvili} {$D$-particle Bound
States and Generalized Instantons}{} {\hh 9803265.}}

\lref\bsh{\paper {L.~Baulieu, S.~Shatashvili} {Duality from Topological
Symmetry}{} {\hh 9811198.}}

\lref\pawu{ {G.~Parisi, Y.S.~Wu,} {}{ Sci. Sinica  {\bf 24} {(1981)} {484}.}}

%%%%%%%%%%%%%%%%
\lref\lbpert{ {L.~Baulieu,}   {\it Pertrubative gauge theories}, {Physics
Reports {\bf 129 } (1985) 1.} {}}

\lref\kugoj{ {T.~Kugo and I.~Ojima,}   {\it Local covariant
operator formalism of non-Abelian gauge theories and quark confinement
problem} {Prog. Theor. Phys. Suppl.
{\bf 66} 1 (1979).} {}}

\lref\nakanoj{ {N.~Nakanishi and I.~Ojima,}   {\it Covariant operator
formalism of gauge theories and quantum gravity} {vol. 27 of Lecture Notes in
Physics (World Scientific 1990).} {}}

\lref\nishijima{ {K. Nishijima,} Czech. J. Phys. {\bf 46}, 1 (1996);
Int. J. Mod. Phys. B {\bf12} 1355 (1998).}

\lref\hirschfeld{ {P.~Hirschfeld,}   {\it } {Nucl. Phys.
{\bf 157} (1979) 37.} {}}

\lref\christlee{ {N. ~Christ and T. ~D. ~Lee,} {\it } {Phys. Rev.
{\bf  D22,} {939} (1980).}}

\lref\fleepr{ {R.~Friedberg, T.~D.~Lee, Y.~Pang, H.~C.~Ren,}   {\it A soluble
gauge model with Gribov-type copies}, {Ann. of Phys. {\bf 246 } (1996) 381.} {}}

\lref\coulomb{ {L.~Baulieu, D.~Zwanziger, }   {\it Renormalizable
Non-Covariant Gauges and Coulomb Gauge Limit}, {Nucl. Phys. B {\bf 548 }
(1999) 527-562.} {\hh 9807024}.}

\lref\coulham{ {D.~Zwanziger, }   {\it Lattice Coulomb hamiltonian and static
color-Coulomb field}, {Nucl. Phys. B {\bf 485 } (1997) 185-240.} {}}

\lref\dzcritlim{ {D.~Zwanziger, }   {\it Critical limit of lattice gauge
theory}, {Nucl. Phys. B {\bf 378 } (1992) 525-590.} {}}

\lref\dznonpertl{ {D.~Zwanziger, }   {\it Non-perturbative Landau gauge and
infrared critical exponents in QCD}, {hep-th/0109224.} {}}

\lref\dztimeind{ {D.~Zwanziger, }   {\it Time-independent stochastic
quantization, DS equations, and infrared critical exponents in QCD},
{hep-th/0206053.} {}}

\lref\rcoulomb{ {D.~Zwanziger, }   {\it Renormalization in the Coulomb
gauge and order parameter for confinement in QCD}, {Nucl.Phys. B {\bf 518
} (1998) 237-272.} {}}

\lref\varadhan{ {S.~R.~S.~Varadhan, }   {\it Diffusion problems and
partial differential equations}, {Tata Institute of Fundamental Research,
Bombay, Springer Verlag, Berlin (1980), pp.~249-251.} {}}

\lref\kogsuss{ {J.~Kogut and L. Susskind, }   {\it } {Phys. Rev. D {\bf 11}
(1975) 395.} {}}

\lref\kugoojima{ {T.~Kugo and I. Ojima, }   {\it Local covariant
operator formalism of non-Abelian gauge theories and quark confinement
problem}, {Suppl. Prog. Theor. Phys. {\bf 66 } (1979) 1-130.} {}}

\lref\horne{ {J.H.~Horne, }   {\it
Superspace versions of Topological Theories}, {Nucl.Phys. B {\bf 318
} (1989) 22.} {}}

\lref\sto{ {S.~Ouvry, R.~Stora, P.~Van~Baal }   {\it
}, {Phys. Lett. B {\bf 220
} (1989) 159;} {}{ R.~Stora, {\it Exercises in   Equivariant Cohomology},
In Quabtum Fields and Quantum Space Time, Edited
by 't Hooft et al., Plenum Press, New York, 1997}            }

\lref\semenov{ {M. Semenov-Tyan-Shanskii and V. Franke},  {\it } {Zap. Nauch.
Sem. Leningrad. Otdeleniya Matematicheskogo Instituta im V. A. Steklov, AN
SSSR, vol 120, p 159, 1982 (English translation: New York: Plenum Press 1986.
{}}{}}

\lref\masknaka{ {T.~Maskawa and H.~Nakajima,}  {\it} {Prog. Theor. Phys.
{\bf 60}, {1526} (1978); {\bf 63}, {641} (1980).}{}}

\lref\dzmin{ {D.~Zwanziger},  {\it} {Phys. Lett. {\bf 114B},
{337} (1982).}{}}

\lref\gfdadzinside{ {G. Dell'Antonio and D. Zwanziger},  {\it All gauge orbits
and some Gribov copies encompassed by the Gribov horizon,} Proceedings of the
NATO Advanced Workshop on Probabilistic Methods in Quantum Field Theory and
Quantum Gravity, Carg\`{e}se, August 21-27, 1989, Plenum (N.Y.), P. Damgaard,
H. H\"{u}ffel and A. Rosenblum, Eds. {\bf} {}}

\lref\dan{ {D.~Zwanziger},  {\it Covariant Quantization of Gauge
Fields without Gribov Ambiguity}, {Nucl. Phys. B {\bf   192}, (1981)
{259}.}{}}

\lref\dzstochquant{ {D. Zwanziger},  {\it Stochastic quantization of gauge
fields,} in {\it Fundamental problems of gauge field theory,} 
International School of Mathematical Physics, (6th : 1985 : 
\`{E}ttore Majorana International Centre for Scientific Culture) at Erice,
Italy, Plenum (N.Y.), 1986, G. Velo and A. S. Wightman, Eds. {\bf} {}}

\lref\danlau{ {L.~Baulieu, D.~Zwanziger, } {\it Equivalence of Stochastic
Quantization and the-Popov Ansatz,
  }{Nucl. Phys. B  {\bf 193 } (1981) {163}.}{}}

\lref\floratos{ {E.~Floratos, J~Iliopoulos, D.~Zwanziger, } {\it A
covariant ghost-free perturbation expansion for Yang-Mills theories,} {Nucl.
Phys. B  {\bf 241 } (1984) {221-227}.}{}}

\lref\dzgribreg{ {D.~Zwanziger},  {\it Non-perturbative modification of the
Faddeev-Popov formula and banishment of the naive vacuum}, {Nucl. Phys. B
{\bf   209}, (1982) {336}.}{}}

\lref\danzinn{  {J.~Zinn-Justin, D.~Zwanziger, } {}{Nucl. Phys. B  {\bf
295} (1988) {297}.}{}}

\lref\bodeker{  {D.~B\"{o}deker,} {}{hep-ph/9905239} {\bf} {}.}

\lref\dzvan{ {D.~Zwanziger, }   {\it Vanishing of zero-momentum lattice
gluon propagator and color confinement}, {Nucl.Phys. B {\bf 364 }
(1991) 127.} }

\lref\horizcon{ {D.~Zwanziger, }   {\it Renormalizability of the critical limit
of lattice gauge theory by BRS invariance,} {Nucl. Phys. B {\bf 399 } (1993)
477.} }

\lref\horizcona{ {D.~Zwanziger, }   {\it Fundamental modular region, Boltzmann
factor and area law in lattice theory }, {Nucl.Phys. B {\bf 412 }
(1994) 657.} }

\lref\horizpt{ {M.~Schaden and D.~Zwanziger, }   
{\it Horizon condition holds pointwise on finite lattice with free boundary
condition,} {hep-th/9410019.} }

\lref\czcoulf{ {A.~Cucchieri and D.~Zwanziger, }   {\it Static color-Coulomb
force}, {Phys. Rev. Lett. {\bf 78 } (1997) 3814} }

\lref\acthrdlan{ {A.~Cucchieri,}   {\it} 
{Phys. Rev. D {\bf 60 } 034508 (1999).} }

\lref\cucchieria{ {A.~Cucchieri,}   {\it} 
{Phys. Lett. {\bf B422 } 233 (1998).}[hep-lat/9709015] }

\lref\acfkppl{ {A.~Cucchieri, F.~Karsch, P~Petreczky}   {\it} 
{Phys. Lett. B {\bf 497 } 80 (2001).} }

\lref\acfkppa{ {A.~Cucchieri, F.~Karsch, P~Petreczky}   {\it} 
{Phys. Rev. D {\bf 64 } 036001 (2001).} }

\lref\cucchierigh{ {A.~Cucchieri,}   {\it  Gribov copies in the minimal
Landau gauge: the influence on gluon and ghost propagators,}  {Nucl. Phys. B
{\bf 508} 353 (1997).} }

\lref\mandulab{ {J.~E.~Mandula,}   {\it } 
{Phys. Rep. {\bf 315 } 273 (1999).} }

\lref\giustietal{ {L.~Giusti et,}   {\it } 
{Int.~J.~Mod.~Phys.~A {\bf 16 } 3487 (2001).} }

\lref\mandula{ {J.~E.~Mandula, and M.~.C.~Ogilvie,}   {\it } 
{Phys. Rev. D {\bf 41 } 2586 (1990).} }

\lref\marinari{ {E.~Marinari, C. Parrinello and R. Ricci,}   {\it } 
{Phys. Rev. D {\bf 41 } 2586 (1990).} }

\lref\deforcrand{ {Ph.~de Forcrand et al,}   {\it }  {Nucl. Phys. B (Proc.
Suppl.) {\bf 20 } 194 (1991).} }

\lref\marenzoni{ {P. Marenzoni and P. Rossi,}   {\it }  {Phys. Lett. B  {\bf
311 } 219 (1993).} }

\lref\actmdzgh{ {A.~Cucchieri, T. Mendes, and D. Zwanziger,}   {\it} 
{Nucl. Phys. B Proc. Suppl. {\bf 106 } 697 (2002).} }

\lref\rengrcoul{ {Attilio Cucchieri, Daniel Zwanziger,} {\it 
Renormalization-group calculation of the color-Coulomb, }{Phys. Rev.
D65 (2001) 014002.}}

\lref\czcoulscen{ {Attilio Cucchieri and Daniel Zwanziger,} {\it 
Gluon propagator and confinement scenario in Coulomb gauge,}{hep-lat/0209068.}}

\lref\greencoul{ {Jeff Greensite and Stefan Olejnik,} {\it 
Coulomb eneargy, vortices and confinement,}{hep-lat/0302018.}}

\lref\acdland{ {A.~Cucchieri,}   {\it Infrared behavior of the gluon
propagator in lattice landau gauge: the three-dimensional case,} 
%{Nucl.Phys. B {\bf 412 } (1994) 657.} 
hep-lat/9902023CHECK IF SAME AS PHYS REV ARTICLE}

\lref\cznumstgl{ {A.~Cucchieri, and D.~Zwanziger,}   {\it Numerical study of
gluon propagator and confinement scenario in minimal Coulomb gauge}, 
{Phys. Rev. D {\bf 65} 014001}}

\lref\attilioctmat{ {A.~Cucchieri, T.~Mendes, and A. R. Taurines,}   {\it SU(2)
Landau gluon propagator on a $140^3$ lattice}, {hep-lat/0302022}}

\lref\czfitgrib{ {A.~Cucchieri, and D.~Zwanziger,}   {\it Fit to gluon
propagator and Gribov formula}, {Phys. Lett. {\bf B524} 123 (2002)}
[hep-lat/0012024]}

\lref\dznonpland{ {D.~Zwanziger,}   {\it Non-perturbative Landau gauge and
infrared critical exponents in QCD},  Phys. Rev. D, {\bf 65} 094039 (2002) and
hep-th/0109224.}

\lref\leinweber{ {D.~B.~Leinweber, J.~I.~Skullerud, A.~G.~Williams,
and C. Parrinello,}   {\it }  Phys. Rev. {\bf D58} (1998) 031501
{\it ibid} {\bf D60} (1999) 094507.}

\lref\bonneta{ {F.~Bonnet, P.~O.~Bowman, D.~B.~Leinweber, A.~G.~Williams,}  
{\it }  Phys. Rev. {\bf D62} (2000) 051501.}

\lref\bonnetb{ {F.~Bonnet, P.~O.~Bowman, D.~B.~Leinweber, A.~G.~Williams
and J. M. Zanotti} {\it }  Phys. Rev. {\bf D64} (2001) 034501.}

\lref\bogolubsky{ {I. L. Bogolubsky and V. K. Mitrjushkin} {\it } 
hep-lat/0204006.}

\lref\nakajima{ {H.~Nakajima and S. Furui,} {\it }  Nucl. Phys. Proc. Suppl.
{\bf 73} 635 (1999).} 

\lref\langfeld{ {K.~Langfeld, H. Reingardt, and J.~Gattnar}   {\it } 
Nucl. Phys. B {\bf 621} (2002) 131.}

\lref\suman{ {H.~Suman, and K.~Schilling}   {\it } 
Phys. Lett. {\bf B373} (1996) 314.}

\lref\alex{ {C. Alexandrou, P. de Forcrand, and E. Follana } {\it The gluon
propagator without lattice Gribov copies,}  Phys. Rev. {\bf D63:} 094504
(2001), and  {\it The gluon propagator with lattice Gribov copies on a finer
lattice,} Phys. Rev. {\bf D65:} 114508, 2002.}

\lref\bowman{ {Patrick O. Bowman, Urs M. Heller, Derek B.
Leinweber, Anthony G. Williams,} {\it Gluon propagator on coarse lattices in
laplacian gauges,} Phys. Rev. {\bf D66:} 074505, 2002.}

\lref\direnzo{ {F.~DiRenzo, L.~Scorzato,}  
{\it Lattice 99,}  Nucl. Phys. B (Proc. Suppl.) {\bf 83-84} (2000) 822.}

\lref\nakamuramiz{ {A. Nakamura and M. Mizutani}   {\it Numerical study of
gauge fixing ambiguity}  Vistas in Astronomy {\bf 37} 305 (1993).}

\lref\nakamuraa{ {M. Mizutani and A. Nakamura}   {\it }  Nucl. Phys. B (Proc.
Suppl.) {\bf 34} (1994) 253.}

\lref\nakamurab{ {F.~Shoji, T.~Suzuki, H.~Kodama, and A.~Nakamura,}  
{\it }  Phys. Lett. {\bf B476} (2000) 199.}

\lref\nakamurac{ {H. Aiso, M. Fukuda, T. Iwamiya, A. Nakamura, T. Nakamura,
and M. Yoshida }   {\it Gauge fixing and gluon propagators,}  Prog. Theor.
Physics. (Suppl.) {\bf 122} (1996) 123.}

\lref\nakamurad{ {H. Aiso, J. Fromm, M. Fukuda, T. Iwamiya, A. Nakamura, T.
Nakamura, M. Stingl and M. Yoshida }   {\it Towards understanding of
confinement of gluons,}  Nucl. Phys. B (Proc. Suppl.) {\bf 53} (1997) 570.}

\lref\nakamurae{ {F.~Shoji, T.~Suzuki, H.~Kodama, and A.~Nakamura,}  
{\it }  Phys. Lett. {\bf B476} (2000) 199.}

\lref\munoz{ { A.~Munoz Sudupe, R. F. Alvarez-Estrada, } {}
Phys. Lett. {\bf 164} (1985) 102; {} {\bf 166B} (1986) 186. }

\lref\okano{ { K.~Okano, } {}
Nucl. Phys. {\bf B289} (1987) 109; {} Prog. Theor. Phys.
suppl. {\bf 111} (1993) 203. }

\lref\baugros{ {L.~Baulieu, B.~Grossman, } {\it A topological Interpretation
of  Stochastic Quantization} {Phys. Lett. B {\bf  212} {(1988)} {351}.}}

\lref\bautop{ {L.~Baulieu}{ \it Stochastic and Topological Field Theories},
{Phys. Lett. B {\bf   232} (1989) {479}}{}; {}{ \it Topological Field Theories
And Gauge Invariance in Stochastic Quantization}, {Int. Jour. Mod.  Phys. A
{\bf  6} (1991) {2793}.}{}}

\lref\bautopr{  {L.~Baulieu, B.~Grossman, } {\it A topological Interpretation
of  Stochastic Quantization} {Phys. Lett. B {\bf  212} {(1988)} {351}};
 {L.~Baulieu}{ \it Stochastic and Topological Field Theories},
{Phys. Lett. B {\bf   232} (1989) {479}}{}; {}{ \it Topological Field Theories
And Gauge Invariance in Stochastic Quantization}, {Int. Jour. Mod.  Phys. A
{\bf  6} (1991) {2793}.}{}}

\lref\bautoprr{  {L.~Baulieu, B.~Grossman, } { } {Phys. Lett. B {\bf  212}
{(1988)} {351}};
 {L.~Baulieu}{ },
{Phys. Lett. B {\bf   232} (1989) {479}}{}; {}{  }, {Int. Jour. Mod.
Phys. A {\bf  6} (1991) {2793}.}{}}
\lref\samson{ {L.~Baulieu, S.L.~Shatashvili, { \it Duality from Topological
Symmetry}, {JHEP {\bf 9903} (1999) 011, hep-th/9811198.}}}{}

\lref\halperna{ {Z. Bern, M.B.~Halpern, L. Sadun, C. Taubes}{}, {Phys. Lett.
{\bf 165B,} 151, 1985.}}

\lref\halpernb{ {Z. Bern, M.B.~Halpern, L. Sadun, C. Taubes}{}, {Nucl. Phys.
{\bf B284,} 1, 1987.}}

\lref\halpernc{ {Z. Bern, M.B.~Halpern, L. Sadun, C. Taubes}{}, {Nucl. Phys.
{\bf B284,} 35, 1987.}}

\lref\halpernd{ {Z. Bern, M.B.~Halpern, L. Sadun}{}, {Nucl. Phys.
{\bf B284,} 92, 1987.}}

\lref\halperne{ {Z. Bern, M.B.~Halpern, L. Sadun}{}, {Z. Phys.
{\bf C35,} 255, 1987.}}

\lref\sadun{ {L. Sadun}{}, {Z. Phys. {\bf C36,} 467, 1987.}}

\lref\halpernr{ {M. B. Halpern}{}, {Prog. Theor. Phys. Suppl. {\bf 111,} 163,
1993.}}

\lref\halpern{ {H.S.~Chan, M.B.~Halpern}{}, {Phys. Rev. D {\bf   33} (1986)
{540}.}}

\lref\bern{ {Z. Bern, H.S.~Chan, M.B.~Halpern}{}, {Z. Phys. {\bf C35}
(1987) {255}.}}

\lref\yue{ {Yue-Yu}, {Phys. Rev. D {\bf   33} (1989) {540}.}}

\lref\neuberger{ {H.~Neuberger,} {\it Non-perturbative gauge Invariance},
{ Phys. Lett. B {\bf 175} (1986) {69}.}{}}

\lref\yangmills{  {C~N.~Yang and R.~L.~Mills,} {}{Phys. Rev. {\bf 96}
(1954) {191}.}{}}

\lref\gribov{  {V.N.~Gribov,} {}{Nucl. Phys. B {\bf 139} (1978) {1}.}{}}

\lref\huffel{ {P.H.~Daamgard, H. Huffel},  {}{Phys. Rep. {\bf 152} (1987)
{227}.}{}}

\lref\creutz{ {M.~Creutz},  {\it Quarks, Gluons and  Lattices, }  Cambridge
University Press 1983, pp 101-107.}

\lref\zinn{ {J. ~Zinn-Justin, }  {Nucl. Phys. B {\bf  275} (1986) {135}.}}

\lref\gozzi{ {E. ~Gozzi,} {\it Functional Integral approach to Parisi--Wu
Quantization: Scalar Theory,} { Phys. Rev. {\bf D28} (1983) {1922}.}}

\lref\singer{
 I.M. Singer, { Comm. of Math. Phys. {\bf 60} (1978) 7.}}

\lref\neu{ {H.~Neuberger,} {Phys. Lett. B {\bf 183}
(1987) {337}.}{}}

\lref\testa{ {M.~Testa,} {}{Phys. Lett. B {\bf 429}
(1998) {349}.}{}}

\lref\martin{ L.~Baulieu and M. Schaden, {\it Gauge Group TQFT and Improved
Perturbative Yang--Mills Theory}, {  Int. Jour. Mod.  Phys. A {\bf  13}
(1998) 985},   hep-th/9601039.}

\lref\ostseil { K.~Osterwalder and E.~Seiler, {\it Gauge field theories on
the lattice} {  Ann. Phys. {\bf 110} (1978) 440}.}

\lref\fradshen { E.~Fradken and S.~Shenker, {\it Phase diagrams of lattice
guage theories with Higgs fields} {  Phys. Rev. {\bf D19} (1979) 3682}.}

\lref\banksrab{ T.~Banks and E.~Rabinovici, {} {  Nucl. Phys. {\bf B160}
(1979) 349}.}

\lref\nielsen{N. K. Nielsen, {\it On The Gauge Dependence Of Spontaneous
Symmetry Breaking In Gauge Theories},
Nucl.\ Phys.\ {\bf B101}, 173 (1975)}

\lref\nadkarni{ S. Nadkarni, {\it The SU(2) adjoint Higgs model in three
dimensions } {  Nucl. Phys. {\bf B334} (1990) 559}.}

\lref\stackteper{ A.~Hart, O.~Philipsen, J.~D.~Stack, and M.~Teper,
{\it On the phase diagram of the SU(2) adjoint Higgs model in 2+1
dimensions } { hep-lat/9612021}.}

\lref\kajantie{K.~Kajantie, M.~Laine, K.~Rummujkainen, M.~Shaposhnikov
{\it 3D SU(N)+adjoint Higgs theory and finite temperature QCD }
{hep-ph/9704416}.}

\lref\batrouni{G. G. Batrouni, G. R. Katz, A. S. Kronfeld, G. P. Lepage,
B. Svetitsky and K. G. Wilson, {} {Phys. Rev. {\bf D32} (1985) 2736}}

\lref\davies{C. T. H. Davies, G. G. Batrouni, G. R. Katz, A. S. Kronfeld,
G. P. Lepage, K. G. Wilson, P. Rossi and B. Svetitsky, {} {Phys. Rev. {\bf
D41} (1990) 1953}}

\lref\fukugita{M. Fukugita, Y. Oyanagi and A. Ukawa, {}
{Phys. Rev. Lett. {\bf 57} (1986) 953;
Phys. Rev. {\bf D36} (1987) 824}}

\lref\kronfeld{A. S. Kronfeld, {\it Dynamics of Langevin Simulation} {Prog.
Theor. Phys. Suppl. {\bf 111} (1993) 293}}

\lref\polyakov{ A.~Polyakov, {} {Phys. Letts. {\bf B59} (1975) 82;
Nucl. Phys. {\bf B120} (1977) 429}; {\it Gauge fields and strings,}
ch. 4 (Harwood Academic Publishers, 1987).}

\lref\thooft{ G.~'t Hooft, {} {Nucl. Phys. {\bf B79} (1974) 276}; {}
Nucl. Phys. {\bf B190} (1981) 455 {}.}

\lref\elitzur{S.~Elitzur, {} {Phys. Rev. {\bf D12} (1975) 3978}}

%\polyakov \nadkarni \stackteper \kajantie

%%%%%%%%%%%%%%%%%%%%%%%%%%%%%%%%%%%%%%%%%%%%%%%%%%%%%%%%%%%%%%%%%
\lref\baugros{ {L.~Baulieu, B.~Grossman, } {\it A topological Interpretation
of  Stochastic Quantization} {Phys. Lett. B {\bf  212} {(1988)} {351}.}}

\lref\bautop{ {L.~Baulieu}{ \it Stochastic and Topological Field Theories},
{Phys. Lett. B {\bf   232} (1989) {479}}{}; {}{ \it Topological Field Theories
And Gauge Invariance in Stochastic Quantization}, {Int. Jour. Mod.  Phys. A
{\bf  6} (1991) {2793}.}{}}

\lref\bautopr{  {L.~Baulieu, B.~Grossman, } {\it A topological Interpretation
of  Stochastic Quantization} {Phys. Lett. B {\bf  212} {(1988)} {351}};
 {L.~Baulieu}{ \it Stochastic and Topological Field Theories},
{Phys. Lett. B {\bf   232} (1989) {479}}{}; {}{ \it Topological Field Theories
And Gauge Invariance in Stochastic Quantization}, {Int. Jour. Mod.  Phys. A
{\bf  6} (1991) {2793}.}{}}

\lref\samson{ {L.~Baulieu, S.L.~Shatashvili, { \it Duality from Topological
Symmetry}, {JHEP {\bf 9903} (1999) 011, hep-th/9811198.}}}{}

\lref\halpern{ {H.S.~Chan, M.B.~Halpern}{}, {Phys. Rev. D {\bf   33} (1986)
{540}.}}

\lref\yue{ {Yue-Yu}, {Phys. Rev. D {\bf   33} (1989) {540}.}}

\lref\neuberger{ {H.~Neuberger,} {\it Non-perturbative gauge Invariance},
{ Phys. Lett. B {\bf 175} (1986) {69}.}{}}

\lref\huffel{ {P.H.~Daamgard, H. Huffel},  {}{Phys. Rep. {\bf 152} (1987)
{227}.}{}}

\lref\creutz{ {M.~Creutz},  {\it Quarks, Gluons and  Lattices, }  Cambridge
University Press 1983, pp 101-107.}

\lref\zinn{ {J. ~Zinn-Justin, }  {Nucl. Phys. B {\bf  275} (1986) {135}.}}

\lref\shamir{  {Y.~Shamir,  } {\it Lattice Chiral Fermions
  }{ Nucl.  Phys.  Proc.  Suppl.  {\bf } 47 (1996) 212,  hep-lat/9509023;
V.~Furman, Y.~Shamir, Nucl.Phys. B {\bf 439 } (1995), hep-lat/9405004.}}

 \lref\kaplan{ {D.B.~Kaplan, }  {\it A Method for Simulating Chiral
Fermions on the Lattice,} Phys. Lett. B {\bf 288} (1992) 342; {\it Chiral
Fermions on the Lattice,}  {  Nucl. Phys. B, Proc. Suppl.  {\bf 30} (1993)
597.}}

\lref\neubergerr{ {H.~Neuberger, } {\it Chirality on the Lattice},
hep-lat/9808036.}

\lref\neubergers{ {Rajamani Narayanan, Herbert Neuberger,} {\it INFINITELY MANY
    REGULATOR FIELDS FOR CHIRAL FERMIONS.}
    Phys.Lett.B302:62-69,1993.
    [HEP-LAT 9212019]}

\lref\neubergert{ {Rajamani Narayanan, Herbert Neuberger,}{\it CHIRAL FERMIONS
    ON THE LATTICE.}
    Phys.Rev.Lett.71:3251-3254,1993.
    [HEP-LAT 9308011]}

\lref\neubergeru{ {Rajamani Narayanan, Herbert Neuberger,}{\it A CONSTRUCTION
OF LATTICE CHIRAL GAUGE THEORIES.}
    Nucl.Phys.B443:305-385,1995.
    [HEP-TH 9411108]}

\lref\neubergerv{ {Herbert Neuberger,}{\it EXACTLY MASSLESS QUARKS ON THE
    LATTICE.}
    Phys. Lett. B417 (1998) 141-144.
    [HEP-LAT 9707022]}

%The first 3 papers deal with chiral fermions in general, while the last
%with the
%particular case of vector like fermions. All these papers are quite well
%known.
%
%If you wish to quote reviews, the review by Shamir is seriously flawed.
%More recent
%reviews are available.  Surprisingly, I happen to like:

\lref\neubergerw{ {Herbert Neuberger,}{\it CHIRAL FERMIONS ON THE
LATTICE.}
    Nucl. Phys. B, Proc. Suppl. 83-84 (2000) 67-76.
    [HEP-LAT 9909042]}

\lref\zbgr {L.~Baulieu and D. Zwanziger, {\it QCD$_4$ From a
Five-Dimensional Point of View},    Nucl. Phys. {\bf B581} 2000, 604;
hep-th/9909006.}

\lref\bgz {P. A. Grassi, L.~Baulieu and D. Zwanziger, {\it Gauge and
Topological Symmetries in the Bulk Quantization of Gauge Theories},
Nucl. Phys. {\bf B597} 583, 2001 hep-th/0006036.}

\lref\bulkq {L.~Baulieu and D. Zwanziger, {\it From stochastic
quantization to bulk quantization; Schwinger-Dyson equations and
the S-matrix}, JHEP 08:016, 2001 hep-th/0012103.}

\lref\bulkqg {L.~Baulieu and D. Zwanziger, {\it Bulk quantization of gauge
theories: confined and Higgs phases}, JHEP 08:015, 2001 and
hep-th/0107074.}

\lref\equivstoch{L.~Baulieu and D. Zwanziger, {\it Equivalence of
stochastic quantization and the Faddeev-Popov Ansatz},
Nucl. Phys. B193 (1981) 163-172.}

 \lref\zbsd {L.~Baulieu and D. Zwanziger, {
\it From stochastic quantization to bulk quantization: Schwinger-Dyson
equations and S-matrix QCD$_4$}, hep-th/0012103.}

\lref\cuzwns {A.~Cucchieri and D.~Zwanziger, {\it Numerical study of
gluon propagator and confinement scenario in minimal Coulomb gauge},
hep-lat/0008026.}

\lref\vanish {D.~Zwanziger, {\it Vanishing of zero-momentum lattice
gluon propagator and color confinement},   Nucl. Phys. {\bf B364}
(1991) 127-161.}

\lref\gribov {V.~N.~Gribov, {\it Quantization of non-Abelian gauge
theories},   Nucl. Phys. {\bf B139} (1978)~1-19.}

\lref\singer {I.~Singer, {\it }   Comm. Math. Phys. {\bf 60}
(1978)~7.}

\lref\feynman {R.~P.~Feynman, {\it The qualitative behavior of
Yang-Mills theory in 2+1 dimensions},   Nucl. Phys. {\bf B188} (1981)
479-512.}

\lref\cutkosky {R.~E.~Cutkosky, {\it}   J. Math. Phys. {\bf 25} (1984)
939; R. E. Cutkosky and K. Wang, Phys. Rev. {\bf D37} (1988) 3024; R. E.
Cutkosky, Czech. J. Phys. {\bf 40} (1990) 252.}

\lref\vanbaal{J. Koller and P. van Baal, Nucl. Phys. {\bf B302} (1988)
1; P. van Baal, Acta Phys. Pol. {\bf B20} (1989) 295;
P. van Baal, Nucl. Phys. {\bf B369} (1992) 259;
P. van Baal and N. D. Hari Dass, Nucl. Phys. {\bf B385} (1992) 185.}

\lref\vanbaalg{P.~van Baal, {\it Gribov ambiguities and the fundamental
domain}  (Cambridge U. \& Leiden U.). INLO-PUB-10-97, Jun 1997. 18pp. 
Talk given at NATO Advanced Study Institute on Confinement, Duality and
Nonperturbative Aspects of QCD, Cambridge, England, 23 Jun - 4 Jul 1997.  In
*Cambridge 1997, Confinement, duality, and nonperturbative aspects of QCD*
161-178; hep-th/9711070.}

\lref\stingl {M. Stingl, {\it Propagation properties and condensate
formation of the confined Yang-Mills field},   Phys. Rev. D {\bf 34} (1986)
3863-3881.}

\lref\smekal{L.~von Smekal, A.~Hauck and R.~Alkofer,  {\it A Solution to
Coupled Dyson-Schwinger Equations in Gluons and Ghosts in Landau Gauge,}  
Ann. Phys. {\bf 267} (1998) 1; L. von Smekal, A. Hauck and R. Alkofer, {\it The
Infrared Behavior of Gluon and Ghost Propagators in Landau Gauge QCD,}  
Phys. Rev. Lett. {\bf 79} (1997) 3591; L. von Smekal {\it Perspectives for
hadronic physics from Dyson-Schwinger equations for the dynamics of quark and
glue,} Habilitationsschrift, Friedrich-Alexander Universit\"{a}t,
Erlangen-N\"{u}rnberg (1998).}

\lref\smekrev {R.~Alkofer and L.~von Smekal, {\it The infrared behavior of
QCD Green's functions},   Phys. Rept. {\bf 353}, 281 (2001).}

\lref\fischalk {C. S. Fischer and R.~Alkofer, {\it Infrared exponents and
running coupling of SU(N) Yang-Mills Theories},   Phys. Lett. B {\bf 536},
177 (2002).}

\lref\fischalkrein{C.~S.~Fischer, R.~Alkofer and H.~Reinhardt,
   {\it The elusiveness of infrared critical exponents in Landau gauge
   Yang-Mills theories,}
   Phys. Rev. D {\bf 65}, 094008 (2002)}

\lref\fischalkqu{C.~S.~Fischer and R.~Alkofer,
   {\it Non-perturbative propagators, running coupling and dynamical quark
mass of Landau gauge QCD,}
   hep-ph/0301094}

\lref\lerche {C. Lerche and L. von Smekal {\it On the infrared exponent for
gluon and ghost propagation in Landau gauge QCD}, hep-ph/0202194}

\lref\atkinsona {D.~Atkinson and J.~C.~R.~Bloch, {\it Running coupling in
non-perturbative QCD}   Phys. Rev. {\bf
D58} (1998) 094036.}

\lref\atkinsonb {D.~Atkinson and J.~C.~R.~Bloch, {\it QCD in the infrared with
exact angular integrations}   Mod. Phys. Lett. {\bf A13} (1998) 1055.}

\lref\brown {N.~Brown and M.~R. Pennington, {\it}   Phys. Rev. {\bf D38} (1988)
2266; Phys. Rev. {\bf D39} (1989) 2723.}

\lref\szczep {A.~P.~Szczepaniak and E.~S.~Swanson, {\it Coulomb Gauge QCD,
Confinement, and the Constituent Representation},   hep-ph/0107078.}

%%%%%%%%%%%%%%%%%%%%CAPTIONS%%%%%%%%%%%%%%%%%%%%%%%%%%%%%%%%%%%%%%%%%%%%%%%

\nfig\compar{The coordinate patch $\cal U$ in $A$-space is the clam-shaped
region viewed edge on.  The Gribov region $\Omega$ is represented by the thick
horizontal line.}

\nfig\compar{The functional DS equation 
\dsghost\ for the complete ghost propagator ${\cal G}(x,y;B)$ in the presence
of the source $B$.   The thin line is the tree-level term.  The heavy line 
with (without) the arrow is the complete ghost (gluon) propagator
${\cal G}(x,y;B)$ (${\cal D}(x,y;B)$) in the presence of the source $B$.
The circle is the complete ghost-ghost-gluon vertex in the presence of the
source~$B$.}

\nfig\compar{The functional DS equation 
\dsglue\ for the complete gluon propagator ${\cal D}(x,y;B)$ in the presence of
the source $B$.   The thin line is the tree-level term.  The heavy line 
with (without) the arrow is the complete ghost (gluon) propagator
${\cal G}(x,y;B)$ (${\cal D}(x,y;B)$) in the presence of the source $B$.
The circles are complete 3- and 4-vertices in the presence of the source~$B$.}

\nfig\compar{The functional DS equation \asdsglue\ for the complete infrared
asymptotic  gluon propagator $\hat{\cal D}(x,y;B)$ in the presence of
the source $B$.   There is no tree term nor any gluon loop, but only the ghost
loop. The heavy line with the arrow is the complete infrared asymptotic ghost
propagator $\hat{\cal G}(x,y;B)$ in the presence of the source~$B$.
The functional DS equation \rasdsghost\ for the complete infrared
asymptotic  ghost propagator $\hat{\cal G}(x,y;B)$ in the presence of
the source $B$ is as in Fig.~2. }

%%%%%%

%\draft

%%%%%%%%%

\Title{\vbox
{\baselineskip 10pt
\hbox{hep-ph/0303028}
%\hbox{CERN-TH-00-??}
%\hbox{LPTHE-00-50}
\hbox{NYU-PH-TH-03.03.03}
 \hbox{   }
}}
{\vbox{\vskip -30 true pt
\centerline{
   }
\medskip
 \centerline{  }
\centerline{Non-perturbative Faddeev-Popov formula}
\centerline{and infrared limit of QCD}
\medskip
\vskip4pt }}
\centerline{
%{\bf Laurent Baulieu}$^{  \dag     }$ and  
{\bf  Daniel Zwanziger}
%$^{ \ddag}$
}
\centerline{
%baulieu@lpthe.jussieu.fr, 
daniel.zwanziger@nyu.edu}
%pag5@nyu.edu
\vskip 0.5cm
%\centerline{\it $^{\dag}$LPTHE, Universit{\'e}s P. \& M. Curie (Paris~VI)
%et D. Diderot (Paris~VII), Paris,  France,}
%{\foot{UMR 7589 associ{\'e}e CNRS et
%Universit{\'e}s P. \&M. Curie (Paris~VI) et D. Diderot (Paris~VII)},
%Boite 126,
%4 place Jussieu, F-75252
%Paris Cedex 05, France.}
%\centerline{\it $^{\dag}$  Dept. of Physics, University
%of Rutgers, New Brunswick, NJ 60637, USA }
%\centerline{\it $^{\S}  $}
\centerline{\it 
%$^{\ddag}$   
Physics Department, New York University,
New-York,  NY 10003,  USA}

\medskip
%\vskip  1cm
\noindent

We show that an exact non-perturbative quantization of continuum gauge theory
is provided by the Faddeev-Popov formula in Landau gauge, 
$\d(\p \cdot A)  \det[-\p \cdot D(A)] \ \exp[-S_{\rm YM}(A)]$,
restricted to the region where the Faddeev-Popov operator is positive 
$-\p \cdot D(A) > 0$ (Gribov region). Although there are Gribov copies inside
this region, they have no influence on expectation-values.  The starting point
of the derivation is stochastic quantization which determines the Euclidean
probability distribution~$P(A)$ by a method that is free of the
Gribov critique.  In the Landau-gauge limit the support of
$P(A)$ shrinks down to the Gribov region with Faddeev-Popov weight.  The
cut-off of the resulting functional integral on the boundary of the Gribov
region does not change the {\it form} of the Dyson-Schwinger equations, because
$\det[-\p \cdot D(A)]$ vanishes on the boundary, so there is no boundary
contribution.  However this cut-off does provide
{\it supplementary conditions} that govern the choice of solution of the DS
equations.  In particular the ``horizon condition", though consistent with the
perturbative renormalization group, puts QCD into a non-perturbative phase.  The
infrared asymptotic limit of the DS equations of QCD is obtained by
neglecting the Yang-Mills action $S_{\rm YM}$.  We sketch the extension to a
BRST-invariant formulation.   In the infrared asymptotic limit, the
BRST-invariant action becomes BRST-exact, and defines a topological quantum
field theory with an infinite mass gap.  Confinement of quarks is discussed
briefly.

\Date{\ }

\def\e{\epsilon}

\def\a{\alpha}
\def\b{\beta}
\def\d{\delta}
\def\c{\gamma}
\def\m{\mu}
\def\n{\nu}
\def\r{\rho}
\def\s{\sigma}
\def\l{\lambda}
\def\x{\xi}

\def\o{\omega}
\def\L{\Lambda}

\def\k{\kappa}

\def\t{\theta}  \def\T{\Theta}

\newsec{Introduction}

	Since the work of Gribov \gribov, a non-perturbative formulation of continuum
gauge theory has appeared problematical due to the existence of Gribov copies.
These are distinct but gauge-equivalent configurations 
$A^{(2)} ={^g} A^{(1)}$
that both satisfy the gauge condition, 
$\p \cdot A^{(1)} = \p \cdot A^{(2)} = 0$,
where ${^g}A_\m = g^{-1}A_\m g + g^{-1}\p_\m g$
is a local gauge transformation.  The difficulty arises when one wishes to
quantize by {\it gauge fixing} namely by taking a single representative
configuration on each gauge orbit.  It has been proven that this cannot be
done in a continuous way when space-time is compactified \singer. 
Geometrically this reflects the intricacy of {\it gauge orbit space}, the
space of configurations $A$ modulo local gauge transformations $g$.  

	There is however an approach that by-passes the difficulties of Gribov copies
by operating directly in $A$-space.  This approach is stochastic
quantization.  For our purposes it is most conveniently expressed by the
time-independent Fokker-Planck equation (given below) that determines the
Euclidean probability distribution $P(A)$.  The geometric structure of the
equation assures that $P(A)$ is correctly weighted.  Although one
cannot solve the Fokker-Planck equation exactly for finite values of the
gauge parameter $a$, one can transform it into a system of Dyson-Schwinger
(DS) equations for the correlation functions, that may be solved
non-perturbatively, as has been done recently
\dztimeind.  However these equations are more cumbersome than the
DS equations in an action formalism. 

	In secs.~2, 3, and 4, we find the exact solution of the time-independent
Fokker-Planck equation in the Landau-gauge limit
$a \to 0$.  The solution is remarkably simple.  It is the familiar
Faddeev-Popov weight, but restricted to the Gribov region $\Omega$,\foot{The
Yang-Mills action is given by
$S_{\rm YM}(A) = (1/4)\int d^4x \  F_{\m\n}^2$ where
$F_{\m\n}^a = \p_\m A_\n^a - \p_\n A_\m^a + g_0f^{abc}A_\m^b A_\n^c$,
and the gauge-covariant derivative by
$[D_\m(A) \o]^a \equiv \p_\m \o^a + g_0 f^{abc}A_\m^b \o^c$.
The Faddeev-Popov operator $M(A) \equiv - \p \cdot D(A)$ is symmetric when
$A$ is transverse, 
$M(A) = - \p \cdot D(A) = - D(A) \cdot \p = M^{\dag}(A)$.
Positivity of $M(A)$ means all its non-trivial eigenvalues $\l_n(A)$ are
positive.   There is a trivial null
eigenvalue with constant eigenvectors $\p_\m \o = 0$, that are generators
of global gauge transformations.  In Appendix C we establish three simple
properties of the Gribov region.} 
\eqn\solution{\eqalign{
P(A) = N \ \d_\Omega(\p \cdot A) 
\ \det[- \p \cdot D(A)] \ \exp[-S_{YM}(A)].
}} 
The Gribov region $\Omega$ is, by definition, the region in $A$-space where
$A$ is transverse, and the Faddeev-Popov operator 
$M(A) \equiv - \p \cdot D(A)$ is positive,
\eqn\gomega{\eqalign{
\Omega \equiv \{A: \ \p \cdot A = 0 \ ; - \p \cdot D(A) > 0 \ \}.
}}
The first factor 
$\d_\Omega(\p \cdot A)$ in \solution\ is the restriction of $\d(\p \cdot A)$
to the region where $M(A)$ is positive.  Observables ${\cal O}(A)$ are
required to be gauge-invariant, ${\cal O}({^g}A) = {\cal O}(A)$ and,
by~\solution,
expectation-values are calculated from
\eqn\expectval{\eqalign{ 
\langle {\cal O}(A) \rangle & = \int dA \ {\cal O}(A) \ P(A)   \cr
& = N \int_\Omega dA^{\rm tr}  \ {\cal O}(A^{\rm tr})
  \ \exp[-S_{YM}(A^{\rm tr})] \ \det[- \p \cdot D(A^{\rm tr})],
}}
where $A^{\rm tr}$ is the transverse part of $A$.  Two comments are in order.

	(i)  {\it Gribov region $\Omega$ vs fundamental modular region $\L$.}  Formula
\expectval\ is paradoxical because the Gribov region~$\Omega$ is {\it
not} free of Gribov copies~\semenov.  The history of this formula is
amusing.  It was originally proposed by Gribov who conjectured in his seminal
work~\gribov\ that there are no Gribov copies in~$\Omega$.  The same
formula was also derived from stochastic quantization \dzstochquant\ by a
method similar to the one presented in the present article (but using globally
defined coordinates instead of coordinates defined only on a coordinate
patch), and was interpreted to mean that the Gribov region $\Omega$ is free of
Gribov copies.  However it was then proven \semenov, with details provided in
\gfdadzinside, that there are Gribov copies inside~$\Omega$.  Moreover
numerical studies \mandula, \marinari, \deforcrand, \nakamuramiz, and
\marenzoni\  revealed that  in general there are many Gribov copies of a
given configuration inside $\Omega$.
Consequently \expectval\ was generally abandoned as an exact formula in favor
of an integration over a region free of Gribov copies, known as the
fundamental modular region $\L$,
\eqn\fundmod{\eqalign{
\langle {\cal O}(A) \rangle = N \int_\L DA^{\rm tr} \det M(A^{\rm tr}) 
\ {\cal O}(A^{\rm tr}) 
\ \exp[-S_{\rm YM}(A^{\rm tr})].
}}
The last formula is certainly correct and appears to contradict \expectval.  It
was subsequently argued nevertheless \horizcona\ that the functional integral 
\expectval\ is in fact dominated by configurations on the common boundary of
$\Omega$ and $\L$.  The derivation given in secs.~2,~3, and~4 shows that
\expectval\ is indeed correct.  This is most fortunate because it is difficult
to give an explicit description of $\L$.  In Appendix A we examine concretely
how the paradox is resolved. The lesson is that the {\it normalized}
pobability distributions over $\L$ and $\Omega$ are equal in the sense
that their moments of finite order~$n$ are equal.  These are the correlation
functions 
$\langle A(x_1) A(x_2)...A(x_n)\rangle$.  This is possible in an
infinite-dimensional space, where the probability distribution may sit on a
lower dimensional subspace such as a boundary or part of a boundary.  This
conclusion is consistent with numerical investigation of ``Gribov noise",
namely the effect on measured quantities of taking different Gribov copies. 
Indeed for the gluon propagator in Landau gauge on reasonably large lattices,
Gribov noise is quite small, of the same magnitude as the numerical accuracy
\cucchierigh,
\mandulab, \giustietal.  The situation is quite different for a
finite-dimensional integral, and the analogous problem for a finite lattice is
also discussed in Appendix~A.  Formula \expectval\ is also supported
by a recent calculation in which the DS equation for the gluon propagator was
derived from the time-independent Fokker-Planck equation at finite gauge
parameter $a$.  It was found to agree with the DS equation for the gluon
propagator in Faddeev-Popov theory  in the Landau gauge limit, $a \to 0$, see
particularly eqs.~(9.4), (10.13), (10.14) and (10.17) of \dztimeind.

	(ii)  {\it The form of the DS equations is unchanged by the cut-off on the
boundary of}~$\Omega$.  The DS equations are a set of equations
for the correlation functions 
$\langle A(x_1) A(x_2)...A(x_n)\rangle$.  We shall derive them 
for the distribution \expectval\ in secs.~5 and~6.  They are compactly
expressed as a single functional differential equation for the 
partition function or generating functional of correlation functions,
\eqn\partition{\eqalign{
Z(J) = N \int_\Omega dA^{\rm tr} \ \det[- \p \cdot D(A^{\rm tr})] 
\ \exp[ \ -S_{YM}(A^{\rm tr}) + (J,A^{\rm tr}) \ ].
}}
The functional DS equation for $Z(J)$ follows from the identity,
\eqn\noboundaryt{\eqalign{
0 = N \int_\Omega dA^{\rm tr} \ { { \d } \over { \d A^{\rm tr}} }
 \Big( \ \det[- \p \cdot D(A^{\rm tr})] 
\ \exp[-S_{YM}(A^{\rm tr}) + (J,A^{\rm tr})] \Big),
}}
which states that the integral of a derivative vanishes when there is
no boundary contribution.  There is in fact no boundary contribution, despite
the cut-off on the boundary $\p \Omega$, defined by the
equation $\l_1(A^{\rm tr}) = 0$, because the Faddeev-Popov determinent 
$\det[ -\p \cdot D(A^{\rm tr})] = \prod_n \l_n(A^{\rm tr}) $ {\it vanishes} on
$\p \Omega$. Thus
the {\it form} of the DS equation is the same as if the integral were extended
to infinity~\dznonpertl.  Again this is most fortunate because it means that
implementing the restriction to the Gribov region causes no complication at
all in the DS equations.  

	Although the restriction to the interior of the Gribov horizon 
does not change the form of the DS equations,	 it does provide {\it
supplementary conditions} that govern the choice of solution.  In fact the
properties that result from the restriction to $\Omega$, in particular the
positivity of the weight
$P(A)$ and of the Faddeev-Popov operator $M(A)$, dictate the natural choice of
solution of the DS equation, that has been implemented previously, without
necessarily invoking explicitly the cut-off at $\p \Omega$, ~\smekal,
\atkinsona,
\atkinsonb, \lerche, \dznonpland, \dztimeind,    
 \fischalkrein, \fischalk,  and reviewed in~\smekrev.
Another property is the {\it horizon condition}~\horizcon.
This is an enhancement,\foot{{\it Entropy} \ favors population
near the boundary, in a configuration space with a high number $N$ of
dimensions, because of the volume element
$r^{N-1}dr$. The boundary $\p \Omega$ of the Gribov region $\Omega$ occurs
where the lowest non-trivial eigenvalue of the Faddeev-Popov operator $M(B)$
vanishes so, for typical configurations~$B$ on a large Euclidean volume~$V$, 
$M(B)$ has a very small eigenvalue.  More precisely, compared to the Laplacian
operator, $M(B)$ has a high density per unit volume
of eigenvalues $\r(\l, B)$ at $\l = 0$ \horizcon. This enhances the ghost
propagator $G(x-y) =  \langle M^{-1}_{xy}(A) \rangle$ in the infrared.}
compared to $1/k^2$, of the ghost propagator
$\tilde{G}(k)$ in the infrared,
$\lim_{k \to 0}[k^2 \tilde{G}(k)]^{-1} = 0$.\foot{The confinement criterion of
Kugo and Ojima \kugoj, \nakanoj, \nishijima\ yields the same condition
in the Minkowskian theory.  However for gauge-non-invariant quantities, 
the relation of the present approach, with a cut-off at the Euclidean
Gribov horizon, to the Minkowskian theory remains to be clarified, perhaps
along the lines of Appendix B.  The relation of numerical gauge fixing by
minimization in (Euclidean) lattice gauge theory to the Minkowskian theory is
also not clear.}  In sec.~7 we show that the horizon condition is most
conveniently expressed as a formula for the ghost-propagator renormalization
constant
$\tilde{Z}_3$.  Although this formula flagrantly contradicts perturbation
theory, it is nevertheless consistent with the perturbative renormalization
group.  The horizon condition puts QCD into a non-perturbative phase.  

	In sec.~8 we deduce the asymptotic infrared limit of~QCD by neglecting the
terms in the DS equations that are subdominant in the infrared.  It is found
that the subdominant terms and only the subdominant terms come from the
Yang-Mills action $S_{\rm YM}(A)$, so the infrared asymptotic limit of QCD is
obtained by setting $S_{\rm YM}(A) = 0$.  This  is a continuum analog of the
strong coupling limit of lattice gauge theory.  The functional integral
with $\exp[-S_{\rm YM}(A)]$ replaced by 1 converges because it is cut off at
the Gribov horizon.  

	In Appendix B we outline the local BRST-invariant formulation of the present
non-perturbative formulation.  This assures that the Slavnov-Taylor identities
hold at the non-perturbative level.  In the infrared asymptotic limit,
obtained by setting 
$S_{\rm YM}(A) = 0$, the BRST-invariant action becomes BRST-exact, and defines
a topological quantum field theory.  As shown in sec.~9, this theory possess an
infinite mass gap in the physical sector.  In sec.~10 the extension to quarks
is sketched out. 

	The starting point of our derivation will be stochastic quantization of
gauge fields.   In the remainder of the Introduction we give a brief review of
this subject so the reader may judge of the well-foundedness of this
approach at the non-perturbative level.

\subsec{Review of stochastic quantization of gauge fields}

	Historically, stochastic quantization originated \pawu\ with the
observation that the formal, unnormalizable Euclidean proabability distribution
$P_0(A) = N \exp[-S_{\rm YM}(A)]$, with 4-dimensional Euclidean 
Yang-Mills action $S_{\rm YM}(A)$, is the equilibrium distribution of the
stochastic process defined by the equation, 
\eqn\fokplan{\eqalign{  
 { { \p P} \over { \p t } } = \int d^4x \ { { \d } \over { \d A_\m^a(x) } } 
     \Big( { { \d P } \over { \d A_\m^a(x) } } 
    + { { \d S_{\rm YM} } \over { \d A_\m^a(x) } } P \Big) 
}}
for the time-dependent probability distribution $P(A,t)$.  This equation
is a continuum analog of the diffusion equation in the presence of
the drift force $K_i$,
\eqn\discretea{\eqalign{
{ { \p P} \over { \p t } } = 
{{\p} \over {\p A^i}}  \Big( {{\p P} \over {\p A^i}} - K_i P \Big) = 0,
}}
that is known as the Fokker-Planck equation.   If the drift force is
conservative, 
$K_i = - {{\p S_{\rm YM}} \over {\p A^i}}$, then
$\exp[-S_{\rm YM}(A)]$ is a time-independent solution.  In
Euclidean quantum field theory, $t$ is an artificial 5th time that
corresponds to the number of sweeps in a Monte-Carlo simulation, and  
that will be eliminated shortly.  The same stochastic process may equivalently
be represented by the Langevin equation
\eqn\langevin{\eqalign{
{ { \p A_\m^a } \over { \p t } } 
= - { { \d S_{\rm YM} } \over { \d A_\m^a } } + \eta_\m^a,
}}
where $A_\m^a = A_\m^a(x, t)$ depends on the artificial 5th time.  Here 
$\eta_\m^a = \eta_\m^a(x, t)$ is Gaussian white noise defined by 
$ \langle \eta_\m^a(x, t)\rangle = 0 $ and 
$ \langle \eta_\n^b(x, t) \eta_\m^a(x, t) \rangle 
  = 2 \d(x-y) \d_{\m\n} \d^{ab} \d(t - t') \rangle$. 
If $N \exp[-S_{\rm YM}(A)]$ were a normalizable probability distribution ---
which it is not --- every normalized solution to \fokplan\ would relax to it as
equilibrium distribution.   However the process defined by \fokplan\ or
\langevin\ does not provide a restoring force in gauge orbit directions, so
probability escapes to infinity along the gauge orbits, and as a result 
$P(A,t)$ does not relax to a well-defined limiting distribution
$\lim_{t \to \infty} P(A,t) \neq  N \exp[-S_{\rm YM}(A)]$ (although
expectation-values of gauge-invariant observables formally do relax to an
equilibrium value).  

	A remedy is provided by the observation \dan\ that
the Langevin equation may be modified by the addition of an infinitesimal gauge
transformation, 
$D_\m^{ac} v^c$,
\eqn\fixlangevin{\eqalign{
{ { \p A_\m^a } \over { \p t } } = - { { \d S} \over { \d A_\m^a } }
    + D_\m^{ac} v^c + \eta_\m^a,
}}
where $v^c$ is at our disposal.  This cannot alter the
expectation-value of gauge-invariant quantities, for only a harmless
infinitesimal gauge-transformation $K_{{\rm gt},\m} = D_\m v$ has been
introduced. In the language of the diffusion equation, we may say that the
additional drift force~$K_{{\rm gt},\m}$ is tangent to the gauge orbit.  
The modified  Langevin  equation is equivalent to the modified Fokker-Planck
equation 
\eqn\fixfokplan{\eqalign{  
 { { \p P} \over { \p t } }  = \int d^4x{ { \d } \over { \d A_\m^a(x) } } 
     \Big( { { \d P } \over { \d A_\m^a(x) } } 
    - K_\m^a(x) P \Big)   \cr  
K_\m^a(x)   \equiv - { { \d S_{\rm YM} } \over { \d A_\m^a(x) } } 
       + D_\m^{ac} v^c(x),
}}
We will choose $v^c(x)$ to make $D_\m^{ac} v^c(x)$ globally
restoring along gauge orbit directions, so every normalized
solution $P(A,t)$ relaxes to a unique equilibrium distribution
$\lim_{t \to \infty} P(A,t) = P(A)$.  

	Stochastic quantization in the time-dependent formulation
has been developed by a number of authors  who have expressed the solution as a
functional integral \gozzi, and demonstrated the renormalizability of this
approach \zinn, \danzinn.  A systematic development is presented in \halperna,
\halpernb, \halpernc, \halpernd, \halperne, \sadun, reviewed in \halpernr, that
includes  the 4-and 5-dimensional Dyson-Schwinger equation for the quantum
effective action, an extension of the method to gravity, and gauge-invariant
regularization by smoothing in the 5th time.  Renormalizability has also been
established by an elaboration of BRST techniques~\bgz,~\bulkqg. Stochastic
quantization may be and has been exactly simulated numerically including on
rather large lattices, of volume $(48)^4$, \nakamuraa, \nakamurab, \nakamurac,
\nakamurad, \nakamurae.

\subsec{Time-independent stochastic quantization}

	When the drift force is globally restoring, $P(A)$ may be calculated directly 
without reference to the artificial 5th time as
the positive normalized solution of the time-independent Fokker-Planck equation
\eqn\tifokplan{\eqalign{        
  H P  \equiv \int d^4x{ { \d } \over { \d A_\m^a(x) } } 
     \Big( - { { \d P  } \over { \d A_\m^a(x) } } + K_\m^a P \Big)  = 0   \cr  
K_\m^a(x)   \equiv - { { \d S_{\rm YM} } \over { \d A_\m^a(x) } } 
       + D_\m^{ac} v^c(x),  
}}  
and Euclidean expectation values are calculated from 
$\langle {\cal O} \rangle = \int dA \ {\cal O}(A) P(A)$.
We call $H$ the ``Fokker-Planck hamiltonian".  (It is {\it
not} the quantum mechanical hamiltonian!).  It has been proven
directly~\dztimeind, without reference to the artificial time, that the
expectation value
$\langle {\cal O} \rangle_v$ of a gauge-invariant observable 
${\cal O}({^g}A) = {\cal O}(A)$, is independent of~$v$.   
Equation \tifokplan\ determines a probability distribution~$P(A)$ directly in
$A$-space, that is correctly weighted at the non-perturbative level.  The
Gribov problem of globally correct gauge-fixing by identifying gauge orbits is
by-passed. By contrast, in the Hamiltonian formulation of gauge theory, 
Gauss's law states that the wave functional 
$\Psi(\vec{A})$ is gauge-invariant and is thus a functional defined on 
the space of gauge orbits \vanbaalg. 

	To ensure that 
$K_{{\rm gt},\m} = D_\m v$ is globally restoring, we 
introduce a minimizing functional 
\masknaka, \dzmin, and \semenov,
and choose $K_{{\rm gt},\m}$
to be in the gauge-orbit direction of steepest descent.  A
convenient choice of minimizing functional\foot{More generally, we may take
for the minimizing function
$\int d^4x A_\m^a(x) \a_{\m\n}A_\n^a(x)$, where $\a_{\m\n}$ is a constant
positive symmetric matrix.  This defines a set of Lorentz-non-covariant but
normalizable gauges that includes the Coulomb gauge as a limiting case
\coulomb.  To include different instanton sectors, one may choose as
minimizing functional $||A - A_n||^2$, where $A_n$ is a fixed configuration of
given instanton number.  An alternative minimizing functional 
suitable for the Higgs phase
was proposed in \bulkqg.} is the Hilbert norm
$||A||^2 = \int d^4x |A|^2$.
For an infinitesimal variation in the gauge-orbit direction 
$\d A_\m = \e D_\m v$, we have 
\eqn\variation{\eqalign{
\d ||A||^2 = 2(A_\m, \d A_\m) = 2\e (A_\m, D_\m v) =  2\e (A_\m, \p_\m v)
= - 2\e (\p_\m A_\m, v),
}}
so steepest descent among gauge orbit directions of the minimizing functional
is provided by $v = a^{-1} \p \cdot A$ with $a > 0$, and the time-independent
Fokker-Planck equation is now specified to within a single gauge parameter,
\eqn\tifp{\eqalign{        
  H P  = \int d^4x{ { \d } \over { \d A_\m^a(x) } } 
     \Big( - { { \d P  } \over { \d A_\m^a(x) } } + K_\m^a P \Big)  = 0   \cr  
K_\m^a(x)   \equiv - { { \d S_{\rm YM} } \over { \d A_\m^a(x) } } 
       + a^{-1} D_\m^{ac} \p \cdot A^c(x),  
}}
(Symmetry and power-counting arguments  also determine 
$v^a = a^{-1}\p_\l A_\l^a = a^{-1}\p \cdot A^a$.)   

	Having introduced the minimizing functional, we note that the
Gribov region $\Omega$ may be characterized as the set of 
{\it relative} minima\foot{At any minimum, this functional is
(a)~stationary, and (b)~the matrix of second derivatives is non-negative. 
These two conditions fix the properties that define the Gribov region:
(a)~transversality, $\p \cdot A = 0$, and (b)~positivity of the
Faddeev-Popov operator $-D(A)
\cdot \p$.   Property (a) follows from
\variation, which states that the first variation of the
minimizing functional is 
$\d ||A||^2 = - 2(\o, \p \cdot A) $.
Property~(b) follows because
the second variation is
$\d^2 ||A||^2 = - 2(\o, \p \cdot D(A) \o)$.}
 with respect to local gauge transformations $g(x)$ of the
minimizing functional $F_A(g) \equiv ||{^g}A||^2$, whereas the fundamental
modular region $\L$ may be characterized as the set of {\it absolute} minima. 
The set of absolute minima is free of Gribov copies, apart from the
identification of gauge-equivalent points on the boundary $\p\L$, and may be
identified with the gauge orbit space.  In a lattice discretization the
minimization problem is of spin-glass type, and one expects many nearly
degenerate local minima on a typical gauge orbit, as is verified by numerical
studies.   Thus $\L$ is a proper subset of $\Omega$, 
$\L \subset \Omega$, but $\L \neq \Omega$. 

\subsec{Region of stable equilibrium of $K_{\rm gt}$}

	The gauge transformation ``force" $K_{\rm gt}$ is not conservative, and cannot
be written, like the first term, as the gradient of some 4-dimensional
gauge-fixing action,
$K_{{\rm gt},\m} = a^{-1}D_\m^{ac} \p \cdot A^c(x) 
\neq - { { \d S_{\rm gf} } \over { \d A_\m^a(x) } }$,
so we cannot write the solution $P(A)$ explicitly in general.  However we shall
solve \tifp\ for
$P(A)$ exactly in the limit $a \to 0$.   In this limit
$P(A)$ gets concentrated in the region of {\it stable equilibrium} of the
force	
$K_{{\rm gt},\m} = a^{-1}D_\m \p \cdot A$.

	{\it Assertion:} The region of stable equilibrium under the gauge
transformation force $K_{{\rm gt},\m} = D_\m \p \cdot A$ is the Gribov region
$\Omega$.  
{\it Proof:} Transversality is a sufficient condition for equilibrium, because
$\p \cdot A = 0$ implies $K_{{\rm gt},\m} = 0$. 
It is also necessary.  For consider the flow under this force,
$\dot{A}_\m = D_\m \p \cdot A$.  We have
$\p ||A||^2 / \p t = 2(A_\m, \dot{A}_\m) = 2 (A_\m, D_\m \p \cdot A)
= 2 (A_\m, \p_\m \p \cdot A) = - 2 ||\p \cdot A||^2 \leq 0$,
which is negative unless $\p \cdot A = 0$.  We conclude that the region
of equilibum under $K_{\rm gt}$, which may be stable or unstable, is the set
of transverse configurations.  To find the region of stable
equilibrium, observe that under this flow, we have
${ {\p } \over {\p t} }\p \cdot A =  \p \cdot \dot{A}
=  \p \cdot D(A) \ \p \cdot A$.  We linearize this equation to first order in 
$\p \cdot A$, which means taking  
$\p \cdot D(A) \to \p \cdot D(A^{\rm tr}) \equiv - M(A^{\rm tr})$,
and we have
${ {\p } \over {\p t} }\p \cdot A =  - M(A^{\rm tr}) \ \p \cdot A$.  Thus the
equilibrium is stable when all eigenvalues of $M(A^{\rm tr})$ are positive,
and it is unstable otherwise.  QED.

\newsec{A well-defined change of variable}

	In order to solve the time-independent Fokker-Planck equation 
\tifp\ in the limit $a \to 0$, we only need the solution for small $a$
{\it in a coordinate patch $\cal U$ in $A$-space that includes the
Gribov region} $\Omega$.  In $\cal U$, we make the change of variable 
$A \to (B, g)$,  defined by the gauge transformation,
\eqn\changevar{\eqalign{
A_\m = A_\m(B,g) = {^g}B_\m = g^{-1}\p_\m g + g^{-1} B_\m g;  \ \ \ \  
{\rm with}
\ \   \p \cdot B = 0  \ \  {\rm and} \ \   M(B) > 0,
}}
where~$B \in \Omega$.  Local gauge transformations are parametrized by 
$g(x) = \exp[t^a \t^a(x)]$ where, for each $x$, the $\t^a(x)$ are coordinates
for the SU(N) group.\foot{Here and below we use the notation $A_\m \equiv
t^a A_\m^a$ and $B_\m \equiv t^a B_\m^a$.  The $t^a$ are set of anti-hermitian
traceless matrices that form the fundamental representation of the Lie algebra
of SU(N), $[t^a, t^b] = f^{abc}t^c$, where the structure constants
$f^{abc}$ are completely anti-symmetric.}  The notation $A = A(B, g)$ is
understood to stand for $A = A(B, \t)$, and we have $B = A(B, 0)$.

	Gribov's critique of the Faddeev-Popov method is that this change of variable
is not well-defined for all transverse~$B$ and~$g$.  We shall show
however that it is well-defined in a coordinate patch~${\cal U}$ that
includes~$\Omega$.  This is true, even though there are Gribov copies within
$\Omega$, because the gauge orbits intersect $\Omega$ transversely.  The
coordinate patch~$\cal U$ must be small enough in the $\t$-directions that the
gauge transformations~$g(\t)$ that relate these Gribov copies are not 
in~$\cal U$.  

	To verify that the gauge-orbits intersect $\Omega$ transversely, 
it is sufficient to show that the change of variables \changevar\ is
invertible for infinitesimal angles
$\t^a(x) = \e^a(x)$ for all $B \in \Omega$.  It follows that it is also
invertible, and thus well-defined, on some finite cordinate patch~$\cal U$ that
includes~$\Omega$. 

	To first order in $\e$, the change of variable \changevar\ is given by
$A_\m = B_\m + D_\m(B)\e$.  The divergence of this equation reads
$\p \cdot A = \p \cdot D(B) \e = - M(B)\e$, which shows that $\p \cdot A$
depends linearly on $\e$.  Note that $\p \cdot A$ is orthogonal to the trivial
null space of $M(B)$, consisting of constant functions, and we specify that
$\e$ is also orthogonal to this null space.\foot{The constant angles $\p_\m
\t^a = 0$ parametrize global SU(N) transformations.  These act {\it
within}~$\Omega$. However we may safely ignore them because they have finite
volume that we normalize to unity.  The spectrum of $M(B)$ is discrete by
quantization in a finite Euclidean volume.}  Since~$B \in \Omega$
by assumption,~$M(B)$ is a strictly positive operator on the orthogonal
space, and thus invertible, and we have 
$\e = - M^{-1}(B) \ \p \cdot A$.  We solve for $B$ in the form
$B_\m = A_\m + D_\m(B) \ M^{-1}(B) \ \p \cdot A$.  To zeroth order in $\e$ we
have $B = A = A^{\tr}$, where 
$A_\m^{\tr} \equiv A_\m - \p_\m \ (\p^2)^{-1} \ \p \cdot A$ is the transverse
part of $A$.  This gives the inversion formulas
$B_\m = A_\m + D_\m(A^{\tr})M^{-1}(A^{\tr}) \ \p \cdot A$
and $\e = - M^{-1}(A^{\tr}) \ \p \cdot A$, valid to first order in
$\e$ or $\p \cdot A$.   Thus for each $A^{\tr} \in \Omega$, the change of
variable \changevar\ is invertible to first order in the small quantity 
$\p \cdot A$.  QED

	Concerning the shape of the coordinate patch
$\cal U$, note that as the configuration $B \in \Omega$ approaches the
boundary 
$\p \Omega$ of the Gribov region, the lowest non-trivial
eigenvalue $\l_1(B)$ of the Faddeev-Popov operator $M(B)$ approaches~0.  
Consequently the width in longitudinal or $\t$-directions of the
coordinate patch $\cal U$ shrinks to zero as the boundary $\p \Omega$ is
approached.  We may picture~$\cal U$ as as a very high-dimensional clam,
shown in Fig.~1.

\newsec{Change of variable in Fokker-Planck equation}

	To change variables in the Fokker-Planck equation, one takes over to
functional variables the standard formulas of differential geometry.  
The mechanics of the calculation are similar to the computation of the Coulomb
hamiltonian by Christ and Lee \christlee, but there the change of
variable was done globally whereas here it is done only in a coordinate patch.
We freely go back and forth from continuum to discrete notation by the
replacements
$A_\m^a(x) \leftrightarrow A^i$ and 
$(B_\m^a(x),\t^a(x)) \leftrightarrow u^\a$.  
In terms of $A^i$, the Fokker-Planck equation reads,
\eqn\discretea{\eqalign{
- H P \equiv
{{\p} \over {\p A^i}} \d^{ij} \Big( {{\p P} \over {\p A^j}} - K_j P \Big) = 0,
}}
and expectation values are given by 
$\langle F \rangle = \int \prod_i dA^i \ F(A) \ P(A) $.  The
coordinates $A^i$ are Cartesian, but the coordinate transformation 
$A = A(B,\t) = A(u)$ is non-linear, and the $u = (B, \t)$ are curvilinear
coordinates.  In terms of these, the Fokker-Planck equation reads
\eqn\discreteu{\eqalign{
-HP = 
{{1} \over {\sqrt G }} {{\p} \over { \p u^\a}} \Big[ \sqrt G
G^{\a \b} \Big( {{\p P} \over {\p u^\b}} - K_\b^{(u)} P \Big) \Big] = 0, 
}}
and expectation-values are given by
$\langle F \rangle = \int \prod_\a du^\a \sqrt G \ F(u) \ P(u) $.  The
metric tensor is given by
$dA^i dA^i = du^\a 
{{\p A^i} \over { \p u^\a}} {{\p A^i} \over { \p u^\b}} du^\b
= du^\a G_{\a \b} du^\b$, with volume element
$\sqrt G = \det {{\p u} \over { \p A}}$.  The covariant and contravariant
components of any Cartesian vector field $K_i$ are given by
$K_\a = {{\p A^i} \over { \p u^\a}}K_i$, and
$K^\a = {{\p u^a} \over { \p A^i}}K_i$.  

	We now calculate these quantities explicitly in functional form.
From $A_\m = g^{-1} B g + g^{-1} \p_\m g$, we obtain
\eqn\diffa{\eqalign{
\d A_\m = g^{-1} \Big( \d B_\m g + \p_\m(\d g g^{-1}) 
+ [B, \d g g^{-1}] \Big) g,
}}
where 
\eqn\cartan{\eqalign{
\o \equiv dg g^{-1} = { {\p g} \over {\p \t^\b } } \ g^{-1} d\t^\b
=   \o_\b d\t^\b = t^a \o_\b^a d\t^\b 
}} 
is the Maurer-Cartan form.  It satisfies
$d\o = dg g^{-1} \wedge dg g^{-1} = \o \wedge \o$ or, in terms of components,
\eqn\maurer{\eqalign{
{ {\p \o_\b^c} \over {\p \t^\a} } - { {\p \o_\a^c} \over {\p \t^\b} }
 = f^{cab}[\o_\a^a, \o_\b^b].
}}
We also have 
$g^{-1} t^a g = R_{ab} t^b$, where the real orthogonal matrices
$R_{ab} = R_{ab}(\t) = R^{-1}_{ba}$ are in the adjoint
representation of the gauge group.  From $A_\m = t^a A_\m$, and
$B_\m = t^a B_\m$, we obtain
\eqn\contdiff{\eqalign{
 \d A_\m^a = R_{ab}^{-1}[ \d B_\m^b + D_\m^{bc} (\o_\a^c \d \t^\a)],
}}	
where $\d B_\m$ is purely transverse, $\p_\m \d B_\m = 0$,
and $D_\m^{ac} \equiv D_\m^{ac}(B)$ is the gauge-covariant derivative with the
connection $B_\m^a$ as argument.  The last expression is the functional form
of 
$\d A^i = {{\p A^i} \over { \p u^\a}}\d u^\a$.  It gives the functional
operator that corresponds to ${{\p A^i} \over { \p u^\a}}$, and we have for the
metric tensor,
\eqn\metric{\eqalign{
ds^2 = \int d^4x \ \d A_\m^a \ \d A_\m^a
= \int d^4x \ [ \d B_\m^b + D_\m^{bc} (\o_\a^c \d \t^\a)] \ 
[ \d B_\m^b + D_\m^{bc} (\o_\a^c \d \t^\a)].
}}

	To calculate $\sqrt G = \det {{\p A} \over { \p u}}$, we start by
writing the linear transformation \contdiff\ as the product of two
transformations, $\d A_\m^a = R_{ab}^{-1} \d C _\m^a$, and 
\eqn\cdiff{\eqalign{
 \d C_\m^a = \d B_\m^b + D_\m^{bc} (\o_\a^c \d \t^\a).
}}
The matrix $R_{ab}$ is orthogonal, so $\det R = 1$, and it is sufficient to
calculate the determinent of the linear transformation \cdiff.  We do this in
two steps.  We first transform from $\d C_\m^a$ to its transverse part
$(\d C)_\l^{{\rm tr,} a} \equiv P_{\l \m}^{\rm tr}(\d C)_\m^a$, and its
divergence $\d L^a \equiv \p_\m \d C_\m^a$, where
$P_{\l \m}^{\rm tr} \equiv \d_{\l \m} - \p_\l (\p^2)^{-1}\p_\m$ is the
projector onto transverse vector fields.  This linear transformation is
independent of the variables $u = (B, \t)$, so its determinent is a constant,
and will be ignored.  The linear transformation from 
$\d B$ and $\d \t$ to $\d C^{\rm tr}$ and $\d L$, is given by
\eqn\tldiff{\eqalign{
 (\d C)_\l^{{\rm tr,} a} & = \d B_\m^b 
+ P_{\l \m}^{\rm tr}D_\m^{bc} (\o_\a^c \d \t^\a)     \cr
\d L^a & = \p_\m D_\m^{bc} (\o_\a^c \d \t^\a),
}}
where we have used the transversality of $\d B_\m^b$.  This linear
transformation is a triangular matrix, and its determinent is the product of
the determinents of its diagonal submatrices.  This gives
\eqn\weight{\eqalign{
 \sqrt G & = \det I \ \det [- \p_\m D_\m(B) \o(\t) ] \cr
  & = \det [- \p_\m D_\m(B)] \ {\rm Det}\o(\t)     \cr
  & = \det M(B) \ \prod_x\det \o(\t(x)),
}}
which contains the Faddeev-Popov determinent $\det M(B)$.  It has
been obtained by a purely local calculation at a fixed point $A = {^g}B$,
without integrating globally over the gauge group.  The volume element $\sqrt
G$ is the product of $\det M(B)$, that depends only on~$B$,
and the functional determinent 
${\rm Det}\o(\t) \equiv \prod_x\det \o(\t(x)) $,
that depends only on~$\t$. Here 
$[\det \o(\t(x)) \prod_\a d\t^\a(x)]$
is the Haar measure of the SU(N) gauge group at~$x$.
It is common to write $\int Dg = \int D\t \ {\rm Det}(\o(\t))$.

	We next find the inverse matrix ${{\p A^i} \over { \p u^\a}}$ by solving for 
$\d B_\m^b$ and  $\d \t^\a$.  From \contdiff\ we obtain
\eqn\solvdiff{\eqalign{
 R_{ba} \d A_\m^a = [ \d B_\m^b + D_\m^{bc} (\o_\a^c \d \t^\a)].
}}
We take the divergence of this equation and use $\p_\m \d B_\m = 0$ to obtain
\eqn\adiff{\eqalign{
 \p_\m (R_{ba} \d A_\m^a) = \p_\m D_\m^{bc} (\o_\a^c \d \t^\a),
}}
which gives the first inverse formula
\eqn\omdiff{\eqalign{
 \d \t^\a = J_c^\a [(\p \cdot D)^{-1}]^{cb}\p_\m (R_{ba} \d
A_\m^a), 
}}
where $J_c^\a(\t) \equiv (\o^{-1})_c^\a(\t)$.
The Faddeev-Popov operator
$M(B) \equiv  - \p \cdot D(B) = - D(B) \cdot \p$\ is symmetric and
positive, so its inverse is well defined. To avoid a proliferation of indices,
we write the last and similar equations in operator notation,
\eqn\omdiff{\eqalign{
 \d \t = J (\p \cdot D)^{-1} \p \cdot (R \d A).
}}
Inserting this into \solvdiff, we obtain the second inverse formula
\eqn\bdiff{\eqalign{
 \d B_\l = [\d_{\l \m} - D_\l (\p \cdot D)^{-1} \p_\m] (R \d A_\m).
}}
One sees that $\d B_\l$ is transverse, $\p_\l \d B_\l = 0$.  
The last two equations give the
operators corresponding to the matrices ${{\p u^\a} \over { \p A^i}}$.  From
them we read off the continuum version of
${ {\p} \over {\p A^i} } 
= { {\p u^\a} \over {\p A_i} }{ {\p} \over {\p u^\a} }$ namely,
\eqn\gradient{\eqalign{ 
{ {\d} \over {\d A_\m} } 
= \tilde{R} \Big( [\d_{\m \l} - \p_\m (D \cdot \p)^{-1} D_\l]
{ {\d} \over {\d B_\l} }  
-  \p_\m (D \cdot \p)^{-1} J(\t) {{\d } \over { \d \t}} \Big),
}}
where $\tilde{R}$ is the transpose of $R$.  The 
$(J{{\d } \over { \d \t}})_b = J_b^\a(\t) {{\d } \over { \d \t^\a}} 
\equiv (\o^{-1})_b^\a(\t){{\d } \over { \d \t^\a}}$ are the angular
momentum or Lie differential operators of the gauge group.  They satisfy
the Lie algebra commutation relations of the local gauge group
\eqn\lie{\eqalign{
 \Big[ J_a^\a(\t(x)) {{\d } \over { \d \t(x)^\a}},
  J_b^\b(\t(y)) {{\d } \over { \d \t(y)^\b}}\Big]
   = - \ \d(x-y) \ f^{abc} \ J_c^\c(\t(x)) {{\d } \over { \d \t(x)^\c}},
}}
that follow from \maurer.

	We need the curvilinear components of the drift force
$K_\m = K_{{\rm YM},\m} + a^{-1}K_{{\rm gt},\m}$
where
$K_{{\rm YM},\m}(A) = - { {\d S} \over {\d A_\m} } = D_\l F_{\l\m}(A)$
and
$K_{{\rm gt},\m} = D_\m \p \cdot A$.  
We shall see that the one-form or covariant $\t$-component of $K_{\rm YM}$
vanishes (because the action $S_{\rm YM}({^g}B) = S_{\rm YM}(B)$ 
is gauge invariant), while the tangent-vector or
contravariant $B$-component of $K_{\rm gt}$ vanishes (because  $K_{\rm gt}$ is
tangent to the gauge orbit).  Thus the Fokker-Planck equation 
\discreteu\ in curvilinear coordinates 
$u = (B,\t)$, reads $HP = 0$, where
\eqn\exph{\eqalign{
H = H_{BB} + H_{B\t} + H_{\t G} + H_{\t \t},
}}
\eqn\expha{\eqalign{
- H_{BB} & \equiv { {1 } \over { \sqrt G } } 
\ { {\p } \over { \p B^\a} } \sqrt G 
\ G_{(B B)}^{\a \b} \ \Big( { {\p } \over { \p B^\b} } 
      - K_{ {\rm YM},\b}^{(B)} \Big)  \cr
- H_{B\t} & \equiv { {1 } \over { \sqrt G } } \ { {\p } \over { \p B^\a} }
\sqrt G \ G_{(B \t)}^{\a \b} \  { {\p } \over { \p \t^\b} }     \cr
- H_{\t B} & \equiv { {1 } \over { \sqrt G } } 
\ { {\p } \over { \p \t^\a} } \sqrt G 
\ G_{(\t B)}^{\a \b} \ \Big( { {\p } \over { \p B^\b} } - 
             K_{ {\rm YM},\b}^{(B)} \Big)  \cr
- H_{\t \t} & \equiv
{ {1 } \over { \sqrt G } } \ { {\p } \over { \p \t^\a} } \sqrt G 
\Big( \ G_{(\t \t)}^{\a \b} \  { {\p } \over { \p \t^\b} } 
- K_{{\rm gt},(\t)}^\a \Big).
}}

	We use the continuum version of the formula
$K_{{\rm YM},i} \d A^i = K_{{\rm YM},\a}^{(B)} \d B^\a
+ K_{{\rm YM},\a}^{(\t)} \d \t^\a$
to obtain the one-form components of $K_{\rm YM}$.  We have
\eqn\consforce{\eqalign{
\int d^4x \ K_{{\rm YM},\m}^a(A) \d A_\m^a 
& = \int d^4x \ D_\l F_{\l\m}^a({^g}B) \d A_\m^a
 = \int d^4x \ R_{ab}^{-1}\ D_\l F_{\l\m}^b(B) \d A_\m^a  \cr
& = \int d^4x \ \ D_\l F_{\l\m}^b(B) 
\ [ \d B_\m^b + D_\m^{bc} (\o_\a^c \d \t^a]   \cr
& = \int d^4x \ \ D_\l F_{\l\m}^b(B) \ \d B_\m^b,
}}
by \contdiff, where we have performed an integration by parts, and used
$(D_\m D_\l F_{\l\m})^a = (1/2)g_0f^{abc}F_{\m\l}^bF_{\l\m}^c = 0$.  Thus the
one-form components of $K_{\rm YM}$ are given by
\eqn\covcons{\eqalign{
K_{{\rm YM},\a} = (K_{{\rm YM},\a}^{(B)}, K_{{\rm YM},\a}^{(\t)})
  = (D_\l F_{\l\m}^b(B), 0).
}}

	We use the continuum version of
$K_{{\rm gt},i} { {\p} \over {\p A^i} } 
= K_{{\rm gt},\a}^{(B)} { {\p} \over {\p B^\a} }
+ K_{{\rm gt},\a}^{(\t)} { {\p} \over {\p \t^\a} } $
to obtain the contravariant or tangent-vector components of 
$K_{{\rm gt},\m} = D_\m \p \cdot A$.
We have
\eqn\diva{\eqalign{
\p_\l A_\l & = \p_\l (g^{-1} B_\l g + g^{-1} \p_\l g)
    = g^{-1}\Big(\p_\l(\p_\l g g^{-1}) + [B, \p_\l g g^{-1}] \Big) \ g  \cr
  & = g^{-1} D_\l(B) (\p_\l g g^{-1}) \ g  
  = g^{-1} D_\l(B)(\o_\a \p_\l \t^\a) \ g,
}}
where we have used 
$ \p_\l g g^{-1} = { {\p g} \over {\p \t_\a} }  g^{-1} \p_\l \t^a
= \o_\a \p_\l \t^\a$,
and $\o$ is again the Maurer-Cartan form.  In
index and operator notation this reads
\eqn\divac{\eqalign{
\p_\l A_\l^a  = \tilde{R}_{ab} \ D_\l^{bc} (\o_\a^c \p_\l \t^\a) \ \ \ \
\leftrightarrow \ \ \ \   
\p_\l A_\l = \tilde{R} \ D_\l (\o \p_\l \t),
}}
where $D_\l \equiv D_\l(B)$.  By the gauge transformation property of the
gauge covariant derivative $D(A) = D({^g}B)$, this gives
\eqn\calcdrift{\eqalign{
D_\m(A) \p_\l A_\l = \tilde{R} D_\m(B) \ D_\l(B) (\o \p_\l \t),
}}
By \gradient\ we obtain
\eqn\cgtdrift{\eqalign{
\int d^4x  \ K_{{\rm gt},\m}^a { {\d} \over {\d A_\m^a} }  
& = \int d^4x \ D_\m(B) \ D_\l(B) (\o \p_\l \t)  \cr 
&  \times \Big( [\d_{\m \n} - \p_\m (D \cdot \p)^{-1} D_\n]
{ {\d} \over {\d B_\n} }  
-  \p_\m (D \cdot \p)^{-1} J(\t) {{\d } \over { \d \t}} \Big).
}} 
We perform an integration by parts and use
$D_\m [\d_{\m \n} - \p_\m (D \cdot \p)^{-1} D_\n]
{ {\d} \over {\d B_\n} } = 0$ to obtain
\eqn\cgtdrifta{\eqalign{
\int d^4x & \ K_{{\rm gt},\m}^a { {\d} \over {\d A_\m^a} }  
= \int d^4x \  \ [D_\l(B) (\o \p_\l \t)]^a    
 \Big[J(\t) {{\d } \over { \d \t}}\Big]_a. 
}} 
Thus the tangent-vector components of $K_{\rm gt}$ are given
by
\eqn\contragtf{\eqalign{
K_{\rm gt}^\a = (K_{\rm gt}^{(B), \a}, K_{\rm gt}^{(\t), \a})
  = (0, \ J_b^\b(\t)[D_\l(B) (\o \p_\l \t)]^b).
}}

	From \gradient\ we obtain the Laplacian operator 
$ { {1 } \over { \sqrt G } }{{\p } \over { \p u^\a}} \sqrt G 
{ {\p u^\a} \over { \p A^i} } { {\p u^\b} \over { \p A^i} } 
 { {\p } \over { \p u^\b} }$
in curvilinear coordinates, 
\eqn\laplace{\eqalign{
 \int d^4x & \ { {1 } \over { \sqrt G } }
\Big( {{\d } \over { \d B_\l}}[\d_{\l\m} - D_\l (\p \cdot D)^{-1} \p_\m] 
+ {{\d } \over { \d \t}}\tilde{J}(\t) (\p \cdot D)^{-1} \p_\m\Big)  \cr
& \times \sqrt G 
\Big( [\d_{\m\n} - \p_\m (D \cdot \p)^{-1} D_\n ] {{\d } \over { \d B_\n}}
- \p_\m (D \cdot \p)^{-1} J(\t) {{\d } \over { \d \t}} \Big).
}}
Putting all terms together, the explicit expressions for the terms 
in \expha\ are
\eqn\hbb{\eqalign{
- H_{BB} 
 = { {1 } \over { \det M(B) } }
\int d^4x \ {{\d } \over { \d B_\l}} \det M(B) & \ 
[\d_{\l\m} - D_\l (\p \cdot D)^{-1} \p_\m]    \cr
& \times  [\d_{\m\n} - \p_\m (D \cdot \p)^{-1} D_\n ] 
\Big[{{\d } \over { \d B_\n}} - D_\k F_{\k\n}(B)\Big],
}}
\eqn\hbth{\eqalign{
- H_{B\t} = { {1 } \over { \det M(B) } } \int d^4x \
{{\d } \over { \d B_\l}} \det M(B) \ [- \p_\l + D_\l (\p \cdot D)^{-1} \p^2] 
\  (D \cdot \p)^{-1} J(\t) {{\d } \over { \d \t}},
}}
\eqn\hthb{\eqalign{
- H_{\t B} = { {1 } \over { {\rm Det}\o(\t) } } \int d^4x \ 
& {{\d } \over { \d \t}} {\rm Det}\o(\t) \  \tilde{J}(\t)   \cr
& \times (\p \cdot D)^{-1}     
[\p_\n - \p^2 (D \cdot \p)^{-1} D_\n ] 
\Big[{{\d } \over { \d B_\n}} - D_\l F_{\l\m}(B)\Big], 
}}
\eqn\hthth{\eqalign{
- H_{\t \t} = { {1 } \over { {\rm Det}\o(\t) } } \int d^4x \ 
{{\d } \over { \d \t}} {\rm Det}\o(\t) \tilde{J}(\t)  
\Big(  \ (\p \cdot D)^{-1}   
 \ (- \p^2) & \ (D \cdot \p)^{-1} J(\t) {{\d } \over { \d \t}} \cr
& - {{1 } \over {a}} 
\ D_\l[\o(\t) \p_\l \t]  \ \Big).
}}

\newsec{Solution in Landau-gauge limit}

	We shall solve the Fokker-Planck equation $HP = 0$ in the limit $a \to 0$. In
this limit the drift force in the gauge-orbit or $\t$-direction is dominant. 
This situation is reminiscent of the Born-Oppenheimer method in molecular
physics.  The $\t$ variables equilibrate rapidly, like the electron
positions in a molecular wave function, and the dependence on the $B$ variable
is determined by an average over the $\t$ variable, like the nuclear
variables.

	We expect that the solution gets concentrated close to $\t = 0$.  We
rescale variable according to $\t = a^{1/2} \T$, and find that $H_{BB}$ is
independent of $a$ and unchanged, whereas
\eqn\hbth{\eqalign{
- H_{B\t} = { {1 } \over { a^{1/2} } }
{ {1 } \over { \det M(B) } } \int d^4x \
{{\d } \over { \d B_\l}} \det M(B) \ [- \p_\l + D_\l (\p \cdot D)^{-1} \p^2] 
\  (D \cdot \p)^{-1}     \cr
\times J(a^{1/2}\T) {{\d } \over { \d \T}},
}}
and
\eqn\hthb{\eqalign{
- H_{\t B} = { {1 } \over { a^{1/2} } }
{ {1 } \over { {\rm Det}\o(a^{1/2}\T) } } \int d^4x \ 
& {{\d } \over { \d \T}} {\rm Det}\o(a^{1/2}\T) \  \tilde{J}(a^{1/2}\T)   \cr
& \times (\p \cdot D)^{-1}     
[\p_\n - \p^2 (D \cdot \p)^{-1} D_\n ] 
\Big[{{\d } \over { \d B_\n}} - D_\l F_{\l\m}(B)\Big], 
}}
are of leading order ${ {1 } \over { a^{1/2} } }$,
while
\eqn\hthth{\eqalign{
- H_{\t \t} = { {1 } \over { a } }
{ {1 } \over { {\rm Det}\o(a^{1/2}\T) } } \int d^4x \ &
{{\d } \over { \d \T}} {\rm Det}\o(a^{1/2}\T) \tilde{J}(a^{1/2}\T)  \cr
& \times \Big(  \ (\p \cdot D)^{-1}   
 \ (- \p^2)  \ (D \cdot \p)^{-1}  J(a^{1/2}\T) {{\d } \over { \d \T}}  \cr
 & \ \ \ \ \ \ \ \ \ \ \ \ \ \ \ \ \ \ \ \ \ \ \ \ \ \ \ \ 
- \ D_\l[\o(a^{1/2}\T) \p_\l \T]  \ \Big)
}}
is of leading order ${ {1 } \over { a } }$.

	The Fokker-Planck hamiltonian has an expansion in $a$ given by
$H = a^{-1} H_0 + a^{-1/2} H_1 + H_2 + O(a^{1/2})$.  We seek a solution
of the form $P = P_0 + a^{1/2} P_1 + a P_2 + ...$, which gives
\eqn\fpina{\eqalign{
(a^{-1} H_0 + a^{-1/2} H_1 + H_2 + ...) 
\ (P_0 + a^{1/2} P_1 + a P_2 + ...) = 0.
}}
To leading order we obtain
\eqn\leading{\eqalign{
- H_0 P_0 =  \int d^4x \
{{\d } \over { \d \T}}    \Big(  \ (\p \cdot D)^{-1}   
 \ (- \p^2)  \ (D \cdot \p)^{-1}   {{\d } \over { \d \T}} 
- \ D \cdot \p \ \T]  \ \Big) \ P_0 = 0,
}}
or    
\eqn\leadingr{\eqalign{
 \int d^4x \
{{\d } \over { \d \T}}    \Big(  \ V   {{\d } \over { \d \T}} 
+ M \T]  \ \Big) \ P_0 = 0,
}}
where $D \equiv D(B)$,
$M \equiv M(B)$.  The operator $V = V(B)$ is defined by $V \equiv
M^{-1}(-\p^2)M^{-1}$.  It is symmetric and positive.

	The last equation is solved by a Gaussian in~$\T$,
\eqn\sol{\eqalign{
P_0(B, \T) & = Q(B) \ N (\det X)^{1/2} \ \exp[- (\T, X\T)/2]   \cr
 & = Q(B) \ N (\det X)^{1/2} \ \exp[- (\t, X\t)/(2a)],
}}
where
$(\t, X\t)  \equiv \int d^4x \ \t^a(x) (X\t)^a(x)$.
Here $X = X(B)$ is a symmetric operator to be determined, and $N$ is fixed by
\eqn\normalize{\eqalign{
\int D\t \ N (\det X)^{1/2} \ \exp[- (\t, X\t)/(2a)] = 1. 
}}
The upper limit on the $\t$ integration actually finite, but this gives a
correction of order $\exp(- 1/a)$ that we neglect.  The solution~\sol\
decreases rapidly as $|\t|$ increases away from
$0$, as expected, with a Gaussian width $|\t| \sim a^{1/2}$.  In the limit $a
\to 0$, the support of the solution $P(B, \t)$
shrinks to $\t = 0$, and is given by
\eqn\limsol{\eqalign{
P(B, \t) = \d(\t) \ Q(B).
}} 

	We now check that \sol\ is actually the solution.  Equation~\leadingr\ yields
two equations for~$X$,
\eqn\condonx{\eqalign{
(\T, XVX\T) - (\T, XM\T) = 0   \cr
{\rm tr}(VX - M) = 0
}}
that hold identically for all $\T$.  The first equation yields
$2XVX = XM + MX$, or 
$MY + YM = 2V$ for $Y \equiv X^{-1}$.  
Moreover when this equation is satisfied, the second equation is automatically
satisfied.  To solve for $Y$, we take
matrix elements in the basis provided by the eigenfunctions of the
Faddeev-Popov operator $M u_n = \l_n u_n$, and obtain
$2 (u_m, Vu_n) = (\l_m + \l_n) (u_m, Yu_n)$, or
\eqn\sola{\eqalign{
(u_m, X^{-1}u_n) = (u_m, Yu_n) = 2(\l_m + \l_n)^{-1} (u_m, Vu_n) \cr
 = 2 \int_0^{\infty} dt \ (u_m, \exp(-Mt) \ V \exp(-Mt) \ u_n).  
}} 
This gives
\eqn\solb{\eqalign{
X^{-1} = Y & = 2\int_0^{\infty} dt \ \exp(-Mt) \ V \exp(-Mt)  \cr
 & = 2 \ M^{-1}\int_0^{\infty} dt \ \exp(-Mt) \ (-\p^2) \exp(-Mt) \ M^{-1}, 
}}
and $X = X(B)$ is indeed a positive operator, as is necessary for the 
normalizability of the Gaussian \sol.

	The coefficient function $Q(B)$ in \sol\ is left undetermined by the equation
$H_0 P_0 = 0$.  Since the leading term in the hamiltonian
$H = { {1} \over {a} } H_0 + ...$ leaves the solution indeterminate, we are in
the case of degenerate perturbation theory, and the lowest order solution is
determined by a higher order perturbation.  To obtain an equation for
$Q(B)$, we integrate the exact equation $HP = 0$ over $\T$, 
\eqn\inttheta{\eqalign{
\int D\T \ {\rm Det}\o(a^{1/2}\T) \ HP = 0,
}}
where, we recall, $H = H_{B B} + H_{B \t}+ H_{\t B} + H_{\t \t}$.
This kills the $H_{\t \t}$ term that is of order 
${ {1} \over {a} }$, for, by \hthth, it is the integral of an exact derivative,
and thus vanishes identically,
$\int D\T \ {\rm Det}\o(a^{1/2}\T) \ H_{\t\t} P 
 = \int D\T { {\d} \over {\d \T} }... = 0$.  For the same reason it kills
the  $H_{\t B}$ term that is of order 
${ {1} \over {a^{1/2}} }$,
$\int D\T \ {\rm Det}\o(a^{1/2}\T) \ H_{\t B} P = 0$.
It also kills the $H_{B \t}$ term that is of order 
${ {1} \over {a^{1/2}} }$ because, by
\hbth, the integral 
$\int D\T \ {\rm Det}\o(a^{1/2}\T) \ H_{B\t} P$ is of the form
\eqn\killhbth{\eqalign{
\int D\T \ {\rm Det}\o(a^{1/2}\T)J(a^{1/2}\T){ {\d} \over {\d \T} } F
 = - \int D\T \ {\rm Det}\o(a^{1/2}\T)
 \ F \ J(a^{1/2}\T){ {\d} \over {\d \T} }1 = 0,
}}
where the explicit form of $F$ is not needed.\foot{The fact that the
integral on $D\T$ surgically kills the $H_{\t B}$ and $H_{B\t}$ terms is the
pay-off for using the curvi-linear coordinates $(B,\t)$.  In a previous
calculation by the author \dztimeind, the time-independent Fokker-Planck
equation was solved using Cartesian coordinates $A^{\rm tr}$ and  $A^{\rm lo}$
instead of $(B,\t)$.  This gave an additional contribution, not
surgically killed by the corresponding integration over $D A^{\rm lo}$, that
was mistakenly neglected, and that was needed to cancel a spurious term,
called $K_2$, in the effective drift force.  Fortunately $K_2$ was
neglected in \dztimeind, so what was thought to be an approximate formula there
is in fact exact, and the calculation reported there is correct.]} The first equality holds by
by the Lie group property that makes
$J(a^{1/2}\T){ {\d} \over {\d \T} }$ anti-hermitian with respect to Haar
measure $\int D\T \ {\rm Det}\o(a^{1/2}\T)$.

	[It is easy to verify that the equation 
$\int D\T \ {\rm Det}\o(a^{1/2}\T) \ H_{B \t} P = 0$
holds in the small-$a$ limit.  This is the same as the small angle
approximation, and we have, to the order required,
$g(\t) = \exp(\t) = 1 + t^\a \t^a + (1/2) (t^\a \t^a)^2$.
For the Maurer-Cartan form
${ {\p g} \over {\p \t^\b} } g^{-1} = t^a \o_\b^a$ 
we obtain, to the order required,
$\o_\b^a =  \d^{a\b} + (1/2) f^{a\g\b} \t^\g
= \d^{a\b} + { { a^{1/2} } \over { 2 } }
f^{a\g\b} \T^\g$.  The second term is an anti-symmetric matrix so for the
Haar measure we get
$\det\o(a^{1/2}\T) = 1 + O(a)$, and for the matrix 
$J_a^\b$, defined by $J_a^\b \o_\b^c = \d_a^c$, we get
$J_a^\b(a^{1/2}\T) = \d^{a\b} 
+ { { a^{1/2} } \over { 2 } } f^{a\g\b} \T^\g + O(a)$.  This gives
$$\int D\T \ {\rm Det}\o(a^{1/2}\T)J_a^\b(a^{1/2}\T){ {\d} \over {\d \T^\b} } 
\ F = \int D\T \ \Big[ \ \Big(\d^{a\b} 
+ { { a^{1/2} } \over { 2 } } f^{a\g\b} \T^\g\Big) 
{ {\d} \over {\d \T^\b} } + O(a) \ \Big] \ F.$$
The term in ${ {\d} \over {\d \T^\b} }$ is an exact derivative because
$f^{a\g\b}$ is anti-symmetric, and gives vanishing contribution.  The leading
term in $F$ is of order ${ { 1} \over {a^{1/2} } }$, so the remainder is of
order $a^{1/2}$ and vanishes in the small-$a$ limit.]

	We conclude that in \inttheta, the only surviving term is $H_{BB}$, given
in \hbb.  It is independent of $a$ and $\T$,
and \inttheta\ simplifies to
\eqn\eqforb{\eqalign{
H_{BB} \ Q = 0.
}} 
From \hbb\ we see that this equation is of the form
$$...[\d_{\m\n} - \p_\m (D \cdot \p)^{-1} D_\n ] 
\Big[{{\d } \over { \d B_\n}} - D_\k F_{\k\n}(B)\Big]Q = 0.$$
The left factor is orthogonal on $\n$ to longitudinal fields, so it may be
written
$$...[\d_{\m\n} - \p_\m (D \cdot \p)^{-1} D_\n ] P_{\n\l}^{\rm tr} 
\Big[{{\d } \over { \d B_\l}} + 
{ { \d S_{\rm YM}(B) } \over {\d B_\l} }\Big]Q = 0,$$
where we have used the fact that functional differentiation with respect to a
transverse field is ordinary functional differentiation with a transverse
projector that comes from 
\eqn\transverse{\eqalign{
{{\d B_\m^b(y) } \over { \d B_\l^a(x)}} = 
P_{\l\m}^{\rm tr}(x-y) \ \d^{ab}. 
}}
Thus the equation, $H_{BB} \ Q(B) = 0$, has the simple solution,
\eqn\boltzmann{\eqalign{
Q(B) = N \exp[-S_{\rm YM}(B)].
}}
In continuum gauge theory, the Gribov region $\Omega$ is convex, as
shown in Appendix C, and therefore it is connected, so the normalization of
the solution \boltzmann\ is unique.  We have obtained the solution in the
coordinate patch $\cal U$, in the limit $a \to 0$, 
\eqn\compsol{\eqalign{
P(B, \t) = N \ \d(\t) \ \exp[-S_{\rm YM}(B)].
}}

	We express the solution $P(B, \t)$
in terms of the original Cartesian coordinates~$A$.  The volume
element is of course $\int dA$.  To first order in $\t$ we have
$A = B + D(B)\t$, and $\p \cdot A = \p \cdot D(B)\t$, so
\eqn\changedel{\eqalign{
\d(\t) = \d(\p \cdot A) \ \det[-\p \cdot D(A)].
}}
Inside the coordinate patch $\cal U$, the solution reads  
\eqn\limsola{\eqalign{
P(A) = N \ \d(\p \cdot A) \det[-\p \cdot D(A)] \ \exp[-S_{\rm YM}(A)].
}}
Its support lies on $\p \cdot A = 0$, and it vanishes with 
$\det[-\p \cdot D(A)]$ on the boundary $\p \Omega$ of the Gribov region.  We
extend it to all of $A$-space by stipulating that it vanishes 
outside~$\cal U$.  For the diffusion equation with a drift force, the
equilibrium distribution is unique~\varadhan.

\newsec{Dyson-Schwinger equation for partition function}

	To be of use, the non-perturbative Faddeev-Popov formula \expectval\ must be
supplemented with a prescription for how the functional integral, restricted
to the Gribov region~$\Omega$, is to be evaluated non-perturbatively.  An
earlier approach \dzcritlim\ is to insert a $\t$-function $\t(\l_1(B))$ that
effects a cut-off at the Gribov horizon. The $\t$-function is given a suitable 
representation as an integral over auxiliary fields with a local effective
action, and one integrates over all~$B$ without restriction and over the
auxiliary fields.  A far simpler approach~\dznonpertl\ rests on
the observation that the Gribov horizon~$\p \Omega$ is {\it a nodal surface}
of the integrand because the Faddeev-Popov determinent, 
$\det M(B) = \prod_{n=1}^\infty \l_n(B)$ vanishes with $\l_1(B)$, that is to
say, on~$\p\Omega$.   The DS equations, which are derived by a
partial integration, do not pick up a boundary term, and would have the same
form if the integral were extended to infinity.  In this
approach we never have to know where the Gribov horizon actually is.

	The partition function for the distribution
\expectval\ is given by
\eqn\partfunct{\eqalign{
Z(J) = N \int_\Omega dB \ \det M(B) 
 \ \exp[-S_{\rm YM}(B) + (J,B)],
}}
where we have written $B \equiv A^{\rm tr}$, and
$(J, B) \equiv \int d^4x \ J_\m^a(x) B_\m^a(x)$.  
Only the transverse
part of
$J$ contributes, and we also take $J$ to be identically transverse, 
$J = J^{\rm tr}$.  
(The extension of the present non-perturbative approach with a cut-off at the 
Gribov horizon to an off-shell gauge condition with a local and BRST-invariant
action is sketched in Appendix~B.) \ The Faddeev-Popov determinent
$\det M(B)$ vanishes on the boundary $\p \Omega$, so the identity 
\eqn\exact{\eqalign{
0  = \int_\Omega dB \  { {\d} \over {\d B_\m^b(x)} } 
\Big( \det M(B)
 \ \exp[-S_{\rm YM}(B) + (J, B)] \Big)      
}}
holds, {\it without any contribution from boundary terms even though the
integral is cut-off at the Gribov horizon $\p \Omega$.}  It is shown in
Appendix~C that the Gribov horizon surrounds the origin at a finite distance in
all directions.

		To derive the functional DS equation for $Z(J)$, we
write $\det M(B) = \exp[ {\rm Tr} \ln M(B)]$, and define the total action
\eqn\totalact{\eqalign{
\Sigma(B) \equiv S_{\rm YM}(B) - {\rm Tr} \ln M(B), 
}}
so \exact\ reads
\eqn\exacta{\eqalign{
0 = \int_\Omega dB \  \Big( J_\m^b(x) 
- { {\d \Sigma(B)} \over {\d B_\m^b(x)} } \Big)
\Big( \det M(B) 
 \ \exp[-S_{\rm YM}(B) + (J,B)] \Big).
}}
Although $\Sigma(B)$ is not local in $B$, we shall derive the same DS
equations as one gets from the usual local action of gluons and ghosts.
We have
\eqn\deract{\eqalign{
{ {\d \Sigma(B)} \over {\d B_\m^b(x)} } = - [D_\l F_{\l\m}(B)]^{b,{\rm tr}}(x)
 - {\cal J}_{{\rm gh},\m}^b(x; B),
}}
by \transverse, where ``tr" means transverse part, 
$[X_\m]^{\rm tr} \equiv X_\m - \p_\m (\p^2)^{-1} \p_\n X_\n$,
and the ghost current is given by
\eqn\derivedet{\eqalign{
{\cal J}_{{\rm gh},\m}^b(x; B) & \equiv
 { {\d \ [{\rm Tr} \ln M(B)}] \over {\d B_\m^b(x)} }
 =  {\rm Tr} \Big( \ { {\d M(B)} \over {\d B_\m^b(x)} } \ M^{-1}(B) \ \Big) 
\cr & = - \int d^4y  \ 
{ {\d [\p^2 \d^{ac} + g_0f^{adc}B_\l^d(y) \p_\l]} 
\over {\d B_\m^b(x)} }  \ (M^{-1})^{ca}(y,z; B)|_{z=y} \   \cr
& = - g_0f^{abc} \int d^4y \ P_{\m\l}^{\rm tr}(x-y) \  
\p_\l (M^{-1})^{ca}(y,z; B)|_{z=y}.
}}
Here and below, derivatives act on the left argument of propagators.  The
identity \exacta\ reads
\eqn\exactb{\eqalign{
0  = \int_\Omega dB \  \Big( J_\m^b(x) 
+ & [D_\l F_{\l\m}^b(B)]^{\rm tr}(x) + {\cal J}_{{\rm gh},\m}^b(x; B)
\Big)  \cr & \times \Big( \det M(B) 
 \ \exp[-S_{\rm YM}(B) + (J,B)] \Big),
}}
and yields the functional DS equation for the partition function $Z(J)$,
\eqn\dsz{\eqalign{
-(D_\l F_{\l\m}^b\Big({ {\d} \over {\d J} }\Big)  )^{\rm tr}(x) \ Z(J)
= [ \ {\cal J}_{{\rm qu.gh},\m}^b(x; J) + J_\m^b(x) \ ] \ Z(J), 
}}
where $D_\l F_{\l\m}^b({ {\d} \over {\d J} })$ is a cubic polynomial in 
${ {\d} \over {\d J} }$.  The quantum ghost current in the presence of the
source $J$ is, by \derivedet,
\eqn\meanghost{\eqalign{
{\cal J}_{{\rm qu.gh},\m}^b(x; J)
& \equiv \langle \ {\cal J}_{{\rm gh},\m}^b(x; B) \ \rangle_J   \cr
 & = - g_0f^{abc} \int d^4y \ P_{\m\l}^{\rm tr}(x-y) \  
\p_\l {\cal G}^{ca}(y, z; J)|_{z = y}.
}}
Here we have introduced the ghost propagator in presence of the source $J$,
\eqn\ghostprop{\eqalign{
{\cal G}^{ca}(x, y; J) \equiv \langle \ (M^{-1})^{ca}(x,y; B) \ \rangle_J,
}}
where $\langle {\cal O} \rangle_J$ denotes the mean value of ${\cal O}(B)$
in the presence of the source $J$, 
\eqn\meancurrent{\eqalign{
\langle {\cal O} \rangle_J = Z^{-1}(J) \ 
N \int_\Omega dB \ \det M(B) \ {\cal O}(B)
 \ \exp[-S_{\rm YM}(B) + (J,B)].
}}

	To obtain a closed system of equations, we need a DS equation for the
ghost propagator 
${\cal G}^{ab}(x, y; J)$.  It contains a term proportional 
to~$\l_1^{-1}(B)$, so {\it we must avoid 
integrating by parts on $B$ or introducing ghost
sources.}  (But see Appendix B.)  Fortunately the functional DS equation for
${\cal G}^{ab}(x, y; J)$ follows from the trivial identity
$I = M(B) \ M^{-1}(B)$, that we average with $P(B)\exp[(J, B)]$,
\eqn\ghostid{\eqalign{
\d(x-y)\d^{ab} \ Z(J) & = \int_\Omega DB 
 \ M^{ac}(B) \ (M^{-1})_{xy}^{cb}(B) \ P(B) \ \exp[(J, B)]     \cr
& = M^{ac}\Big({ {\d} \over {\d J} } \Big) 
\int_\Omega DB \ (M^{-1})_{xy}^{cb}(B)  \ P(B) \ \exp[(J, B)],      
}}
where
$M^{ac}({ {\d} \over {\d J} }) 
= - \p^2 \d^{ac} - g_0f^{abc} { {\d} \over {\d J_\m^b} }  \p_\m$.
Here $P(B) = \det M(B) \exp[- S_{\rm YM}(B)]$ is the probability distribution,
although the form of the DS equation for the ghost propgator is independent of
$P(B)$.  This gives the DS equation for the ghost propagator
\eqn\dsghost{\eqalign{
 M^{ac}\Big({ {\d} \over {\d J} } \Big) 
\ [ \ {\cal G}^{cb}(x,y;J) \ Z(J) \ ] = \d(x-y)\d^{ab} \ Z(J).   
}}
Equations \dsz\ and \dsghost\ and formula \meanghost\ provide a
complete system of functional DS equations for the partition function $Z(J)$
and the ghost propagator ${\cal G}^{cb}(x,y;J)$.

\newsec{Functional DS equation for gluon and ghost propagators}

	We change variable from $Z(J) = \exp W(J)$ to the ``free energy" $W(J)$.  For
the ghost propagator we obtain
\eqn\wghost{\eqalign{  
M^{ac}\Big({ {\d W} \over {\d J} } + { {\d} \over {\d J} } \Big) 
\ {\cal G}^{cb}(x,y;J) = \d(x-y)\d^{ab}.   
}}
We again change variables by Legendre transformation from the free energy
$W(J)$ to the quantum effective action
\eqn\legendre{\eqalign{ 
\Gamma(B_{\rm cl}) = J_x B_{{\rm cl}, x} - W(J),
}}
where the new variable $B_{{\rm cl}, \m}^a(x)$ is defined by
\eqn\aclassical{\eqalign{ 
B_{{\rm cl}, \m}^a(x; J) \equiv 
 { { \d W(J) } \over {\d J_\m^a(x)} } 
 = { { 1 } \over {Z} } { { \d Z(J) } \over {\d J_\m^a(x)} }
 = \langle B_\m^a(x) \rangle_J.
}}
It is identically transverse, 
$B_{{\rm cl}, \m} = B_{{\rm cl}, \m}^{\rm tr},$ 
and takes values in~$\Omega$ because
$B_{\rm cl} (J) = \langle B \rangle_J$ 
is an average with a positive probability, 
$N \ \det M(B) \ \exp(B, J)$,
over the convex region~$\Omega$. 
Inversion of 
$B_{\rm cl} = B_{\rm cl}(J)$ to obtain $J = J(B_{\rm cl})$
is possible because the gluon propagator in the
presence of the source~$J$, 
\eqn\twopt{\eqalign{
{\cal D}_{xy}(J) & \equiv 
\langle \ (B_x - \langle B_x\rangle_J) 
 \ (B_y - \langle B_y\rangle_J) \ \rangle_J 
 = {{\p^2 W  } \over {\p J_x \p J_y}} = {{ \p B_y(J) } \over {\p J_x}},
}} 
is a positive matrix.   
The gluon propagator is expressed in terms of the Legendre-transformed
variables $B$ and $\Gamma(B)$ by
\eqn\propgamma{\eqalign{
{{\cal D}^{-1}}_{xy}(B) 
= {{\p^2 \Gamma(B)  } \over {\p B_x \p B_y}}. 
}}
Here and below, we write $B$ instead of $B_{\rm cl}$.  The gluon propagator
and its inverse are identically transverse, 
$\p_\l {\cal D}_{\l \m}(x,y;B) = 0$.

	Under the Legendre transformation, derivatives transform according to
\eqn\transder{\eqalign{
{ {\d} \over {\d J_\l^a(x)} }  = 
\Big( {\cal D}{ {\d} \over {\d B} } \Big)_\l^a(x) 
\equiv \int d^4y \ {\cal D}_{\l\m}^{ab}(x,y;B){ {\d} \over {\d B_\m^b(y)} },
}}
as one sees from \twopt.  In terms of the Legendre transformed variables, the
DS equation \wghost\ for the ghost propagator reads
\eqn\legendregh{\eqalign{
\d(x-y)\d^{ab} & = M^{ac}\Big(B + {\cal D}{ {\d} \over {\d B} } \Big) 
\ {\cal G}^{cb}(x,y;B)        \cr 
& = M^{ac}(B){\cal G}^{cb}(x,y;B) - g_0f^{adc} \int dz \ 
{\cal D}_{\m\n}^{de}(x,z;B){ {\d} \over {\d B_\n^e(z)} }  
\p_\m{\cal G}^{cb}(x,y;B),  
}}
where 
${\cal G}^{cb}(x,y;B) \equiv {\cal G}^{cb}(x,y;J(B))$
is the ghost propagator expressed in terms of the source $B$.
Finally, instead of ${\cal G}^{ab}(x,y;B)$, we take as new
unknown variable the inverse ghost propagator $\G_{\rm gh}^{ab}(x,y;B)$
defined by
\eqn\inverseghost{\eqalign{
\G_{{\rm gh},xy}(B) \equiv {\cal G}^{-1}_{xy}(B) 
\ \ \ \  \leftrightarrow \ \ \ \ 
\int dy \ \G_{\rm gh}^{ab}(x,y;B) \ {\cal G}^{bc}(y,z;B) = \d(x-z) \d^{ac}.
}}
We substitute
\eqn\inverseghost{\eqalign{
{ {\p} \over {\p B_z} } {\cal G}_{xy}(B) 
= - {\cal G}_{xu}(B) { {\p \G_{{\rm gh},uv}(B)} \over {\p B_z} } 
 {\cal G}_{vy}(B). 
}}
into the previous DS equation, and multiply on the right by
the matrix $\G_{{\rm gh},yw}$ to obtain the functional DS equation for the
inverse ghost propagator 
\eqn\dsghost{\eqalign{
\G_{\rm gh}^{ab}(x,y;B) = & M^{ab}(B)\d(x-y)    \cr
& + g_0f^{adc} \int dz du \ 
{\cal D}_{\m\n}^{de}(x,z;B) \p_\m{\cal G}^{cf}(x,u;B)
{ {\d \G_{\rm gh}^{fb}(u,y;B)} \over {\d B_\n^e(z)} } , 
}}
that is represented diagrammatically in Fig.~2.
Here 
${ {\d \G_{\rm gh}^{fb}(u,y;B)} \over {\d B_\n^e(z)} }$
is the complete ghost-ghost-gluon vertex in the presence of the source.  

	We make the same changes of variable in the functional DS equation  \dsz\ for
$Z(J) = \exp[W(J)]$.  We evaluate\foot{In this section we write
$\nabla_\m^{ac}(A) = \p_\m \d^{ac} + g_0 f^{abc}A_\m^b$ for the gauge-covariant
derivative instead of $D_\m^{ac}(A)$ to avoid confusion with the
gluon propagator $\cal D$.}
\eqn\dsgamma{\eqalign{
\nabla_\l F_{\l\k}^a\Big( & { {\d} \over {\d J} } \Big)(x)  \ \exp[W(J)]
     \cr
& = \exp[W(J)] \Big[ \p_\l 
\Big( \p_\l  B_\k^a - \p_\k  B_\l^a 
+ g_0f^{abc}(B_\l^bB_\k^c + {\cal D}_{\l\k}^{bc}(x,x,B) \Big)   \cr
& \ \ \ \ \ \ \ \ \ \ \ \ \ \ \ \ \ \ \ \ \ \
+ g_0f^{abc} \Big( B_\l^b + { {\d} \over {\d J_\l^b} } \Big) 
g_0f^{cde} (B_\l^dB_\k^e + {\cal D}_{\l\k}^{cd}(x,x,B) \Big) \Big]  \cr
&  = \exp[W(J)] \Big[\nabla_\l F_{\l\k}^a(B) 
+ {\cal J}_{{\rm qu.gl,}\m}^b(x; B) \Big], 
}}
where
$B_\m = { {\d W(J)} \over {\d J_\m} }$ and
${\cal D}_{\m\n}(x,y;B) = { {\d B_\n(x)} \over {\d J_\m(y)} }$.
The quantum gluon current in the presence of the source~$B$ is defined by
\eqn\quantumgl{\eqalign{
{\cal J}_{{\rm qu.gl,}\k}^b(x; B)
\equiv & \ \Big( g_0f^{abc} \ (\d_{\l\m}\d_{\n\k} 
+ \d_{\l\k}\d_{\m\n} - 2\d_{\l\n}\d_{\k\m})
 \ \nabla_\l^{bd} {\cal D}_{\m\n}^{dc}(x,z;B)|_{z=x}  
- g_0^2f^{abc}f^{cde}      \cr
& \times \int dydzdw \ {\cal D}_{\l\r}^{bf}(x,y;B) 
 {\cal D}_{\l\s}^{dg}(x,z;B)
 {\cal D}_{\k\tau}^{eh}(x,w;B)
 \ \G_{\r\s\tau}^{fgh}(y,z,w;B)\Big)^{\rm tr}.
}}
Here
\eqn\tripleglue{\eqalign{
\G_{\r\s\tau}^{fgh}(y,z,w;B) \equiv 
{ { \d^3 \G(B) } \over {\d B_\r^f(y)\d B_\s^g(z)\d B_\tau^h(w)} }
}}
is the complete triple-gluon vertex in the presence of the source $B$.
This gives the functional DS equation for $\G(B)$,
\eqn\dsgamma{\eqalign{
{ {\d \G(B)} \over {\d B_\m^a(x)} } 
= & - \nabla_\l F_{\l\m}^a(B)(x)
- {\cal J}_{{\rm qu.gh,}\m}^b(x; B) 
- {\cal J}_{{\rm qu.gl,}\m}^b(x; B),
}}
where, by \meanghost,
\eqn\quantumgh{\eqalign{
{\cal J}_{{\rm qu.gh,}\m}^b(x; B)
& \equiv {\cal J}_{{\rm qu.gh,}\m}^b(x; J(B))   \cr
 & = - g_0f^{abc} \int d^4y \ P_{\m\l}^{\rm tr}(x-y) \  
\p_\l {\cal G}^{ca}(y, z; B)|_{z=y}
}}
is the quantum ghost current in the presence of the
source $B$.  

	A more explicit form of this equation, is obtained by differentiating with
respect to
$B_\tau^g(u)$, which yields a functional DS equation for the inverse gluon
propagator,
\eqn\dsglue{\eqalign{
{ {\d^2 \G(B)} \over {\d B_\k^a(x) \d B_\tau^g(u)} } =
& \Big( - \d_{\k\tau} (\nabla_\l \nabla_\l)^{ag} 
+ (\nabla_\k \nabla_\tau)^{ag} 
 - 2g_0f^{acg} F_{\k\tau}^c\Big)^{\rm tr} \d(x-u)   \cr
	& + ({\rm ghost \ loop}) + ({\rm 1 \ gluon \ loop}) + ({\rm tadpole}) 
+ ({\rm 2 \ gluon \ loops}) 
}}
where
\eqn\ghostloop{\eqalign{ 
({\rm ghost \ loop}) \equiv - g_0f^{abc} \Big(
\int dydz \  \p_\k {\cal G}^{bd}(x,y;B)
 \  {\cal G}^{ce}(x,z;B)
 \ { { \d \G_{\rm gh}^{de}(y,z; B) } \over {\d B_\tau^g(u)} } \Big)^{\rm tr},
}}
\eqn\gluonloop{\eqalign{ 
({\rm 1 \ gluon \ loop}) \equiv  
\int dydz &\ \Big( g_0f^{abc} \ 
(\d_{\l\m} \d_{\k\n} + \d_{\k\l} \d_{\m\n} - 2 \d_{\l\n} \d_{\k\m})   \cr
& +  \ \nabla_\l^{bd}  {\cal D}_{\m\r}^{de}(x,y;B)
 \  {\cal D}_{\n\s}^{cf}(x,z;B)    \Big)^{\rm tr}   
\G_{\r\s\tau}^{efg}(y,z,u;B) ,  
}}
\eqn\tadpole{\eqalign{ 
({\rm tadpole}) \equiv  
 &\  g_0^2f^{abc} f^{bgd}\ 
\Big( \ (\d_{\l\m} \d_{\k\n} + \d_{\k\l} \d_{\m\n} - 2 \d_{\l\n} \d_{\k\m})   
 \ \d(x-u)    \  {\cal D}_{\m\n}^{dc}(x,x;B)  \  \Big)^{\rm tr},  
}}
where superscript ``tr" means projection onto transverse parts on 
$(x, \k)$ and $(u,\tau)$.  The complete ghost-ghost-gluon vertex in
the presence of the source $B$, 
${ { \d \G_{\rm gh}^{de}(y,z; B) } \over {\d B_\tau^g(u)} }$,
reappears in \ghostloop, and the complete triple-gluon vertex
$\G_{\r\s\tau}^{efg}(y,z,u;B)$ in the presence of the source $B$ 
is defined in \tripleglue.
We do not write out explicitly the two-loop term, but all terms
are expressed graphically in Fig.~3.  

	The pair of equations \dsghost\ and \dsglue\ are a complete system of
functional DS equations for the quantum effective action $\G(B)$, and for the
inverse ghost propagator $\G_{\rm gh}^{ab}(x,y;B)$.  These functional
equations are converted to equations for the coefficient functions by
differentiating an arbitrary number of times with respect to $B$, and then
setting $B = 0$.

\newsec{Horizon condition and renormalization}

Solutions are subject to the {\it supplementary conditions} that both the gluon
and ghost inverse propagators
${ {\d^2 \G(B)} \over {\d B_\k^a(x) \d B_\l^b(y)} }$
and $\G_{\rm gh}^{ab}(x,y;B)$ be positive matrices.  Another supplementary
condition results from the fact, discussed in Appendix A, that in a space of
high-dimension, entropy favors a high concentration of population very near
the boundary $\p \Omega$ of the bounded region $\Omega$.  The boundary occurs
where the lowest non-trivial eigenvalue of the Faddeev-Popov operator $M(B)$
vanishes. Thus, for typical configurations $B$, the positive operator
$M(B)$ has a very small eigenvalue and, in fact, it has a
high density of eigenvalues $\r(\l, B)$ at $\l = 0$, 
per unit Euclidean volume $V$, as
compared to the Laplacian operator~\horizcon.  This makes the
ghost propagator, 
$G(x-y) \d^{ab} = \langle (M^{-1})^{ab}(x,y; B) \rangle$, 
long range, so in momentum space it is enhanced at $p = 0$ compared to
$1/p^2$, $\lim_{p \to 0}[p^2 \tilde{G}(p)]^{-1} = 0$,~\gribov,~\horizcon.
This property will provide a non-perturbative formula for the ghost-propagator
renormalization constant~$\tilde{Z}_3$ that moreover is consistent with the
perturbative renormalization-group.

	The gluon and ghost propagators, with source $B = 0$, are given
in momentum space by 
\eqn\momprop{\eqalign{ 
D_{\m\n}(x) & = (2\pi)^{-4} \int d^4k \ \tilde{D}_{\m\n}(k) \ \exp(ik\cdot x)
 \   \cr
G(x)  & = (2\pi)^{-4} \int d^4p \ \tilde{G}(p) \exp(ip \cdot x),
}}
and the
ghost-gluon vertex by
\eqn\momvertex{\eqalign{ 
f^{abc} \ \G_\m(x-y,y-z) & \equiv
{ {\d \G_{\rm gh}^{ac}(x,z;B)} \over {\d B_\m^b(y)} }|_{B = 0}  \cr
& = f^{abc} \ (2\pi)^{-8} \int d^4p \ d^4q \ \tilde{\G}_\m(p,q).
}} 
The DS equation for the ghost propagator $\tilde{G}(p)$, obtained
from \dsghost\ by setting $B = 0$, reads  
\eqn\dsghostp{\eqalign{ 
\tilde{G}^{-1}(p) = p^2 - N g_0 \  ip_\m \ (2\pi)^{-4} \
\int d^4k \ \tilde{D}_{\m\n}(k) \ \tilde{G}(p-k) \ \tilde{\G}_\n(p-k,p).
}}
All quantities are unrenormalized, and we have used
$f^{abc} f^{cde} = N \d^{ae}$ for SU(N).  

	Factorization of the external ghost momentum is a well-known special property
of the Landau gauge that makes it less divergent than other gauges.  To make
it explicit, we note that the ghost-ghost-gluon vertex $\tilde{\G}_\m(p,q)$
is a function of two linearly independent 4-vectors.  It is also transverse,
$(p-q)_\m\tilde{\G}_\m(p,q) = 0$, because the transversality condition is
imposed on-shell, so it may be written  
\eqn\transvertex{\eqalign{ 
\tilde{\G}_\m(p,q) = -ig_0 \ P_{\m\n}^{\rm tr}(k) \ p_\n \ V(p^2,k^2, q^2), 
}}
where $k \equiv q-p $.  The scalar vertex function,
$V(p^2,k^2, q^2)$ is symmetric $V(p^2,k^2,q^2) = V(q^2,k^2, p^2)$
in consequence of the symmetry 
${\cal G}^{ac}(x,z;B) = {\cal G}^{ca}(z,x;B)$.  The DS equation
for the ghost propagator reads,  
\eqn\factorext{\eqalign{ 
\tilde{G}^{-1}(p) = p^2 - N g_0^2 \ p_\m p_\n \ (2\pi)^{-4} \
\int d^4k \ \tilde{D}_{\m\n}(k) \ \tilde{G}(p-k) \ V((p-k)^2, \ k^2, \ p^2),
}}
where the factorization of the two external ghost momenta
$p_\m$ and $p_\n$ is now explicit. 

	This equation is divergent and must be renormalized.  In perturbative
renormalization theory, quantities renormalize
according to
\eqn\renormconst{\eqalign{ 
D_{\m\n}  = Z_3 D_{R,\m\n}; \ \ \ \ \ \ \   G  = \tilde{Z}_3 G_R;      
\ \ \ \ \ \ \ V  = \tilde{Z}_1^{-1} V_R; 
\ \ \ \ \ \ \ \ \   
g_0   = \tilde{Z}_1(\tilde{Z}_3 Z_3^{1/2})^{-1} g_R,
}}
and in Landau gauge the additional special
property
\eqn\landau{\eqalign{ 
   \tilde{Z}_1  = 1; \ \ \ \ \ \ \ \ \ \ \ \    V  = V_R;
\ \ \ \ \ \ \ \ \ \ \ \   g_0   = (\tilde{Z}_3 Z_3^{1/2})^{-1} g_R    
}}
holds.  In terms of renormalized quantities, the DS equation for the
ghost propagator reads,  
\eqn\renormdsgh{\eqalign{ 
\tilde{G}_R^{-1}(p) = p^2 \tilde{Z}_3 - N g_R^2 \ p_\m p_\n \ (2\pi)^{-4} \
\int d^4k \ \tilde{D}_{R,\m\n}(k) \ \tilde{G}_R(p-k) 
\ V_R((p-k)^2, \ k^2, \ p^2).
}}

	To avoid infrared difficulties, the ghost propagator is usually
renormalized at some finite renormalization mass $\m$.  However the horizon
condition, $\lim_{p^2\to 0} [p^2G(p)]^{-1} = 0$,
allows us to renormalize at $p = 0$.  It tells us that in the DS equation
\renormdsgh, the first term, $p^2\tilde{Z_3}$,  must be cancelled by the term
of order $p^2$ in the second term.  This gives a renormalization condition at
$p = 0$,  in the form of an equation for $\tilde{Z}_3$,   
\eqn\zeethree{\eqalign{ 
\tilde{Z}_3 = N g_R^2 \  (2\pi)^{-4} \
\int_{|k| < \L} d^4k \ \hat{p}_\m \hat{p}_\n \tilde{D}_{R,\m\n}(k) \
\tilde{G}_R(k) 
\ V_R(k^2 \ k^2, \ 0), 
}}
where $\L$ is an ultraviolet cut-off.  We have set $p = 0$ in the integrand,
and the integral is independent of the direction~$\hat{p}$.  This 
statement of the horizon condition shows that it is flagrantly
non-perturbative because, in perturbation theory, the left hand side is of
order~1, but the right hand side is of leading order~$g_R^2$.  

	The last equation gives the renormalization-group flow,
\eqn\renormgr{\eqalign{ 
\L { { \p \tilde{Z}_3} \over {\p \L} } & = 
N g_R^2 \ (p_\m p_\n/p^2) \ (2\pi)^{-4} \  \L^4
 \ \tilde{D}_R(\L) \ \tilde{G}_R(\L) \ V_R(\L^2,\L^2, 0)   \cr
 & \ \ \ \ \ \ \ \ \ \ \ \ \ \ \ \ \ \ \ \ \ \ \ 
\times \int d^3 \hat{k} \ (\d_{\m\n} - \hat{k}_\m\hat{k}_\n)    \cr
& = N g_R^2 (4\pi)^{-2} (3/2)
 \ \L^4 \ \tilde{D}_R(\L) \ \tilde{G}_R(\L) \ V_R(\L^2,\L^2, 0).
}}
As a check, we
note that if we take the tree values 
$\tilde{D}_R(\L) = \tilde{G}_R(\L) = 1/\L^2$, and
$V(p^2,k^2, q^2) = 1$, we obtain
\eqn\prenormgr{\eqalign{ 
\L { { \p \tilde{Z}_3} \over {\p \L} } = (4\pi)^{-2} (3/2)N g_0^2 
+ O(g_0^4).
}}
The term of order $g_0^2$ is scheme-independent, and agrees with the standard
one-loop expression in Landau gauge.  Thus the horizon condition provides a
normalization condition for the ghost propagator at $p = 0$ that is in
flagrant disagreement with perturbation theory, but nevertheless satisfies the
perturbative renormalization-group flow equation.  

	We substitute \zeethree\ into
the DS equation \renormdsgh\ for the ghost propagator, and obtain
\eqn\renormgh{\eqalign{ 
\tilde{G}_R^{-1}(p) = N g_R^2 \ p_\m p_\n \ & (2\pi)^{-4} \
 \int d^4k \  \tilde{D}_{R,\m\n}(k) \cr
& \times [\tilde{G}_R(k) \ V_R(k^2,k^2,0) 
- \tilde{G}_R(p-k) \ V_R((p-k)^2,k^2, p^2)]. 
}} 
This is a finite, renormalized DS equation for the ghost propagator.
It is invariant under the renormalization group in the sense that it
is form-invariant under the transformation \renormconst\ and \landau\ 
of perturbative renormalization theory in Landau gauge. This equation, from
which the tree term $k^2$ has been eliminated by the horizon condition, 
gives the ghost propagator an infrared anomalous dimension $a_G$, so it behaves
likes $G(k) \sim (\m^2)^{a_G}/(k^2)^{1+a_G}$ in the infrared.  This puts QCD
into a non-perturbative phase.

\newsec{Exact infrared asymptotic limit of QCD}

	Recent solutions of the truncated coupled DS equations for the gluon and ghost
propagators yield ghost propagators that are enhanced in the infrared, and
gluon propagators that are infrared suppressed
 \smekal, \atkinsona, \atkinsonb, \lerche, \dznonpland, \fischalkrein,
\fischalk, and \dztimeind.  Typical values for the infrared asymptotic form
of the gluon and ghost propagators \lerche\ and \dznonpland\ are,
\eqn\propvalues{\eqalign{ 
D^{\rm as}(k) & = 
\m^{2a_D}/(k^2)^{1+a_D} \approx (k^2)^{0.187}/(\m^2)^{1.187} \cr
G^{\rm as}(k) & = 
\m^{2a_D}/(k^2)^{1+a_G} \approx (\m^2)^{0.595}/(k^2)^{1.595}, }}
$a_G = (93-\sqrt1201)/98 \approx 0.595$, $a_D = -2a_G$,
where $a_D$ and $a_G$ are the infrared critical exponents of the ghost and
gluon. The gluon propagator $\tilde{D}(k)$ is
so strongly suppressed at $k = 0$ that it vanishes $\tilde{D}(0) = 0$.
With $D(x-y) = \langle A(x) A(y)\rangle$, this corresponds to suppression of
the low-frequency modes of $A(x)$ in the functional integral.  The actual
values of the infrared critical exponents do not depend too strongly on the
truncation scheme~\lerche.  The salient infrared features are
easily understood.  The cut-off of the functional integral at the Gribov
horizon is implemented in the DS equations by the horizon condition.  It
states that the ghost propagator $G(k)$ is enhanced in the infrared
or, equivalently, that the infrared critical exponent of the ghost positive,
$a_G > 0$.  The DS equations yield $a_D = - 2a_G$, so enhancement of the ghost
causes suppression of the gluon in the infrared.  This is the expression in
the DS equations of the proximity of the Gribov horizon in infrared directions.

	The results of calculation with the DS equations are in at least qualitative
agreement with numerical evaluations of gluon and ghost propagators
\suman, \leinweber, \cucchierigh, \actmdzgh, \bonneta, \langfeld, 
which, on sufficiently large lattices, yield a gluon propagator $D(k)$ that
turns over and decreases as $k$ decreases 
\acthrdlan, \acfkppl, \acfkppa, \cucchieria, 
\bonnetb, \nakajima, \bogolubsky, \czfitgrib\ 
(possibly extrapolating to $D(0) = 0$ at infinite lattice volume),
with a turn-over point $k_{max}$ that scales like a physical
mass~\attilioctmat. The only explanation for this counter-intuitive turn-over
is the strong suppression of infrared components by the proximity of the Gribov
horizon in infrared directions.  The agreement of DS and numerical
calculations gives us confidence that we have a reliable picture of the gluon
and ghost propagators including, in particular, in the infrared region.  

	One may use the above expressions for the asymptotic propagators to estimate
the convergence and magnitude of the various terms on the right hand side of
the DS equations, simply by counting powers of momentum.  The dominant
terms in the infrared region are the ones that contain the most ghost
propators $G(k)$ in the loop integrals.  The infrared limit of the 
truncated DS equations are found to have the following remarkable properties:

	(i)  {\it The infrared limit of the DS equations decouples from
the degrees of freedom associated with finite momentum and is free of
ultraviolet divergences.}  Technically, what is found is that when the
external momenta $k_e$ are small compared to $\L_{\rm QCD}$, then the internal
loop momenta $k_i$ scale like the $k_e$, and the contribution when the 
$k_i$ are large compared to $k_e$ may be neglected.  As a result,
when the $k_e$ are small, one may replace the propagators and vertices in
internal loops by their infrared asymptotic forms
$G^{\rm as}(k)$ and $D^{\rm as}(k)$ etc.  The loop contributions that
are dominant in the infrared are the ones that contain the most ghost
propagators $\tilde{G}(k)$.\foot{For the ghost-propagator equation \dsghostp\
or \renormgh,  both terms on the right-hand side are dominant, and both
originate from the action $- {\rm Tr} \ln M(B)$. The gluon-propagator
equation~\dsglue, with source $B = 0$, reads
\eqn\sdgluon{\eqalign{
\tilde{D}_{\m \n}^{-1}(k) = \ & (\d_{\m \n}k^2 - k_\m k_\n)  
   + {\rm (gluon \  loops)} \cr
& + N g^2 (2 \pi)^{-d} \int d^d p \ \ \tilde{G}(p+k) (p+k)_\m 
  \ \tilde{G}(p) \ \Gamma_\n(p,p+k) . 
}}
The tree term, of order $k^2$, is subdominanant in the infrared compared to
$(D^{\rm as})^{-1}(k) \sim (k^2)^{-0.187}$.
The dominant term on the right-hand
side is the ghost loop that originates from the action 
$- {\rm Tr} \ln M(B)$, whereas the subdominant terms --- namely the tree
term, the gluon loop and the two-loop term --- all
originate from the Yang-Mills action $S_{\rm YM}(B)$.}
The asymptotic infrared limit of the DS equations is highly convergent in the
ultraviolet because $G^{\rm as}(k)$ is strongly suppressed there.  In fact the
asymptotic gluon equation, given below, is finite without renormalization, and
the asymptotic ghost equation is finite with the renormalization
\renormgh.  We conclude that the DS equations possess an infrared asymptotic
limit that is well-defined, and decoupled from propagators and vertices at
finite momentum.       

(ii)  {\it The terms that are dominant in the
infrared limit come from the action $- {\rm Tr} \ln M(B)$, whereas the
subdominant terms come from Yang-Mills action $S_{\rm YM}(B)$.}  It is
instructive to classify terms that are dominant or subdominant on the right
hand side of the DS equations according as they originate with the action,
$- {\rm Tr} \ln M(B)$, or with the Yang-Mills action, $S_{\rm YM}(B)$.
Because the ghost propagator is enhanced in the infrared while the gluon
propoagator is suppressed, one finds that all subdominant terms and only
the subdominant terms disappear if one sets 
$S_{\rm YM}(B) = 0$ in the derivation of the DS equations given in secs.~5
and~6.

	Because the solutions of the truncated DS equations are consistent with
numerical evaluations of the gluon and ghost propagators, the effects of
truncation should not be too drastic.  We therefore expect that
properties~(i) and~(ii) of the truncated DS equations hold also for the
solutions of the exact, untruncated, DS equations, that is, that
{\it there exists an exact infrared asymptotic limit
of the DS equations that is obtained by setting} 
$S_{\rm YM}(B) = 0$. This implies that the cut-off at the Gribov horizon 
suffices to make the functional integral over $B$ converge, even though
$\exp[-S_{\rm YM}(B)]$ is replaced by~1. 

	We now write in functional form the exact infrared asymptotic DS equations
(without truncation!) that are obtained by setting $S_{\rm YM}(B) = 0$. 
We designate the generating functionals where the coefficient functions are
given their asymptotic forms by~$\hat{\G}(B)$ and $\hat{\G}_{\rm gh}(x,y;B)$. 
The infrared asymptotic gluon and ghost propagators are designated 
$(\hat{{\cal D}})_{xy}^{-1}(B) 
= { { \p^2 \hat{\G}(B) } \over { \p B_x \p B_y} }$
and
$(\hat{{\cal G}})^{-1}_{xy}(B) = \hat{\G}_{{\rm gh},xy}(B)$.  
The functional DS equation \dsghost\ for the
ghost propagator is unchanged in form, as represented in Fig.~2,
\eqn\asdsghost{\eqalign{
\hat{\G}_{\rm gh}^{ab}(x,y;B) = 
& (-\p^2 \d^{ab} - g_0f^{acb}B_\m^c \p_\m) \ \d(x-y)    \cr
& + g_0 f^{adc} \int dz du \ 
\hat{\cal D}_{\m\n}^{de}(x,z;B) \p_\m\hat{{\cal G}}^{cf}(x,u;B)
{ {\d \hat{\G}_{\rm gh}^{fb}(u,y;B)} \over {\d B_\n^e(z)} }.
}}
In the infrared asymptotic limit, only the ghost loop contributes to the
functional DS equation for the gluon propagator
\dsglue, 
which reads
\eqn\asdsglue{\eqalign{
(\hat{\cal D}^{-1})_{\m\n}^{ag}(x,y;B) =
 - g_0f^{abc} \Big(\int dzdu \  \p_\m \hat{{\cal G}}^{bd}(x,u;B)
 \  \hat{{\cal G}}^{ce}(x,z;B) \
{ { \d \hat{\G}_{\rm gh}^{de}(u,z; B) } \over {\d B_\n^g(y)} }
\Big)^{\rm tr},   
}}
and is diagrammed in Fig.~4.  

	An enormous simplification is apparent here, because {\it the last equation
allows an exact elimination} of the asymptotic functional gluon propagator 
$\hat{\cal D}_{\m\n}^{ab}(x,y;B)$.  The {\it one remaining unknown} is the
inverse ghost propagator  $\hat{\G}_{{\rm gh},xy}(B)$.

	When supplemented by the horizon condition \zeethree, eqs.~\asdsghost\ and
\asdsglue\ are a finite system when expressed in terms of renormalized
quantities.  Indeed, with
$\tilde{Z}_1 = 1$, renormalization of the exact functional asymptotic equations
is accomplished by writing
\eqn\asrenorm{\eqalign{ 
B & = Z_3^{1/2} B_R;  \ \ \ \ \ \ \ \ \ \ \ \ \ \ \   
\hat{\G}(B) = \hat{\G}_R(B_R); \ \ \ \ \ \ \ \ \ \ \ \ 
\hat{\G}_{\rm gh}(x,y; B) 
= \tilde{Z}_3^{-1}\hat{\G}_{{\rm gh},R}(x,y; B_R);      \cr  
g_0 & = (\tilde{Z}_3 Z_3^{1/2})^{-1} g_R;      
\ \ \ \ \ \ \ 
\hat{{\cal G}}(B) = \tilde{Z}_3 \hat{{\cal G}}_R(B_R);
\ \ \ \ \ \ \ \ \ 
\hat{{\cal D}}_{\m\n}(x,y;B)  
  = Z_3 \hat{{\cal D}}_{R,\m\n}(x,y;B_R).
}}
Upon making these substitutions, the functional equation for the ghost
propagator reads,
\eqn\rasdsghost{\eqalign{
\hat{\G}_{{\rm gh},R}^{ab}(x,y;B_R) = 
& \ (-\p^2 \d^{ab} \tilde{Z}_3 - g_Rf^{acb}B_{R,\m}^c \p_\m) \ \d(x-y)    \cr
& + g_R f^{adc} \int dz du \ 
\hat{{\cal D}}_{R,\m\n}^{de}(x,z;B_R) \p_\m\hat{{\cal G}}_R^{cf}(x,u;B_R)
{ {\d \hat{\G}_{{\rm gh},R}^{fb}(u,y;B_R)} \over {\d B_{R,\n}^e(z)} } , 
}}
where $\tilde{Z}_3$ is given in \zeethree, and the renormalized infrared
asymptotic functional gluon propagator is given by
\eqn\rasdsglue{\eqalign{
(\hat{\cal D}_R^{-1})_{\m\n}^{ag}(x,y;B_R) =
- g_Rf^{abc} \Big(\int dzdu \  \p_\m \hat{{\cal G}}_R^{bd}(x,u;B_R)
 \  \hat{{\cal G}}_R^{ce}(x,z;B_R) 
{ { \d \hat{\G}_{R,{\rm gh}}^{de}(u,z; B_R) } \over {\d B_{R,\n}^g(y)} }
\Big)^{\rm tr}.    
}}
When \rasdsghost\ is expanded
in a functional power series in~$B$, 
$\tilde{Z}_3$ appears only in the equation of order $(B)^0$
that determines the ghost propagator with source B = 0.  This equation is
finite as in the preceding section.  All the higher order equations are
independent of $\tilde{Z}_3$ and finite.  Equations \rasdsghost\ and
\rasdsglue\ are a complete system of functional DS equations, diagrammed in
Figs.~4 and~2, that are free of divergences, and that define the asymptotic
infrared theory.  The gluon propagator may be eliminated exactly from
\rasdsglue, and the asymptotic infrared theory is defined by the functional
inverse ghost propagator $\hat{\G}_{{\rm gh},R}(x,y; B_R)$.

	With $\tilde{Z}_3$ given in \zeethree, these equations are invariant under
the finite renormalization-group transformations
\eqn\finrenorm{\eqalign{ 
B_R & = z_3^{1/2} B'_R;  \ \ \ \ \ \ \ \  \ \ \ \ \
\hat{\G}_R(B_R) = \hat{\G}'_R(B'_R); \ \ \ \ \ \ \ \ \ 
\hat{\G}_{{\rm gh},R}(x,y; B_R)
= \tilde{z}_3^{-1}\hat{\G}'_{{\rm gh},R}(x,y; B_R);      \cr  
g_R & = (\tilde{z}_3 z_3^{1/2})^{-1} g'_R;      
\ \ \ \ \ \   
\hat{{\cal G}}_R(B_R) = \tilde{z}_3 \hat{\cal G}'_R(B'_R);
\ \ \ \ \ \ \  \hat{\cal D}_{\m\n}(x,y;B_R)  
  = z_3 \hat{\cal D}'_{R,\m\n}(x,y;B'_R).
}}
The quantity $g_R^2 D_R(k) G_R^2(k) = g_0^2 D(k) G^2(k)$ is invariant
under the renormalization \renormconst\ and \landau. Consequently a
scheme-independent running coupling constant, 
characteristic of the Landau gauge, may be defined  \smekal\ by
$\a_{\rm land}(k) \equiv (4\pi)^{-1}g_0^2 D(k) G^2(k) (k^2)^3.$  
The asymptotic infrared theory is characterized, in addition to the infrared
critical exponents $a_G$ and $a_D$, by $\a_{\rm land}(0) \approx 8.915/N$, for
color SU(N) \lerche.\foot{A scheme-independent running
coupling constant may be defined in the Coulomb gauge~\rengrcoul\ by, 
$\a_{\rm coul}(k) = (4\pi)^{-1}[12N/(11N - 2N_f)] k^2 \tilde{V}(k)$,
with $N_f$ quarks in the fundamental representation,
where $\tilde{V}(|\vec{k}|) \equiv g_0^2\lim_{k_4 \to \infty}
D_{44}(\vec{k},k_4)$. By contrast with $\a_{\rm land}(k)$ that is finite at $k
= 0$,  it appears that $\a_{\rm coul}(k)$ diverges like $1/k^2$ at small $k$,
in a realization of infrared slavery that features a string tension,
$V(r) \sim \s_{\rm coul}r$ at large $r$~\czcoulscen\ and~\greencoul.}

	The limit, in which the Yang-Mills action $S_{\rm YM}(B)$ is systematically
neglected, is a continuum analog of the lattice strong-coupling limit. 
Indeed if one rescales the gauge connection by the change of variable
$A' \equiv g_0 A$, the effective action, from which the DS equations were
derived,  reads
\eqn\strongcouple{\eqalign{ 
\Sigma(A)  & = -  {\rm Tr} \ln M(A) + S_{\rm YM}(A)   \cr
 & = -  {\rm Tr} \ln M'(A') + (g_0^2)^{-1}S'_{\rm YM}(A'),  
}}
where $M'(A')$ and $S'_{\rm YM}(A')$ are independent of $g_0$.
Neglect of $S_{\rm YM}(A)$ is the same as setting $g_0^{-2} = 0$ or, after
renormalization, $g_R^{-2} = 0$.
The asymptotic infrared limit is described by the effective action 
\eqn\aseffectact{\eqalign{ 
\hat{\Sigma} = -  {\rm Tr} \ln M(A).
}}

	If one extends the non-perturbative
formulation to a BRST-invariant theory, as outlined in Appendix B, the
BRST-invariant local action reads
\eqn\totalaction{\eqalign{ 
S = \int d^4x \ [s(\p_\m\bar{c} A_\m) + S_{\rm YM}(A)] ,
}}
where the BRST operator acts according to
\eqn\bsrtacts{\eqalign{ 
sA_\m = D_\m c; \ \ \ \ \ \ \  sc = - c^2;  
\ \ \ \ \ \ \  s\bar{c} = \l;  \ \ \ \ \ \ \  s\l = 0.
}}
The asymptotic infrared limit is described by the local BRST-invariant action  
\eqn\exactaction{\eqalign{ 
\hat{S} \equiv \int d^4x \ s(\p_\m\bar{c} A_\m) 
= \int d^4x \ (- \ \p_\m\bar{c} \ D_\m c + \p_\m\l \ A_\m),  
}}  
and the infrared asymptotic correlators satisfy the Slavnov-Taylor identities.

\newsec{Mass gap}
	The action $\hat{S}$ that describes the infrared asymptotic theory
is not only BRST-invariant, it is BRST-exact, $\hat{S} = sX$, and  
defines a topological quantum field theory.  To see what its properties
may be, recall that $\hat{S}$ describes the asymptotic infrared limit,
in which external momenta $k$ were small compared to $\L_{\rm QCD}$, so it
is the limit $\L_{\rm QCD} \to \infty$.  If QCD is a theory with a mass gap of
order $\L_{\rm QCD}$, then physical correlation lengths should vanish in the
asymptotic theory, $R \sim \L_{\rm QCD}^{-1} \to 0$.  

	To show this, consider a gauge-invariant correlator, for example
\eqn\physcor{\eqalign{
C(x) & = \langle F_x^2(A) \ F_0^2(A)\rangle \cr
	& = N\int_\Omega dA dc d\bar{c} d\l \ F_x^2(A) \ F_0^2(A) \ \exp(- \hat{S}),
}}
with $x \neq 0$, where Lorentz indices are suppressed 
$F^2(x) \to F_{\k\l}^a(x)F_{\m\n}^a(x)$, and the connected part is
understood.  Since the action is topological, we may make any transformation
that commutes with $s$, without changing expectation values.  As an example,
consider the change of variable corresponding to
a coordinate transformation $x'_\m = x'_\m(x)$ of 
$A$ and $c$, leaving $\bar{c}$ and $\l$ unchanged,
\eqn\coordtrans{\eqalign{
A'_\m(x') = { {\p x^\l} \over {\p x'^\m} }  A_\l(x); \ \ \ \
c'(x') = c(x); \ \ \ \ 
\bar{c}'(x) = \bar{c}(x); \ \ \ \ 
\l'(x) = \l(x).
}}
(The result is the same if $\bar{c}$ and $\l$ are also transformed.)
The infinitesimal form of this change of variable, with 
$x'^\m = x^\m - \x^\m(x)$, is given by
\eqn\rescale{\eqalign{
A_\m(x) & \to A'_\m(x) = A_\m(x) + \d A_\m(x) 
= A_\m(x) + \x^\l \p_\l A_\m(x) + \p_\m \x^\l A_\l(x)  \cr
c(x) & \to c'(x) = c(x) + \d c(x) =   c + c(x) + \x^\l \p_\l c(x)  \cr   
\bar{c}(x) & \to \bar{c}'(x) = \bar{c}(x); \ \ \ \ \ \ \ \ 
\l(x) \to \l'(x) = \l(x).
}}
Upon making this change of variable in the functional integral, we obtain
\eqn\rephyscor{\eqalign{
C(x) = \ N\int_{\Omega'} dA dc d\bar{c} d\l 
\ F_x^2(A') \ F_0^2(A') \ \exp[- \hat{S}(A',c',\bar{c}, \l)], 
}}
where $A' \equiv A + \d A$, and
\eqn\exactacta{\eqalign{ 
\hat{S}(A',c',\bar{c}, \l)  
= \int d^4x \ (- \ \p_\m\bar{c} \ D_\m(A') c' + \p_\m\l \ A'_\m).  
}}
The integration in $A$-space is cut-off at the Gribov horizon $\p \Omega'$
corresponding to $M(A')$.  Integration over the ghost fields gives $\det
M(A')$ which vanishes on the boundary $\p \Omega'$.  One may
change the cut-off to the Gribov horizon $\p\Omega$ corresponding to $M(A)$
because the error is only of order $\x^2$.  Moreover $F_x^2(A')F_0^2(A')$ is
the coordinate transform of 
$F_x^2(A)F_0^2(A)$, which we write
as
\eqn\coordid{\eqalign{
F_x^2(A') \ F_0^2(A') = [1 + \x^\l \p_\l + L(\p \x)]F_x^2(A) \ F_0^2(A),
}}
where $L(\p \x)$ is a numerical matrix that is linear in $\p_\l \x_\m$ and
acts on the tensorial indices of
$F_x^2(A) \ F_0^2(A)$, and we have
\eqn\recoordtrans{\eqalign{
C(x) = [1 + \x^\l \p_\l + L(\p \x)]
\ N\int_{\Omega} dA dc d\bar{c} d\l 
\ F_x^2(A) \ F_0^2(A) \ \exp[- \hat{S}(A',c',\bar{c}, \l)]. 
}}
One may verify that
$s$-operator commutes with the coordinate transformation,
$sA' = D(A')c'$, so
\eqn\exactactb{\eqalign{ 
\hat{S}(A',c',\bar{c}, \l)  
= \int d^4x \ s(\p_\m\bar{c} A'_\m) 
= \hat{S}(A,c,\bar{c}, \l) + s \d X,  
}}
where $\d X = \int d^4x \ \p_\m\bar{c} \ \d A_\m$.  Thus the variation of
$\hat{S}$ is also $s$-exact, and we have
\eqn\physcora{\eqalign{
C(x) & =  [1 + \x^\l \p_\l + L(\p \x)]
\ N\int_{\Omega} dA dc d\bar{c} d\l 
\ F_x^2(A) \ F_0^2(A) \ ( 1 - s \d X) \
 \exp[- \hat{S}(A,c,\bar{c}, \l)]       \cr
& = [1 + \x^\l \p_\l + L(\p \x)] \ 
\langle \ F_x^2(A) \ F_0^2(A) \ ( 1 - s \d X) \ \rangle .
}}
Gauge-invariant operators are $s$-invariant, 
\eqn\physcora{\eqalign{
\langle \ F_x^2(A) \ F_0^2(A) \ s \d X \ \rangle
= \langle \ s[F_x^2(A) \ F_0^2(A) \  \d X] \ \rangle = 0,
}}
which vanishes because it is the expectation-value of an $s$-exact
observable.  This gives
$C(x) = [1 + \x^\l \p_\l + L(\p \x)]C(x)$, 
so $C(x)$ is invariant under arbitrary coordinate
transformation.  Thus it is a number independent of $x$.  It vanishes for 
$x = \infty$, so $G(x) = 0$ for $x \neq 0$. 
The argument holds for a generic gauge-invariant correlator.  

	We have shown that the correlation length $R$ of gauge-invariant observables
vanishes in the gauge-invariant, physical sector of asymptotic theory defined
by $\hat{S}$.  In other words, the mass gap is infinite, $M = 1/R = \infty$, in
the physical sector of the asymptotic theory.  It is tempting to conclude from
this that there is a finite mass gap in the physical sector of the exact
non-asymptotic theory, for otherwise we would have obtained non-zero
correlators in the infrared limit.  However local gauge-invariant observables
like $F^2(x)$ are composite operators, and so far we have discussed only the
correlators of elementary fields.  To establish that the mass gap in the
non-asymptotic theory is finite, one should check that the correlators of
local gauge-invariant operators in the limit of large separation are also
given by the infrared asymptotic theory defined by~$\hat{S}$.

\newsec{Quarks}

	So far we have neglected quarks, but they may be included in the
time-independent Fokker-Planck equation \dztimeind.  The derivation of the
non-perturbative Faddeev-Popov formula, including quarks, proceeds as in
secs.~2 -- 4, by changing quark variables according to $\psi = g^{-1}\Psi$ and
$\bar{\psi} = \bar{\Psi}g$.  The result is that the quark action
$S_{\rm qu} = \int d^4x \ \bar{\psi} (\g_\m D_\m +  M) \psi$ 
gets added to the gluon action $\Sigma$ or $S$.  According to the latest DS
calculations that include $N_f = 3$ flavors of dynamical quarks, the
quark-loop term in the DS-equation for gluons is subdominant in the
infrared~\fischalkqu.  Provided that the effects of truncation are not too
drastic, the quark contribution will also be subdominant in the infrared limit
of the exact functional DS equation for the gluon propagator.  In this case the
inclusion of quarks does not disturb the simplicity of the gluon sector
described by $\hat{S}$.

	If the intrinsic mass of the quarks is finite, then the quark sector does not
appear in the asymptotic infrared limit.  If the instrinsic mass of the quarks
is zero, the pion is a massless Goldstone boson associated with spontaneous
breaking of chiral symmetry. However even in this case, in the (truncated)
DS equation for the quark propagator given in~\fischalkqu, the infrared limit
of the quark propagator does {\it not} decouple from the degrees of freedom
associated with finite momentum (in contrast to the gluon).  This is to be
expected because the parameters that characterize the dynamics of massless
quarks, 
$\langle \bar{\psi}\psi \rangle$ and $f_\pi$, are finite multiples of $\L_{\rm
QCD}$, but the infrared asymptotic limit corresponds to 
$\L_{\rm QCD} \to \infty$.  Nevertheless one may ask if chiral symmetry is
broken in the asymptotic infrared theory.  The chiral-symmetry breaking
parameter is given by 
$\langle \bar{\psi}\psi \rangle = \pi \langle \r(0, A) \rangle$, 
where $\r(\l, A)$ is density of eigenvalues~$\l$, per unit volume, of the
Dirac operator 
$i\g \cdot D(A)$ in the configuration~$A$.  In the infrared asymptotic limit,
the expectation-value $\langle \r(\l, A) \rangle$ is evaluated in the theory
defined by the action $\hat{S}$.  One would expect that it gives $\langle
\bar{\psi}\psi \rangle = \infty$, since this corresponds to 
$\L_{\rm QCD} = \infty$.  Thus in the theory defined by $\hat{S}$,
the average density of levels per unit volume 
$\langle \r(0, A) \rangle$ of the Dirac operator
$i\g \cdot D(A)$ should be infinite at $\l = 0$.  

	The infrared asymptotic theory is far simpler than full QCD and provides a
valuable model in which the characteristic features of the confining phase, as
described in the Landau gauge, are revealed.  To understand confinement in the
asymptotic theory, note that while the infrared components of $A(x)$ are
severely suppressed by the cut-off at the Gribov horizon, its
short-wave-length components fluctuate wildly because the factor 
$\exp[-S_{\rm YM}(A)]$ is replaced by~1.  Indeed, the infrared asymptotic gluon
propagator 
$D^{\rm as}(k)$, eq.~\propvalues, is strongly enhanced in the ultraviolet. This
suggests a picture of confinement in the infrared asymptotic theory in which
the short-wave-length fluctuations of $A^a(x)$ in color directions cause the
decoherence of any field that carries a color charge.  Indeed transport of a
color vector $q(\tau)$ along a path $z_\m(\tau)$, is described by
$P\exp(g_0 \int A_\m \dot{z}^\m dt)$.
In a highly random field $A_\m^b(x)$, superposition of different paths is
incoherent, so a field that bears a color charge does not propagate.  In full
QCD in Landau gauge, the dominant flucuations of $A(x)$ responsible for
confinement should be on the length scale~$\L_{\rm QCD}^{-1}$.  This picture
of confinement is quite different from the scenario in Coulomb gauge, where
confinement of color charge is attributed to a realization of infrared slavery
by an instantaneous, long-range color-Coulomb
potential~\rcoulomb,~\czcoulscen\ and~\greencoul.

\newsec{Conclusion}

	We briefly review the salient features of the non-perturbative continuum
Euclidean formulation of QCD developed here.  

	(i)~In Landau gauge one may integrate the Faddeev-Popov
weight over the Gribov region $\Omega$ instead of over the fundamental modular
region $\L$.  

	(ii)~The {\it form} of the Dyson-Schwinger equations  
is unchanged by the cut-off of the functional integral on the boundary
$\p\Omega$ of the Gribov region, because the Faddeev-Popov determinent vanishes
there.  This simplicity makes the DS equations the method of choice for
non-perturbative calculations in QCD.  

  (iii)  The restriction to the Gribov region provides {\it supplementary
conditions} that govern the choice of solution of the DS equations.  Two
conditions are the positivity of the gluon and ghost propagators.  Another is
the {\it horizon condition} which is the statement that the ghost propagator
$G(k)$ is more singular than
$1/k^2$ in the infrared, 
$\lim_{k \to 0} [k^2 G(k)]^{-1} = 0$.  This fixes the
ghost-propagator renormalization constant $\tilde{Z}_3$ to the value
\zeethree.  Although~\zeethree\ is in flagrant disagreement with the
perturbative expression for $\tilde{Z}_3$, nevertheless it is consistent with
the perturbative renormalization group.  

	(iv) Implementation of the horizon condition in the DS equations 
puts QCD into a non-perturbative phase.

	(v) Recent solutions of the truncated DS equations possess an asymptotic
infrared limit that is obtained by systematically neglecting the terms in the
DS equations that come from the Yang-Mills action 
$S_{\rm YM}(A)$, but keeping the Faddeev-Popov determinant and the cut-off at
the Gribov horizon.  If the effects of truncation are not too drastic, this
also gives an exact asymptotic infrared limit of QCD that is a continuum
analog of the strong-coupling limit in lattice gauge theory.  This is possible
because convergence of the $A$-integration without the Yang-Mills factor
$\exp[-S_{\rm YM}(A)]$ may be assured by the cut-off at the Gribov horizon.

	(vi) The asymptotic infrared limit of QCD is defined by the functional DS
equations \rasdsghost\ and \rasdsglue.  The gluon propagator may be eliminated
exactly from \rasdsglue, and the asymptotic infrared theory is completely
characterized by the functional inverse ghost propagator~$\hat{\G}_{\rm
gh}(x,y; B)$.

	(vii) There exists a local BRST-invariant extension of the present
non-perturbative formulation, sketched out in Appendix~B.  This ensures that
the Slavnov-Taylor identities hold in the non-perturbative theory.  The
asymptotic infrared limit of QCD, valid at distances large compared to
$1/\L_{\rm QCD}$, is described by the BRST-exact action,
$\hat{S} =  \int d^4x \ s \ (\p_\m \bar{c}A_\m)$, 
that defines a topological quantum field theory with an infinite mass gap.  

(viii) The extension of the non-perturbative formulation to include the
quark action $\int d^4x \ \bar{\psi} (\g_\m D_\m + m) \psi$ is immediate.
The presence of quarks does not disturb the asymptotic infrared limit
of the gluon sector.

(ix)  The asymptotic infrared theory provides a simple model in the Landau
gauge in which the characteristic features of confinement may be understood.
A picture of confinement of color charge emerges, in which the highly
random fluctuations of the gluon field $A$ cause the superposition from the
transport of color charge along different paths to interfere incoherently, so
the fields that bear a color charge do not propagate.

\vskip .5cm
{\centerline{\bf Acknowledgments}}

It is a pleasure to thank Reinhard Alkofer, Laurent Baulieu, Alexander
Rutenburg, S.~R.~S.~Varadhan, and Lorenz von Smekal for valuable discussions
and correspondence.  This research was partially supported by the National
Science Foundation under grant PHY-0099393.

\appendix A{Resolution of paradox}

		At first sight it is surprising that expectation-values taken over the
fundamental modular region $\L$ and the Gribov region $\Omega$ are equal.  In
this Appendix we show how this paradox is resolved.

\subsec{Argument of Semenov-Tyan-Shanskii and Franke}

	The proof by	Semenov-Tyan-Shanskii and Franke \semenov\  that the Gribov
region~$\Omega$ and the fundamental modular region~$\L$ are different,
substantiated by instances are given in~\gfdadzinside, was 
long considered to disprove~\expectval.  We review the
argument of~\semenov.  Let
$g(t) = \exp(t \o)$ be a one-parameter subgroup of the local gauge group with
generator $\o = \o(x)$.  To be definite, we normalize $\o$ to $(\o, \o) = V$,
where $V$ is the Euclidean volume.  Let 
$A_\m(t, \o, B) \equiv g(t)^{-1} B_\m g(t) + g(t)^{-1} \p_\m g(t)$, 
be the gauge-transform of $B_\m$ under
$g(t) = \exp(t\o)$, so $A(0, \o, B) = B$,
and let $F_B(t, \o)$ be the Hilbert square norm of
$A(t, \o, B)$, regarded as a function of $t$ and $\o$ for fixed $B$,
\eqn\hilbert{\eqalign{ 
	F_B(t, \o) = ||A(t, \o, B)||^2 = \int d^4x \ |A_\m(t, \o, B)|^2.
}}
The fundamental modular region $\L$ is the set of $B$ such that $F_B(0, \o)$
is an absolute minimum, $F_B(0, \o) \leq F_B(t, \o)$ for all $\o$ and $t$.  The
Gribov region
$\Omega$ is the set of $B$ for which $F_B(0, \o)$ is a relative minimum
$F_B(0, \o) \leq F_B(t, \o)$ for all~$\o$ and sufficiently small~$t$.

	We differentiate $F_B(t, \o)$ with respect to $t$, and use 
$A_\m' = D_\m(A)\o \equiv D_\m\o$,  
\eqn\derivatives{\eqalign{ 
F'_B(t, \o) & = 2 \ (D_\m\o, A_\m) = 2 (\p_\m \o, A_\m)  
= - 2 (\o, \p_\m A_\m)    \cr
F''_B(t, \o) & = 2 \ (\p_\m \o, D_\m\o)
=  - 2 \ (\o, \p_\m D_\m\o)  \cr
F'''_B(t, \o) & = 2 \ (\p_\m \o, D_\m\o \times \o)   \cr
F''''_B(t, \o) & = 2 \ (\p_\m \o, (D_\m\o \times \o) \times \o),
}}
where $X \times Y = [X, Y]$ is the commutator in the Lie algebra.
These formulas show that the interior of $\Omega$ consists of all transverse
configurations $B$, $\p \cdot B = 0$, such that all non-trivial eigenvalues
of $M(B) = - \p_\m D_\m(B)$ are strictly positive, 
$\l_n(B) > 0$.  Moreover for $B$ on the boundary
$\p\Omega$, $M(B)$ has at least one non-trivial eigenvalue that vanishes,
$\l_1(B) = 0$.

We specialize to the SU(2) group, so the commutator $X \times Y$ is
the ordinary 3-vector cross product.  The vector triple product gives
\eqn\fourthder{\eqalign{
F''''_B(t, \o) & = 2 \ (\p_\m \o,  \ (\o \cdot D_\m\o \  \o - \o^2 \ D_\m\o) \
)  \cr & = 2 \ (\p_\m \o, \ \o \cdot \p_\m \o \ \o) 
+ 2 \ (\o, \p_\m(\o^2 D_\mu \o))    \cr
& = 2(\o \cdot \p_\m \o, \o \cdot \p_\m \o)  
+ 2(\o,  \p_\mu( \o^2) D_\m \o) + 2(\o, \o^2 \p_\m D_\m \o)  \cr
& = (1/2)( \p_\m(\o^2) , \p_\m(\o^2) ) 
+ 2(\o \cdot \p_\m \o, \p_\m (\o^2) ) + 2(\o^2, \o \cdot \p_\m D_\m \o)  \cr
& = (3/2) ( \p_\m(\o^2) , \p_\m(\o^2) )+ 2(\o^2, \o \cdot \p_\m D_\m \o) , 
}}
where the dot is contraction on color indices.

		Let $B$ be a point on
the Gribov horizon~$\p \Omega$, so $B$ is transverse $\p_\mu B_\mu = 0$,
and the Faddeev-Popov operator $-\p_\mu D_\mu(B)$ is non-negative, but with at
least one non-trivial null eigenvalue,
$\p_\mu D_\mu(B) \o_0 = 0$, for some $\o_0$.  By~\derivatives, we have
$F'_B(0, \o_0) = F''_B(0, \o_0) = 0$.  For $B$ on $\p \Omega$, it follows
that in general $F_B(0, \o)$
is {\it not} a local minimum on the gauge orbit through $B$ because, in
general, $F'''_B(0, \o_0) \neq 0,$ so $F_B(t, \o_0) - F_B(0, \o_0)$
changes sign at $t = 0$.  By continuity
this implies that nearby points inside the Gribov region $\Omega$ cannot
be absolute minima, even though they are relative minima.  They are Gribov
copies inside $\Omega$.  This is the argument of~\semenov, and examples for
which $F'''_B(0, \o_0) \neq 0,$ are given in~\gfdadzinside. 

	But let's evaluate the 4th derivative at $t = 0$, in the direction $\o_0$.
With $\p_\m D_\m(B) \o_0 = 0$, we have from \fourthder,
\eqn\evfourth{\eqalign{
F''''_B(0, \o_0) = (3/2) \int d^4x \ [\p_\m(\o_0^2)]^2.
}}
This is the integral of a positive density, and we expect
that $F''''_B(0, \o_0)$ is {\it large and positive}.

	The relevant question for comparing the expectation values over
$\Omega$ and over $\Lambda$ is not whether these regions coincide --- they do
not --- but whether the normalized averages over these sets are equal in the
thermodynamic limit. Here we implicitly suppose a lattice discretization, and
configurations that are sampled from the Wilson ensemble.  In the
thermodynamic limit, the probability may get concentrated on a subset that
consists of a boundary or part of a boundary.  The boundaries of~$\L$
and~$\Omega$ may approach each other in the thermodynamic limit for {\it
typical} configurations on the boundary.  If $F''''_B(0, \o_0)$ is large, and 
$F'''_B(0, \o_0)$ is small, then there is a local minimum near 
$B$, which could be the absolute minimum on the gauge orbit.  If the distance
to the absolute minimum vanishes in the thermodynamic limit for a typical
configuration, then the argument of \semenov\ does not disprove~\expectval.

	We normalize $\o_0$ to $(\o_0, \o_0) = V$, where $V$ is the volume of
Euclidean space.  We estimate quantities using this normalization, and
we shall verify that the conclusions do not depend on the normalization of
$\o_0$. With this normalization, we estimate that $\o_0(x) = O(1)$.  Since 
$F''''_B(0,\o_0)$ is the integral of a positive local density over a volume
$V$, we estimate that
$F''''_B(0, \o_0) = O(V)$, for a typical configuration $B$ on the Gribov
horizon.  On the other hand, the density that appears in $F'''_B(0, \o_0)$ has
no definite sign.  For a typical configuration, sampled from the Wilson
ensemble, we make the crudest statistical estimate namely random density, so
$F'''_B(0, \o_0) = O(V^{1/2})$.  This is small compared to
$F''''_B(0, \o_0)$.  We seek a nearby minimum on the gauge orbit through~$B$. 
For simplicity we assume that all non-trivial eigenvalues of $M(B)$ are
strictly positive, apart from the zero eigenvalue belonging to $\o_0$, which
is the only dangerous direction.  We write 
$F(t) \equiv F_B(t, \o_0)$, and we have
\eqn\expandf{\eqalign{
F(t) = F(0) + (1/3!)F'''(0)t^3 + (1/4!)F''''(0)t^4,
}}
with neglect of higher order
terms.  The minimum is found at $F'(t_{\rm cr}) = 0$, which gives
$t_{\rm cr} = - 3 F'''(0)/F''''(0)$, and one has,
\eqn\fcrit{\eqalign{
F(t_{\rm cr}) = F(0) - (9/8){ {[F'''(0)]^4} \over {[F''''(0)]^3} }.
}}
This is lower than $F(0)$, in agreement with the argument of \semenov.
This expression is independent of the normalization of $\o_0$, as one sees from
\derivatives, so our estimate for this quantity is independent of the
normalization of $\o_0$. By the above estimates, the second term is of
order $(V^{1/2})^4/(V)^3 = V^{-1}$.  It is small compared to the first term,
$F(0) = ||B||^2$, which is of order $V$. The configuration at the nearby
minimum is
\eqn\evfourth{\eqalign{
B_\m(x,t_{\rm cr}) & = B_\m(x) + t_{\rm cr}[D_\m(B)\o_0](x)   \cr
& = B_\m(x) - \ { {3F'''(0)} \over {F''''(0)} } \  [D_\m(B)\o_0](x),
}}
which is again independent of the normalization of $\o_0$.  According to the
above estimates, the second term is of order $V^{-1/2}$.  Thus in the
thermodynamic limit of lattice gauge theory, $V \to \infty$, the nearby minimum
approaches the point $B$ on the Gribov horizon.  In actuality, 
the problem of minimizing the functional $F_A(g) = ||{^g}A||$ on the lattice
is a problem of spin-glass type, so one expects many, nearly degenerate,
relative minima, and the one found here is not necessarily the absolute
minimum.  Nevertheless the point remains that~\semenov\ does not disprove the
equality of expectation-values on $\L$ and $\Omega$ in the thermodynamic limit.
  
\subsec{Many Gribov copies inside the Gribov region from numerical simulations}

	We now consider the fact that in numerical gauge-fixing to
Landau gauge in lattice gauge theory, there are
many local minima (i.e.Gribov copies inside the Gribov region,~$\Omega$),
on a typical gauge orbit,
\mandula, \marinari,
\deforcrand, \nakamuramiz, \marenzoni.  Their number grows with the lattice
size as is characteristic of a spin-glass.  In this
sense $\Omega$ is very large compared to~$\L$. However  the number of
dimensions of configuration space is high, and our geometrical intuition
from 3-space may be misleading.  Indeed, on a lattice of Euclidean volume $V$,
the dimension
$D$ of configuration space is $D = fV$, where $f$ is the number of degrees of
freedom per lattice site, and the dimension $D$ of configuration space diverges
with the Euclidean volume $V$.
	Ê

	In continuum gauge theory $\L$ and $\Omega$ are both convex and bounded in
every direction~\semenov. 
By simple entropy considerations, the population in a bounded
region of a high-dimensional space gets concentrated on the boundary.  For
example inside a sphere of radius $R$ in a $D$-dimensional space, the radial
density is given by $r^{D-1}dr$, and for $r \leq R$ is highly concentrated
near the boundary
$r = R$.  To take the simplest example, consider two spheres (in
configuration space), the first of radius $R$, and the second of radius 
$R + cV^{-1/2}$.  In the spirit of the previous estimates, these would be
the radii of $\L$ and $\Omega$.  The ratio of the radii $(R + cV^{-1/2})/R$
approaches unity, in the limit $V \to \infty$, so all $n$-th moments,
$\langle r^n \rangle$ for finite $n$, of the two spheres become equal. 
On the other hand the ratio of their volumes is given by 
$[(R + cV^{-1/2})/R]^{D} = [(R + cV^{-1/2})/R]^{fV},$  
where $D = fV$ is the dimension of configuration space.  
For large V the ratio of the volumes of the two spheres is thus
$\exp(afV^{1/2}/R),$ which diverges exponentially like $V^{1/2}$. 
In this example the ratio of the volumes of the two spheres diverges with V,
but all {\it finite} moments of the two spheres become equal!  In field theory
the $n$-th moments of the distribution are the $n$-point
functions $\langle A(x_1)...A(x_n)$.  So again, the fact that there are many
Gribov copies inside~$\Omega$, does not disprove that averages calculated
over~$\L$ or~$\Omega$ are equal.

\subsec{Gauge theory on a finite lattice}

	For a finite lattice the paradox becomes acute.  Stochastic quantization
may also be defined in lattice gauge theory \nakamuramiz.  As in the continuum
theory, a drift force $a^{-1}K_{\rm gt}$ tangent to the gauge orbit may be
chosen in the direction of steepest descent of a suitable minimizing function,
and is globally resoring.  It appears that one may solve the lattice
Fokker-Planck equation in the limit $a \to 0$ on a finite lattice, by the
method used in secs.~2 to~4, for it depends only on general geometrical
properties that are common to lattice and continuum gauge theories.
If so, one would again be led to the conclusion that the weight
inside the Gribov region is given by the lattice analog of
\boltzmann\ namely $N\exp[- S_{\rm W}(U)]$, where $S_W(U)$ 
is the Wilson action, and~$U$ is a configuration in the lattice Gribov
region~$\Omega$.  However on a finite lattice the distinction between the
fundamental modular region $\L$ and the Gribov region $\Omega$ can surely not
be ignored.  The resolution of this paradox would appear to be that in lattice
gauge theory the Gribov region
$\Omega$ is made of disconnected pieces $\Omega_i$.  In each piece, the
solution is indeed given by 
$Q_i(U) = N_i \exp[-S_{\rm W}(U)]$, for $U \in \Omega_i$, where the
normalizations $N_i$ are left indeterminate by the method of secs.~2 to~4. 
Presumably, the average with the lattice Faddeev-Popov weight over all the
disconnected pieces $\Omega_i$ of the Gribov region, with the correct
normalization $N_i$ in each piece,  will agree with with same integral over
the fundamental modular region~$\L$.

\appendix B{BRST-invariant formulation}

	New issues arise when the non-perturbative approach is extended to a
theory with a local BRST-invariant action.

\subsec{Off-shell transversality condition}

	To obtain a local action, on must take the transversality condition
``off-shell". The off-shell partition function is given by
\eqn\offshell{\eqalign{
Z(J,L) \equiv \int_\Omega DA D\l \ \det M(A) 
\ \exp[- S_{\rm YM}(A) + i(\l, \p \cdot A) + i(J,A) + i(L,\l)],
}}
where $\l$ is the Nakanishi-Lautrup Lagrange multiplier field
that enforces the gauge condition $\p \cdot A = 0$,
and $L$ is its source.  This reduces to \partfunct\ for $L = 0$.
It is not immediately obvious what region~$\Omega$ to integrate over because
$A$ is not transverse for $L \neq 0$, so $M(A) = - \p \cdot D(A)$ is not a
symmetric operator.  One must also take the Gribov horizon~$\p\Omega$ off
shell when the gauge condition is off-shell.  If we effect the $\l$
integration, the last integral becomes,
\eqn\outlambda{\eqalign{
Z(J,L) = \int_\Omega DA \ \det M(A) \ \d(\p \cdot A + L) 
\ \exp[- S_{\rm YM}(A) + i(J,A)].
}}
Only configurations $A$ of the form $A = B - \p (\p^2)^{-1}L$ are relevant,
where $B$ is transverse.   We regard the partition function~$Z(J,L)$ as
a formal power series in the source~$L$.  Both the lowest non-trivial
eigenvalue 
$\l_1[B - \p (\p^2)^{-1}L]$ of the Faddeev-Popov operator, 
$M[B - \p(\p^2)^{-1}L]$, and the points $B_0(L)$  where it vanishes, may be
calculated  by formal perturbation theory as a power series in~$L$.  
Here $B_0(0)$ is a point on the on-shell horizon.
In this way we may take the Gribov horizon $\p \Omega$ off-shell.   

\subsec{Faddeev-Popov ghosts}  

One may make the action local  
by  writing
\eqn\outlambda{\eqalign{
 \det M(B) = \int Dc D\bar{c} \ \exp(\bar{c}, M(B)c),
}}
where $c$ and $\bar{c}$ are anti-commuting ghost and anti-ghost fields.
Grassmannian sources, $\eta$ and $\bar{\eta}$,
are then introduced, so this gets replaced by
\eqn\grassmanns{\eqalign{
\int Dc D\bar{c} \ \exp[(\bar{c}, M(B)c) + (\bar{\eta},c) + (\bar{c}, \eta)]
= \det M(B) \exp (\bar{\eta}, M^{-1}(B) \eta) .
}}
This expression does not vanish on the boundary $\p \Omega$.  For by an
eigenfunction expansion of $M^{-1}(B)$, we obtain for the last
expression
\eqn\grassmanns{\eqalign{
\prod_n \l_n \ \exp\Big[\sum_n { {1} \over {\l_n} }\bar{\eta}_n \eta_n \Big]
= \prod_n \l_n \prod_n \Big(1 + { {1} \over {\l_n} }\bar{\eta}_n \eta_n \Big)
= \prod_n \Big(\l_n + \bar{\eta}_n \eta_n \Big).
}}
It does not contain as a factor $\l_1(B)$ that vanishes on $\p \Omega$. 
For this reason, we did not use Faddeev-Popov ghost fields and their
sources in the derivation of the DS equations in secs.~5 and~6.  
Nevertheless {\it we obtained the same DS equations, including the ghost
propagators, that we would have obtained if we had introduced the ghost fields
and their sources.}  For this reason, and by use of the off-shell 
Gribov horizon, it should be possible to extend the non-perturbative approach
to the theory defined by the familiar BRST-invariant local action
\totalaction, integrated over the off-shell Gribov region.

\appendix C{Properties of the Gribov region}

	We note three properties of the Gribov region~$\Omega$ defined in~\gomega.
(i)~$\Omega$ contains the origin $A = 0$.  (ii)
It is bounded in every direction.  (iii) It is convex.  We
give the one-line proofs of these properties \dzgribreg.  They follow from
the expression
$M(A) = M_0 + M_1(A)$, where $M_0^{ac}(A) = - \p^2 \d^{ac}$, and
$M_1^{ac}(A) = - g_0f^{abc}A_\m^b \p_\m$, 
where $A$ is transverse.  Property~(i) is obvious since $M_0 = -\p^2 \d^{ac}$
is strictly positive.   To establish~(ii), note that $M_1(A)$  has zero trace,
since it is traceless on color indices $f^{aba} = 0$.  Thus, for any
given~$A$, there exists a state~$\o$ for which the expectation value of
$M_1(A)$ is is negative, $E \equiv (\o, M_1(A) \o) < 0 $.  Moreover 
$M_1(A)$ is linear in~$A$,  $M_1(\l A) = \l M_1(A)$, so upon replacing
$A$ by $\l A$, where $\l$ is a positive number, we have
$(\o, M(\l A) \o) = (\o, M_0 \o) + \l (\o, M_1(A) \o)
= (\o, M_0 \o) + \l E$.  By taking $\l$ sufficiently large and positive,
the expectation value is negative $(\o, M(\l A) \o) < 0 $.  This
establishes~(ii).  To establish convexity, we must show that 
$M(\a A_1 + \b A_2)$ is a strictly positive operator when $M(A_1)$ and $M(A_2)$
are both strictly positive operators, for all positive $\a$ and $\b$, with $\a
+ \b = 1$.  This is immediate because $M_1(A)$ depends linearly on $A$, and
we have 
$M(\a A_1 + \b A_2) = \a M(A_1) + \b M(A_2)$.  QED

\footatend\vfill\supereject\immediate\closeout\rfile\writestoppt
\baselineskip=14pt\centerline{{\bf References}}\bigskip{\frenchspacing%
\parindent=20pt\escapechar=` \input refs.tmp\vfill\eject}\nonfrenchspacing

%%%%%%%%%%%%%%%%%%%% FIGURES %%%%%%%%%%%%%%%%%%%%%%%%

\vfill\eject\immediate\closeout\ffile{\parindent40pt
\baselineskip14pt\centerline{{\bf Figure Captions}}\nobreak\medskip
\escapechar=` \input figs.tmp\vfill\eject}

%%%%%%%%%%%%%%%%%%%%%%%%%%%%%%%%%%%%%%%%%%%%%%%%%%%%%%%%%%%%%%%%%%%

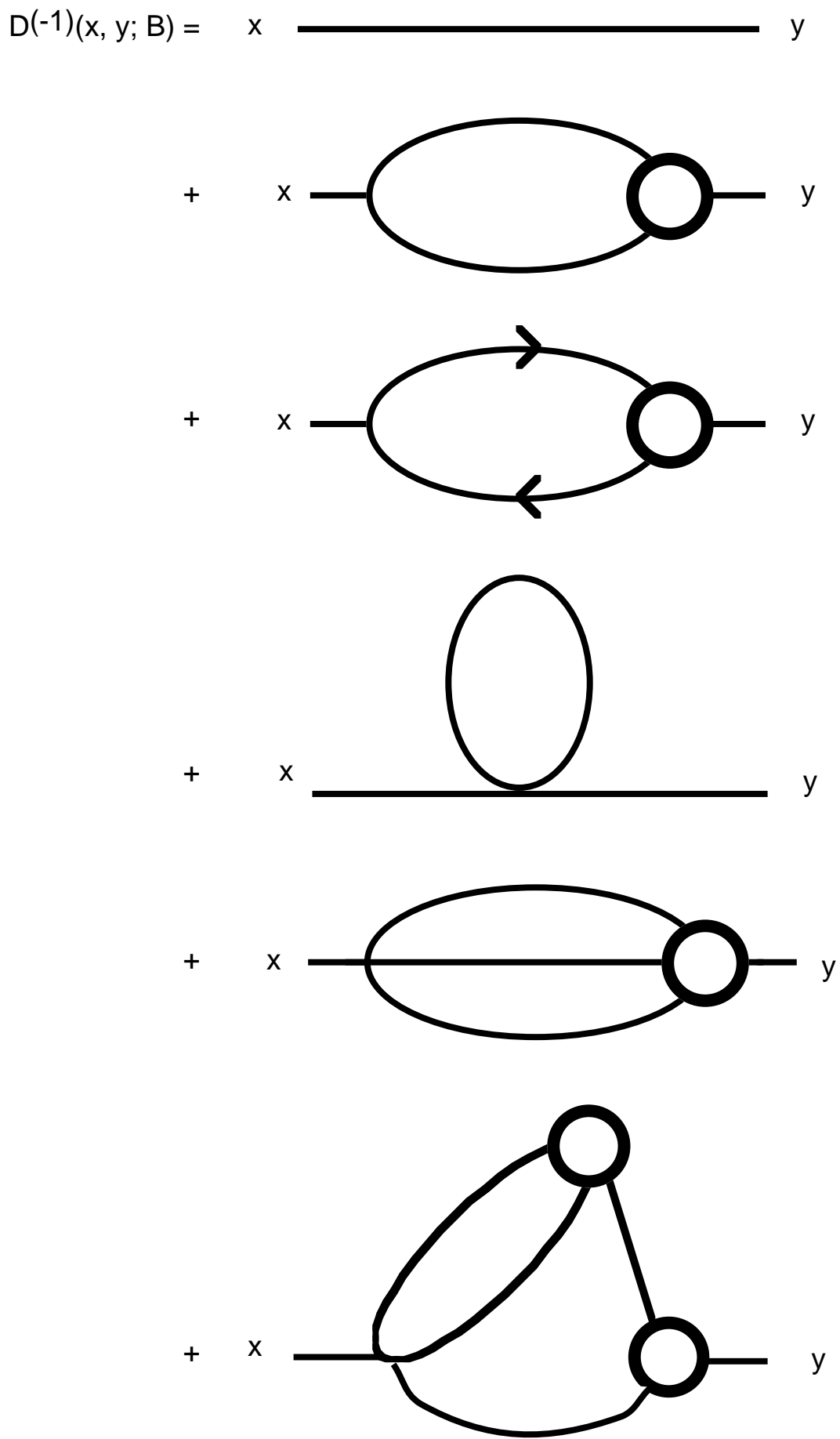
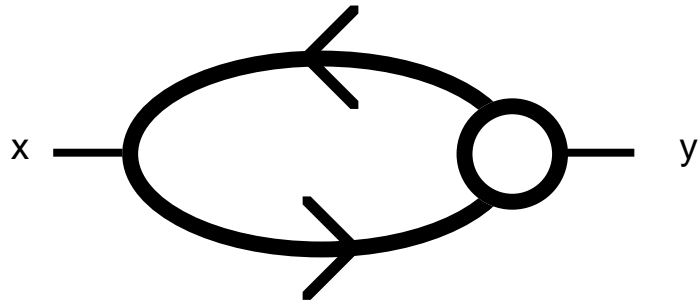
\bye